\documentclass[letter]{cas-dc}
\usepackage[numbers]{natbib}
\usepackage{graphicx} 
\usepackage{fancyhdr}            

\usepackage{amssymb} 
\usepackage{mathrsfs} 
\usepackage{amsthm} 
\usepackage{amsmath}
\usepackage{mathtools}
\usepackage{amsfonts}
\usepackage{eso-pic}  
\usepackage{graphpap}  
\usepackage[noend]{algpseudocode} 
\usepackage{algorithm}

\usepackage{hyperref}

\usepackage{siunitx}
\usepackage[version=3]{mhchem} 

\usepackage{rotating}

\usepackage{lettrine}
\usepackage{empheq}
\usepackage{lastpage}
\usepackage{fancyhdr}
\usepackage{color}
\usepackage{stmaryrd} 
\usepackage{caption}
\usepackage{tabularx}

\usepackage{arydshln} 

\DeclareSIUnit\Molar{\textsc{m}}
\DeclareSIUnit\calorie{cal}
\DeclareSIUnit\debye{D}
\graphicspath{{figs/}}
\newsavebox{\saveit}

\DeclareMathAlphabet{\mathpzc}{OT1}{pzc}{m}{it}

\newcommand{\red}[1]{\textcolor{red}{#1}}

\newcommand{\Assert}[1]{\State \textbf{assert} #1}
\makeatletter
\newcommand{\thickhline}{%
  \noalign{\ifnum0=`}\fi
  \hrule \@height 1.8\arrayrulewidth
  \futurelet\reserved@a\@xhline}
\makeatother
\graphicspath{{figs/}}
\DeclareMathOperator{\grad}{\nabla}

\DeclareMathOperator{\Rng}{Rng}
\begin{document}

\let\WriteBookmarks\relax
\def\floatpagepagefraction{1}
\def\textpagefraction{.001}
\shorttitle{Optimized Force Field for TBP}
\shortauthors{Hatami and de Almeida}
\title[mode = title]{Molecular Dynamics Force Field Genetic Optimization for Tri-n-butyl Phosphate Liquid}
%
%
\author[1]{Faranak Hatami}[
                           ]
%
\ead{fhatami@uic.edu}
\address[1]{University of Illinois Chicago, Dept. of Chemistry, Chicago, IL 60607, USA}
%
\author[2]{Valmor F. {de Almeida}}[
                        orcid=0000-0003-0899-695X,
                        linkedin=<valmor-f-de-almeida>
                        ]
\cormark[1]
\ead{dealmeidavf@cortix.tech}
\ead[url]{https://cortix.tech}
\address[2]{Cortix Tech, Lowell, MA 01854, USA}
%
\cortext[cor2]{Corresponding author}
%
%
%
%
\begin{abstract}
 An iterative optimization algorithm with MD simulations in the loop is developed and
 applied to optimize Lennard-Jones (LJ) parameters specific for liquid
 tri-n-butyl phosphate (TBP). The optimization loop uses non-dominated sorting
 genetic algorithms to obtain LJ parameters that reproduce key properties such
 as mass density, electric dipole moment, heat of vaporization, self-diffusion
 coefficient (SDC), and shear viscosity. Errors relative to
 experimentally measured properties lead to a multi-objective function optimization
 problem stated in terms of a Pareto-optimal set. A systematic application of the
 optimization algorithm to cases involving single- and multi-objective functions was
 carried out in this work, establishing a framework for atomistic TBP property
 predictions. We demonstrate the use of a neural network property model to amortize the
 high cost of MD simulations in the optimization loop and to allow for large populations
 and more generations to be used in the genetic algorithms. In our previous study of
 finding the best force field for TBP property predictions as judged by the
 aforementioned thermophysical properties, we found the Polarized AMBER-MNDO force field
 to be the best overall showing a \num{74}\% relative deviation from experimental values.
 However, in this study, we show optimized values of the LJ parameters that improve the
 overall deviation from experimental data to \num{23}\% when using the NN NSGA-III
 algorithm. Despite this large improvement, the accurate prediction of the transport
 properties, SDC and shear viscosity, remains difficult since improvements in one of them
 worsen the other, and vice versa. Finally, finding optimized LJ parameters
 that further minimize the error of all transport properties is still possible via
 additional computing power used with larger population of parameter vectors and more
 generations in the genetic algorithms employed here.
\end{abstract}
%

   

  

%
%
%
%
%
%
%
%
%

\maketitle

%
%


\hypersetup{
  pdftitle = { Cermet Nuclear Waste Forms: Phase Identification and Connectivity
             },
  pdfsubject = { Manuscript for Chemical Engineering Science
               },
  pdfauthor = { Valmor F. de Almeida },
  pdfkeywords = {pdflatex, hyperref}
            }
%

%
%
%
\newcommand{\I}{\RomanNumber{1}}
\newcommand{\II}{\RomanNumber{2}}
\newcommand{\III}{\RomanNumber{3}}
\newcommand{\Fmod}{\ensuremath{\mathcal{F}}}
\newcommand{\Emod}{\ensuremath{\mathcal{E}}}
\newcommand{\Pmod}{\ensuremath{\mathcal{P}}}
\newcommand{\Ttensormod}{\ensuremath{\pmb{\mathcal{T}}}}
\newcommand{\VectorSpace}{\ensuremath{\boldsymbol{V}}}
\newcommand{\nnmap}{\ensuremath{\boldsymbol{\mu}}}
\newcommand{\xvec}{\ensuremath{\boldsymbol{x}}}
\newcommand{\zvec}{\ensuremath{\boldsymbol{z}}}
\newcommand{\yvec}{\ensuremath{\boldsymbol{y}}}
\newcommand{\vvec}{\ensuremath{\boldsymbol{v}}}
\newcommand{\avec}{\ensuremath{\boldsymbol{a}}}
\newcommand{\uvec}{\ensuremath{\boldsymbol{u}}}
\newcommand{\wvec}{\ensuremath{\boldsymbol{w}}}
\newcommand{\fvec}{\ensuremath{\boldsymbol{f}}}
\newcommand{\Fvec}{\ensuremath{\boldsymbol{F}}}
\newcommand{\pvec}{\ensuremath{\boldsymbol{p}}}
\newcommand{\rvec}{\ensuremath{\boldsymbol{r}}}
\newcommand{\vvector}{\ensuremath{\boldsymbol{v}}}
\newcommand{\pvector}{\ensuremath{\boldsymbol{p}}}
\newcommand{\rvector}{\ensuremath{\boldsymbol{r}}}
\newcommand{\zerovec}{\ensuremath{\boldsymbol{0}}}
\newcommand{\deltaI}{\ensuremath{\delta\!I}}
\newcommand{\deltaIh}{\ensuremath{\deltaI^h}}
\newcommand{\Istar}{{\ensuremath{I^{*}}}}
\newcommand{\Istarh}{\ensuremath{{I^*}^h}}
\newcommand{\Istarhp}{\ensuremath{{I^*_{+}}^h}}
\newcommand{\Istarhm}{\ensuremath{{I^*_{-}}^h}}
\newcommand{\Ihm}{\ensuremath{I^h_{-}}}
\newcommand{\Ibnd}{\ensuremath{I_{b^-}}}
\newcommand{\GammaBar}{\ensuremath{\overline{\Gamma}}}
\newcommand{\wBar}{\ensuremath{\overline{w}}}
\newcommand{\velocity}{\ensuremath{\boldsymbol{v}}}
\newcommand{\ibasis}{\ensuremath{\boldsymbol{i}}}
\newcommand{\jbasis}{\ensuremath{\boldsymbol{j}}}
\newcommand{\kbasis}{\ensuremath{\boldsymbol{k}}}
\newcommand{\ebasis}{\ensuremath{\boldsymbol{e}}}
\newcommand{\angular}{\ensuremath{\boldsymbol{\omega}}}
\newcommand{\xpoint}{\ensuremath{\boldsymbol{x}}}
\newcommand{\ypoint}{\ensuremath{\boldsymbol{y}}}
\newcommand{\Xpoint}{\ensuremath{\boldsymbol{X}}}
\newcommand{\gpt}{\ensuremath{\boldsymbol{g}}}
\newcommand{\xposition}{\ensuremath{\boldsymbol{x}}}
\newcommand{\Ttensor}{\ensuremath{\boldsymbol{T}}}
\newcommand{\Dtensor}{\ensuremath{\boldsymbol{D}}}
\newcommand{\Itensor}{\ensuremath{\boldsymbol{I}}}
\newcommand{\Ptensor}{\ensuremath{\boldsymbol{P}}}
%
\newcommand{\Domain}{\ensuremath{\Omega}}
\newcommand{\normal}{\ensuremath{\boldsymbol{n}}}
\newcommand{\dvol}{\ensuremath{d\xpoint}}
\newcommand{\darea}{\ensuremath{d\!a}}
\newcommand{\Reals}{\ensuremath{\mathbb{R}}}
\newcommand{\Positives}{\ensuremath{\mathbb{P}}}
\newcommand{\abs}[1]{\bigl\lvert{#1}\bigr\rvert}
\newcommand{\absi}[1]{\lvert{#1}\rvert}
\newcommand{\Sspace}{\ensuremath{\mathcal{S}}}
\newcommand{\norm}[1]{\bigl\lVert{#1}\bigr\rVert}
\newcommand{\Lin}[2]{\mathcal{L}\bigl({#1},{#2}\bigr)}
\newcommand\Ltwo[1]{{\mathrm{L}}^{2}\!\left({#1}\right)}
\newcommand\Ltwomap[2]{{\mathrm{L}}^{2}\!\bigl({#1};{#2}\bigr)}
\newcommand\Ltwonorm[2]{{\norm{#1}}_{\Ltwo{#2}}}
\newcommand{\Ltwozero}[1]{{\mathrm{L}}_0^{2}\!\left({#1}\right)}
\newcommand{\Uset}{\ensuremath{{\mathrm{U}}}}
\newcommand{\Paretoset}{\ensuremath{{\mathscr{P}}}}
\newcommand{\Frontset}{\ensuremath{{\mathscr{F}}}}
\newcommand{\Zset}{\ensuremath{{\mathscr{Z}}}}
\newcommand{\Sset}{\ensuremath{{\mathscr{S}}}}
\newcommand{\Xset}{\ensuremath{{\mathscr{X}}}}
\newcommand{\Pset}{\ensuremath{{\mathscr{P}}}}
\newcommand{\Bset}{\ensuremath{{\mathscr{B}}}}
\newcommand{\Uh}{\ensuremath{\Uset^h}}
\newcommand{\Amtrx}{\ensuremath{\boldsymbol{\mathsf{A}}}}
\newcommand{\alphavector}{\ensuremath{\boldsymbol{\alpha}}}
\newcommand{\bvector}{\ensuremath{\boldsymbol{\mathsf{b}}}}
\newcommand{\Hilbone}[1]{{\mathrm{H}}^{1}\!\left({#1}\right)}
\newcommand{\Hilbonezero}[1]{{\mathrm{H}}_0^{1}\!\left({#1}\right)}
\newcommand{\Hilbnorm}[2]{{\norm{#1}}_{ {\mathrm{H}}^{1}\!\left({#2}\right)}}
\newcommand{\aform}[2]{\mathrm{a}\bigl(#1\,,\,#2\bigr)}
\newcommand{\aformbig}[2]{\mathrm{a}\bigl(#1\,,\,#2\bigr)}
\newcommand{\Ke}{\ensuremath{{\mathcal{K}_e}}}
\newcommand{\Kezero}{\ensuremath{{\mathcal{K}\!{\scriptstyle e}_{\scriptstyle 0}}}}
\newcommand{\Ne}{\ensuremath{{\mathcal{N}_e}}}
\newcommand{\Kf}{\ensuremath{{\mathcal{K}_f}}}
\newcommand{\Kd}{\ensuremath{{\mathcal{K}_d}}}
\newcommand{\Kj}{\ensuremath{{\mathcal{K}_j}}}
\newcommand{\Kbar}{\ensuremath{\overline{\mathcal{K}}}}
\newcommand{\Kebar}{\ensuremath{\overline{\K_e}}}
\newcommand{\Th}{\ensuremath{{\mathcal{T}}^{h}}}
\newcommand{\Xho}{\ensuremath{{\mathcal{X}}^{h}_{0}}}
\newcommand{\Pk}{\ensuremath{{\mathcal{P}}_{k}}}
\newcommand{\Pone}{\ensuremath{{\mathcal{P}}_{1}}}
\newcommand{\Pzero}{\ensuremath{{\mathcal{P}}_{0}}}
%

%



%
%
%
\section{Introduction}\label{sec:intro}
%
 Tri-\emph{n}-butyl phosphate (TBP) has been a pivotal component in hydrometallurgical
 solvent extraction for decades, playing a crucial role in nuclear materials separation
 processes such as PUREX-like extraction~\cite{bertelsen2022electrochemistry}. Despite its
 extensive history, TBP remains an important model in the study of interfacial transport
 processes. Widely employed in methods like extraction of uranium~\cite{chiarizia2003third,
 moyer2008proceedings, zilberman2001extraction}, plutonium~\cite{plaue2006small}, and
 zirconium~\cite{chiarizia2004extraction}, it has unique features that make it an economic
 solvent for separating metals, ions and radioactive compounds. Research on TBP provides
 valuable insights for modeling and simulation, where a detailed understanding of
 molecular interactions is critical for optimizing processes and devising more effective
 separation approaches. A significant challenge in the broader context of nuclear power
 generation is the effective disposal of radioactive waste. Aqueous reprocessing is one
 option for waste treatment wherein TBP plays a practical role in solvent extraction
 processes. Recent experimental studies~\cite{basu2013volumetric, billah2018densities,
 fang2008densities, tian2007densities, schulz1990science} measuring TBP properties offer
 calibration data for molecular models, while molecular dynamics (MD) simulations serve as
 an effective tool for understanding and visualizing the molecular behavior of
 extraction~\cite{ye-etal2013:art, ye-etal2010:art, ye-etal2009:art}.
\par
 In the pursuit of accurate and transferable molecular force fields for MD simulations,
 various computational studies~\cite{cui2012molecular, cui2014molecular, mu2016comparative,
 vo2018microscopic, vo2015computational, benay2014liquid, beudaert1998theoretical} have
 focused on determining thermodynamic and transport properties of TBP. The precision of
 the force field employed significantly influences the simulation outcomes. In our
 previous research~\cite{hatami-de_almeida25:art}, we conducted a comprehensive
 examination, exploring different force field models applied to pure TBP. The analysis
 calculated thermodynamic and transport properties, including mass density, heat of
 vaporization (HOV), electric dipole moment (EDM), self-diffusion coefficient (SDC),
 and shear viscosity. Additionally, we introduced
 polarizability~\cite{hatami-de_almeida25:art} into TBP models to enhance specific
 properties and models, and showed that this improvement comes at a computational cost and
 increased model complexity.  Furthermore, each model yields distinct results, each with
 its own set of advantages and disadvantages, in which there is no single TBP model at the
 moment that excels in describing all properties equally well.
\par
 For instance, for predicting the aforementioned thermodynamic
 properties~\cite{hatami-de_almeida25:art}, the Non-Polarized (NP) AMBER-DFT force field
 produced
 the best results (\SI{4}{\percent} overall deviation from experimental data), for
 predicting the transport properties, the Polarized (P) OPLS2005 force field produced best
 results (\SI{63}{\percent} overall deviation), and for all properties combined, the
 P AMBER-MNDO force field produced best results (\SI{74}{\percent} overall deviation).
\par
 Moreover, for each individual property~\cite{hatami-de_almeida25:art}, namely, mass
 density, HOV, EDM, shear viscosity, and SDC, the
 best force field was, respectively, NP OPLS2005-DFT (\SI{0}{\percent}),
 Polarized (P) AMBER-DFT (\SI{-2}{\percent}), NP OPLS-MNDO (\SI{-1}{\percent}), NP OpenFF
 (\SI{-36}{\percent}), and P OPLS2005 (\SI{-17}{\percent}); values in parenthesis are the
 relative deviation from the corresponding average experimental values.
\par
 From the foregoing, it is concluded that any attempt to improve the prediction of one
 specific property leads to the deterioration of the prediction of another property (or
 multiple properties). Hence, the current capability of open-source force fields for
 predicting transport properties of TBP is unsatisfactory. However there is much room for
 improvement since force fields have numerous parameters that can be optimized using
 domain-specific computational advances and strategies. This state of affairs carries over
 to other solvent extraction molecules of interest in solvent extraction processes in
 general, hence progress made in developing force field optimization for TBP could be
 applicable to other solvents.
\par
 \emph{
 This was the goal of this research, that is, develop a force field parameter optimization
 strategy for TBP that can be transferable to other solvent molecules. Here we use TBP as
 a molecular model because of its prevalence in the nuclear waste aqueous processing
 domain.}
\par
 The calibrated parameterization of force fields is computationally challenging,
 particularly for systems like TBP with numerous parameters. Initiatives such as
 ForceBalance and OpenFF~\cite{wang2013systematic, wang2014building, boothroyd2022open}
 aim to automate this process, but these tasks involve differentiation of functions with
 respect to force field parameters or system geometry, resulting in limited efficiency and
 poor scalability. While machine learning (ML) models such as
 BPNN~\cite{behler2007generalized}, Grappa~\cite{seute2025grappa},
 DeepPotential~\cite{zhang2018deep}, PhysNet~\cite{unke2019physnet}, and
 EANN ~\cite{zhang2019embedded}, have demonstrated success for homogeneous and hard
 material systems, they exhibit reduced reliability in organic, flexible molecular
 systems, due to lack of bonding topology information, insufficient treatment of
 long-range interactions and complexity of the molecular systems.  Consequently, our
 focus in this research is on optimizing covalent-bonded force fields for TBP using an
 intuitive approach, striving to achieve accurate predictions for various thermophysical
 properties compared to experimental values.
\par
 There exists different approaches for parameter set optimization, they typically involve
 either the use of a single objective function or multiple objective
 functions~\cite{faller1999automatic, wang2001automatic, bourasseau2003new,
 hulsmann2010grow, hulsmann2010assessment, deublein2013automated} whose critical point
 (either a maximum or a minimum) is at the sought after optimum parameter set. For the
 application at hand, objective functions for optimizing force field parameters naturally
 emerge as the error of computed thermophysical properties relative to experimental
 values. One or more relative error functions can be posed depending on one or more
 available experimental value for thermophysical properties.
\par
 While multiple objective functions can be combined into a single one, say by linear
 combination, it is more instructive to use the relative error objective function of
 individual thermophysical properties in a multi-objective optimization framework which
 simultaneously considers conflicting objective functions, providing a set of parameters
 that represent optimal compromises. The use of multi-objective functions optimization for
 TBP thermophysical properties provides additional insight into the trade-off between
 different parameter set candidates that could produce acceptable errors for all
 thermophysical properties involved.
\par
 In this work, we consider a multi-objective optimization method that involves identifying
 a set of optimal parameters, known as the Pareto-optimal set. Within this set, each one
 represents the best compromise, that is to say, improving one objective function comes at
 the cost of the deterioration of another. Because the objective functions for the LJ
 parameters depend on MD ensemble averages, employing optimization methods built on
 differentiation of the objective function with respect to parameters, is impractical, if
 not impossible. Therefore, we employ differentiation-free, genetic algorithms for the
 calculation of a population of parameter sets, combined with searching and ranking of
 parameter solutions. This combination is known as non-dominated search genetic
 algorithms, NSGA-II~\cite{deb-etal2002:art} and NSGA-III~\cite{deb-jain2014:art,
 jain-deb2014:art} (the latter was developed for a larger number of multi-objective
 functions).
\par
 Optimizing parameters through the NSGA-II, and NSGA-III algorithms involves running a
 molecular dynamics simulation for each evaluation of the objective function at a new
 parameter set. This will result in a high computational cost. To mitigate this cost, we
 use a neural network fit of previously obtained thermophysical properties as a function
 of LJ parameters. The use of the multi-dimensional data fit allows for evaluating
 objective functions at a much lower cost than an MD simulation. However this does not
 eliminate the need for new MD simulations once the fit falls outside the parameter space
 of training. Therefore a hybrid approach is proposed where the neural network fit is an
 accelerator and enabler for large-generation genetic algorithm minimization.
\par
 This work investigates the improvement of Lennard-Jones (LJ) parameters of TBP
 force field models~\cite{hatami-de_almeida25:art}. These parameters control the
 intermolecular interactions of non-electrostatic origin. All other parameters are kept
 immutable so we
 develop a progressive understanding of how LJ parameters affect thermophysical properties.
 Since LJ parameters differ for various models we have studied, namely, AMBER, OPLS,
 OPLS1005, GAFF, and OpenFF, this work attempts to find optimized LJ parameters that
 improve the prediction of thermophysical properties. In this analysis, we investigate the
 sensitivity of thermophysical properties of TBP with respect to individual LJ parameters.
 Section~\ref{sec:oc} describes the multi-objective optimization components: mathematical
 problem statement, iterative workflow, force
 field parameters to be optimized, molecular dynamics simulation core, objective
 functions, and genetic algorithms. Section~\ref{sec:rd} describes the results obtained
 and analysis made highlighting the benefits of the employed optimization technique and
 optimization of the LJ parameters. Finally, section~\ref{sec:co} summarizes our findings,
 emphasizing the potential of multi-objective optimization in refining atomistic force
 fields and advancing our understanding of how to predict TBP fluid properties.
%

%
%
%
\section{Optimization components} \label{sec:oc}
 This section outlines key components of our optimization framework, including the
 definition of multi-objective functions, the optimization problem statement, solution
 existence and uniqueness, the employed optimization algorithms, the MD simulation
 protocol, and the integration of a neural network data fitting method as a cheaper
 alternative to a full MD simulation within the optimization loop.
%
\subsection{Mathematical problem statement}\label{subsec:mps}
 Let $N$ be the number of LJ parameters in any TBP force field model, and $\vvec \in \Positives^N$ a 
 vector of ordered, positive-valued LJ parameters (sec.~\ref{subsec:mds}). Let the 
 thermophysical property vector-valued function of interest be denoted 
 $\yvec:\vvec\in\Positives^N\rightarrow \pvec\in\Positives^M$ where the $M=5$ properties
 of interest in this work are: mass density, HOV, EDM, shear viscosity, and SDC,
 respectively. Given a set of LJ parameters, $\vvec$, a significant computational effort
 is required to compute the property vector $\pvec=\yvec(\vvec)$ via MD simulations
 (sec.~\ref{subsec:mds}). For concreteness, $p_1 =  \langle \rho \rangle_\text{NPT}$ for
 mass density, $p_2 = \bigl\langle \Delta H_\text{vap} \bigr\rangle_\text{NPT}$ for heat
 of vaporization, and so on for other properties \cite{hatami-de_almeida25:art}.
\par
 We denote the experimentally measured positive values of thermophysical properties, 
 $\pvec^\text{(exp)} \in \Positives^M$. Therefore there is great practical interest in the signed relative 
 error of a predicted thermophysical property, that is, 
 $E(p_k) := \frac{p_k - p_k^\text{(exp)}}{p_k^\text{(exp)}}, \ k=1,\ldots,M$.
 In particular, one is interested in finding the best vector of parameters, $\vvec^*$, that makes the 
 magnitude of the error the smallest possible for every property $k$.
\par
 To that end, we define the $k^\text{th}$ positive-valued thermophysical property objective function 
 $F_k:\vvec\in\Positives^N\rightarrow \Positives$ with rule 
 $F_k(\cdot) := E^2 \circ y_k(\cdot)$. Hence the objective function for each $k$ is the square of the
 error of the $k^\text{th}$ thermophysical property prediction relative to its experimental value,
 which gives
 \begin{equation}\label{eqn:obj}
     F_k(\vvec) = E^2\circ y_k(\vvec) = 
     \biggl(\frac{y_k(\vvec) - p_k^\text{(exp)}}{p_k^\text{(exp)}} \biggr)^2
     \quad \forall \quad k = 1,\ldots,M.
 \end{equation}
 for any vector $\vvec$. By design $\abs{E\bigl(y_k(\vvec)\bigr)} = \sqrt{F_k(\vvec)} $, thus
 minimizing any objective function associated to a thermophysical property, reduces the magnitude of the 
 corresponding signed relative error. Why not make the objective function equal to the magnitude of the 
 error itself has to do with differentiation approaches to find a minimum of the objective function;
 there is also a statistical justification for this choice when the error is normally distributed.
 It is more attractive to differentiate the quadratic form \eqref{eqn:obj} than
 differentiate an absolute value function, particularly when $y_k(\cdot)$ is linear in $\vvec$; definitely
 not the case here. This work does not use any differentiation approach, hence the magnitude of the error 
 could have been used directly as an objective function; nevertheless we used \eqref{eqn:obj} for legacy 
 reasons.
\par
 A vector-valued thermophysical objective function $\Fvec:\vvec\in\Positives^N\rightarrow\Positives^M$ is 
 naturally induced by \eqref{eqn:obj} where $\bigl(\Fvec\bigr)_k\equiv F_k$.
 From \eqref{eqn:obj} we can now formulate an optimization problem for minimizing a vector-valued objective 
 function (multi-objective function) as follows. Find the best LJ parameter vector 
 $\vvec^* \in \Positives^N$ such that
 \begin{equation}\label{eqn:minproblem1}
  F_k(\vvec^*) = \min_{\vvec\,\in\,\Positives^N} 
     \biggl(\frac{y_k(\vvec) - p_k^\text{(exp)}}{p_k^\text{(exp)}} \biggr)^2
     \quad \forall \quad k = 1,\ldots,M.
 \end{equation}
 This posed mathematical problem is notorious for having objective conflicts where the
 individual minimum of each objective function component $F_k$ is found at different
 points $\vvec^*_k$ in parameter space (sec.~\ref{subsec:single}); that is to say, for
 most realistic applications, the problem has no solution as stated. We hope for
 physically realistic force field potential models not to fall into this category but in
 practice most empirical force fields do \cite{hatami-de_almeida25:art}.
\par
\begin{figure}
 \begin{center}
  \graphicspath{{figs/}}
  \includegraphics[width=3.0in]{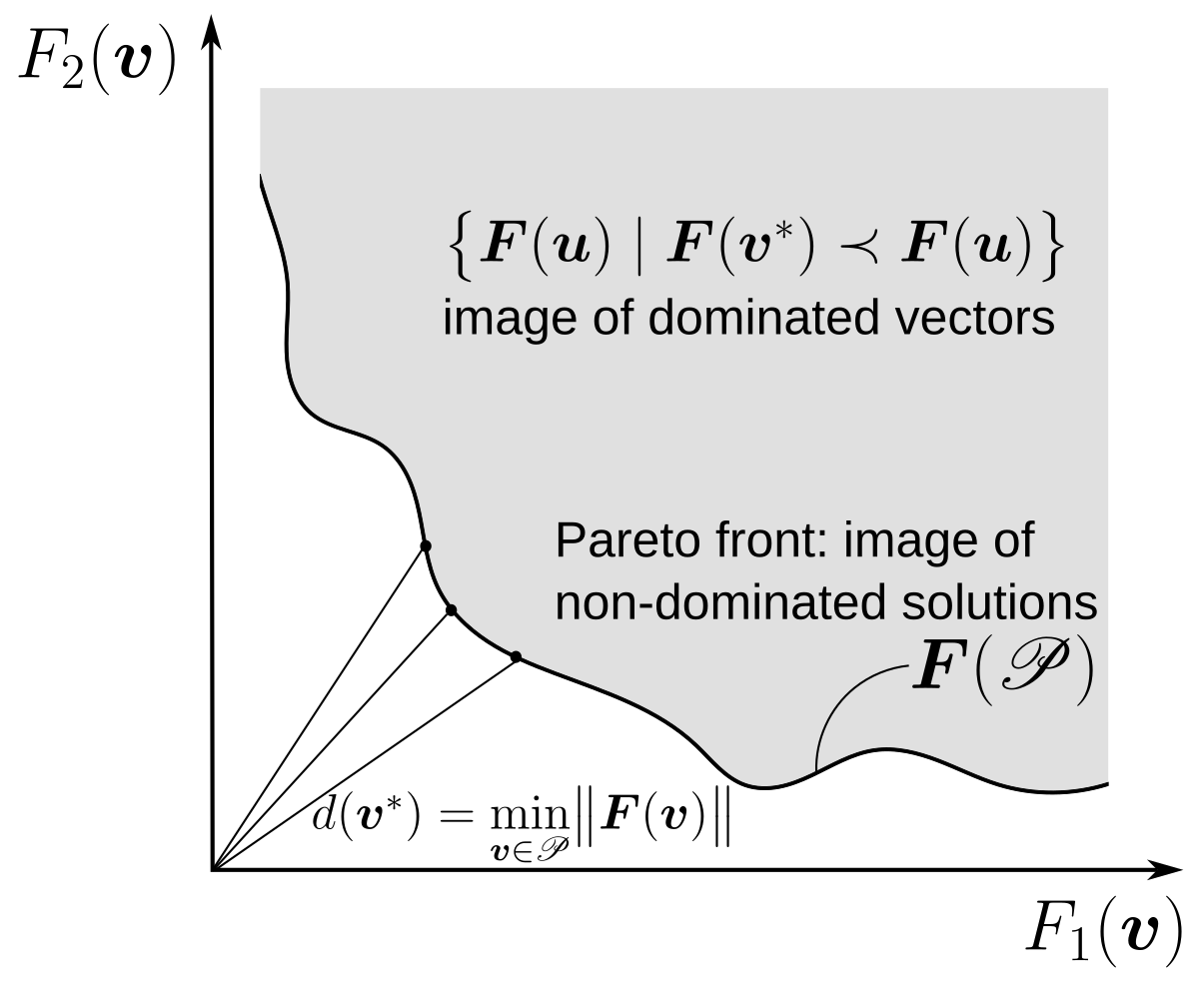}
 \end{center}
 \caption
  []  
  {Sketch of 2D objective space and its key elements.
   Pareto front $\Fvec(\Paretoset)$, \emph{i.e.} image of the Pareto-optimal set 
   $\Paretoset$. Non-convex range of the vector-valued objective function $\Rng(\Fvec)$.
   Pareto-optimal solution(s), $\vvec^*$, selected from their images on the Pareto front
   closest to the origin.}
  \label{fig:pa-sol}
\end{figure}

%
 Given this state of affairs one can find a compromise between different objectives by
 computing special vectors called Pareto-optimal vectors $\vvec^*$, where a more
 forgiving and useful problem can be formulated as a search-and-decide approach. That is,
 find the Pareto-optimal vector(s) $\vvec^*\in\Paretoset$ such that
 \begin{equation}\label{eqn:minproblem2}
 d(\vvec^*) = \min_{\vvec\,\in\,\Paretoset}\norm{\Fvec(\vvec)},
 \end{equation}
 for the Pareto-optimal set defined as
 \begin{equation}\label{eqn:paretoset}
  \Paretoset :=
     \bigl\{ \vvec^* \in \Positives^N \mid \nexists\, \uvec \in \Positives^N:
     \Fvec(\uvec)\prec\Fvec(\vvec^*)
     \bigr\},
 \end{equation}
 where the precedence operator, $\prec$, indicates that:
 \begin{itemize}
  \item $F_k(\vvec^*) \le F_k(\uvec) \quad \forall \quad k=1,\ldots,M$,
  \item $F_k(\vvec^*) < F_k(\uvec) \quad \text{for at least one} \ k$.
 \end{itemize}
 That is to say, we look for optimal parameter vectors that are the best compromise for all $F_k$ objectives.
 We say that $\vvec^*$ dominates $\uvec$, and all $\vvec^*\in\Paretoset$ are non-dominated because they
 are equally attractive from a stand-point of optimal parameters. Note that in general an uncountable
 number of Pareto-optimal vectors $\vvec^*$ exists, and no component of $\Fvec(\vvec^*)$ can be made
 smaller without increasing the value of at least one other component, hence the existence of a Pareto
 front $\Fvec(\Paretoset)$ in objective function space (fig.~\ref{fig:pa-sol}) wherein the range,
 $\Rng(\Fvec)$, is typically non-convex.
\par
 Pareto-optimal vectors represent optimal trade-off of errors of thermophysical
 properties. Which Pareto-optimal vector is ultimately selected from $\Paretoset$ depends
 on the decision step of the solution approach. Here we will select the Pareto-optimal
 vector(s) corresponding to the Pareto front point(s) (fig.~\ref{fig:pa-sol}) closest to
 the origin of the objective function space \eqref{eqn:minproblem2}.
\par
 We now have a clearly formulated mathematical minimization problem 
 \eqref{eqn:minproblem2}--\eqref{eqn:paretoset} with solution(s) $\vvec^*$ 
 (fig.~\ref{fig:pa-sol}) to be computed approximately 
 by the algorithms described next. Note that if only one thermophysical property is optimized, then the 
 Pareto-optimal set has only one vector, that is, the solution of \eqref{eqn:minproblem1} for $k=1$ which 
 reduces to solving a scalar minimization problem.
\par
 Problem \eqref{eqn:minproblem2}--\eqref{eqn:paretoset} requires a complete search method for the 
 construction of $\Paretoset$ which can be tackled by heuristic genetic algorithms (GA) since they work 
 with a population of solutions, and can compute many of the Pareto-optimal vectors at once. An additional 
 feature of GA that makes them suitable for the problem at hand is the differentiation-free approach since 
 the thermophysical property vector function $\yvec$ does not allow for an analytical Jacobian matrix,
 $\partial_{\vvec}\yvec$, to be obtained; moreover, numerical differentiation is prohibitively expensive.
\par
 Finally, \eqref{eqn:minproblem2} measures the overall inaccuracy of an empirical force
 field model to represent the selected thermophysical properties of TBP (or any other
 molecule). If this measure is not small enough, the force field is not realistic.
 The conjecture here is that in practice some $\vvec^*$ can be found so that
 $d(\vvec^*)\approx 0$ (fig.~\ref{fig:pa-sol}).
 The foregoing framework is a tool that helps analyze which
 thermophysical property predictions have competing objectives and which parameters are involved so
 improvements on force fields can be made to make \eqref{eqn:minproblem2} sufficiently small.
%
\subsection{Optimization loop}\label{subsec:oploop}
 In this work, we refine the values of a given set of TBP LJ parameters employing a
 differentiation-free, genetic algorithm optimization, which attempts to reduce the
 deviation between MD simulation predictions of thermophysical properties and the
 corresponding experimental data while avoiding the complexity and high computational
 cost of traditional differentiation-based optimization methods.
 Although LJ parameters can be derived from \emph{ab initio} simulations, the computational
 complexity and cost of quantum mechanical calculations are even higher than any optimization method of
 empirically calibrated classical force fields. Therefore our choice for improved calibration of 
 MD force fields is aimed at developing a practical framework which scales with commonly available 
 parallel computing hardware.
\par
\begin{figure}
 \begin{center}
  \graphicspath{{figs/}}
  \includegraphics[width=3.4in]{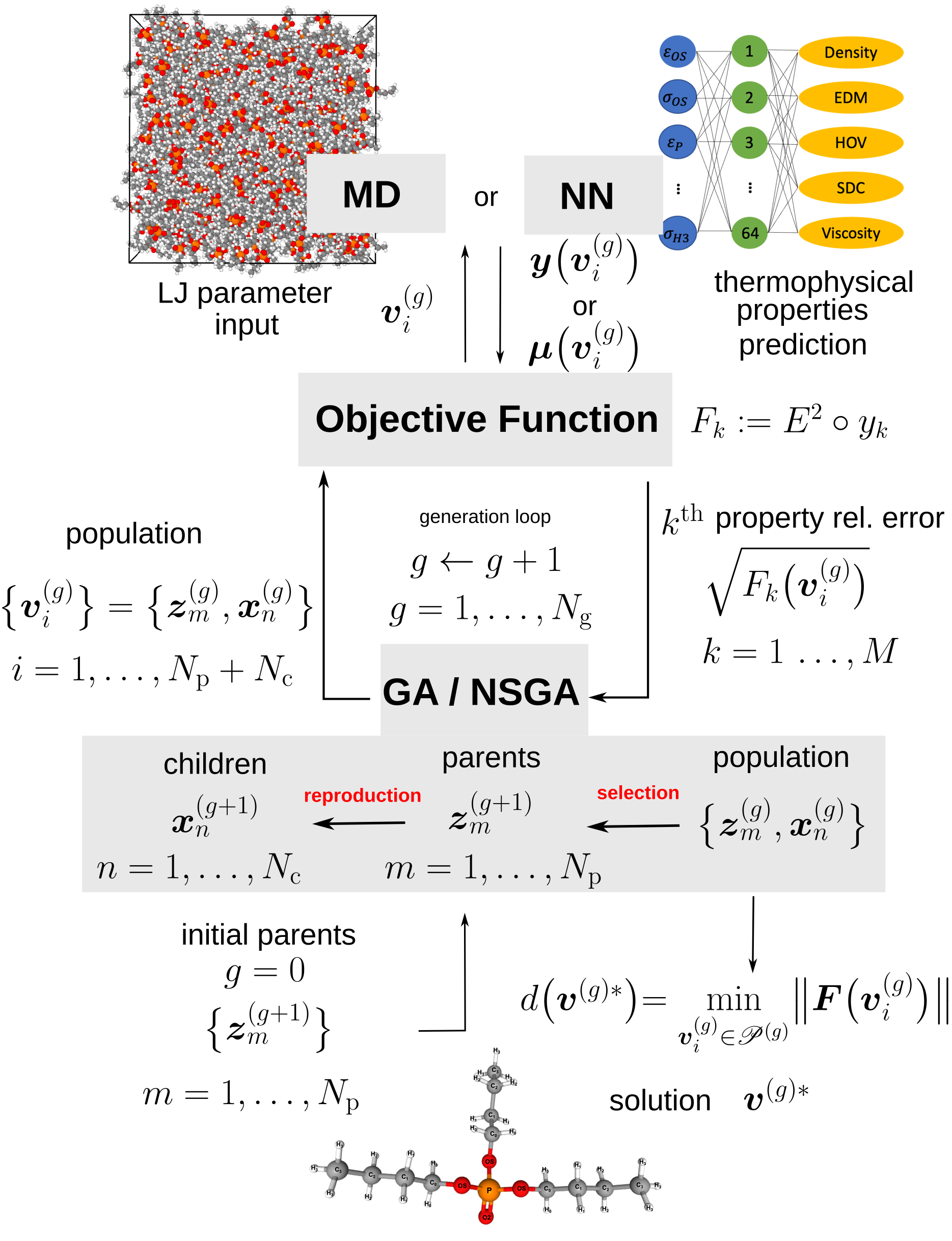}
 \end{center}
 \caption
  []
  {Genetic algorithm loop (algo.~\ref{alg:evolve}) developed for force field
  multi-parameter (multi-objective) optimization. The cost of the MD simulation in the
  loop can be mitigated by using a neural network mapping constructed with accumulated
  data from previous MD simulations.}
  \label{fig:ga-oploop}
\end{figure}

 The solution of problem \eqref{eqn:minproblem2}--\eqref{eqn:paretoset} requires the construction
 of $\Paretoset$ which typically has an uncountable number of Pareto-optimal vectors. Hence any search 
 algorithm with a finite number of steps can only compute a sub-set of $\Paretoset$. 
 To this end, we develop a GA optimization loop (fig.~\ref{fig:ga-oploop}) to build a subset of 
 $\Paretoset$ from which we compute the solution via \eqref{eqn:minproblem2}. 
\par
 The iteration begins with a set of positive-valued LJ parameter vectors called the
 parents, $\bigl\{\zvec^{(1)}_m\in\Positives^N \mid  m = 1,\ldots,N_\text{p}\bigr\}$, of
 the first generation, $g=1$,  of a population of vectors, where $N$ is the number of LJ
 parameters, and $N_\text{p}$ is the number of parents (algo.~\ref{alg:evolve}, line~1).
 Then the loop iteratively evolves a population of parents and their children in
 successive generations that have improved values of the LJ parameters as follows.
\par
 With given parents, children vectors are generated via \emph{reproduction} (\emph{i.e.},
 crossover and mutation operators, sec.~\ref{subsec:iofo}), denoted
 $\bigl\{\xvec^{(1)}_n\in\Positives^N \mid  n = 1,\ldots,N_\text{c}\bigr\}$, where
 $N_\text{c} > N_\text{p}$ is the number of children
 (algo.~\ref{alg:evolve}, lines~5--7). The union of children and parents vectors forms the
 population of the first generation 
 $\bigl\{\vvec^{(1)}_i\bigr\} := \bigl\{\zvec^{(1)}_m,\xvec^{(1)}_n\bigr\}$ with
 $i=1,\ldots,N_\text{p}+N_\text{c}$ (algo.~\ref{alg:evolve}, line~8).
\par
 In order to evolve this population to the next generation, the thermophysical properties 
 and the relative errors (objective functions) \eqref{eqn:obj} need to be computed for 
 all population vectors
 (fig.~\ref{fig:ga-oploop}, algo.~\ref{alg:evolve}, line~9). This computational step 
 requires an equilibrium
 MD simulation for all population vectors of LJ parameters, $\vvec^{(1)}_i$, an expensive
 undertaking of $N_\text{c}$ equilibrium MD simulations (sec.~\ref{subsec:mds}); note that
 the thermophysical properties of the parents are known from the previous generation.
\par
 Once the objective functions of the population are all available, the objective space
 $L_2$ norm of the population is computed (algo.~\ref{alg:evolve}, line~10) and the 
 smallest
 norm compared to a tolerance, $\varepsilon$. When this tolerance is reached, the
 iterations stop if the number of generations has not exceeded the maximum. Otherwise
 the iteration loops back (algo.~\ref{alg:evolve}, line~3), and a \emph{selection} of
 $N_\text{p}$ top vectors is made to be considered as the parents of the next generation
 (line~5), $\bigl\{\zvec^{(2)}_n\mid n=1,\ldots,N_\text{p}\bigr\}$, and the process
 repeats itself for all generations $g=2,\ldots,N_\text{g}$
 (fig.~\ref{fig:ga-oploop}, algo.~\ref{alg:evolve}).
\par
 The selection of the parents for the next generation 
 $\bigl\{\zvec^{(g+1)}_m\mid m=1,\ldots,N_\text{p}\bigr\}$ is what distinguishes GA
 variants (GA, sec.~\ref{subsec:iofo}, and NSGA, sec.~\ref{subsec:mofo}). Basically one
 can have a selection grounded on a scalar (individual) objective function, or a vector
 (multi-objective) functions with other types of selection strategy which is an internal
 implementation (algo.~\ref{alg:evolve}, \textsc{selection} function, line~5). In any
 case, the evolutionary algorithm (algo.~\ref{alg:evolve}) applies to both scalar and
 vector-valued objective functions with the former being any one component of a
 vector-valued objective function.
\begin{algorithm}
 \caption{Vector-valued objective function minimization for MD force field parameter optimization.}
 \label{alg:evolve}
 \begin{algorithmic}[1]
 \Function{evolve}{$\bigl\{ \zvec^{(1)}_m \bigr\}, \Fvec, \varepsilon, N_\text{g}, N_\text{p}, N_\text{c}$}
   \State $g \gets 1; \ d_{\vvec} \gets \infty$ \Comment{Init. generation count; tol.}
   \While{$d_{\vvec} > \varepsilon \ \text{and} \ g \le N_\text{g}$} \Comment{Tol.; iter. limit}
     \If{$g > 1$}
       \State $\bigl\{\zvec_\text{n}^{(g)}\bigr\} \gets \Call{selection}{\bigl\{\vvec^{(g)}_i\bigr\},\Fvec,N_\text{p}}$ \EndIf
     \State $\bigl\{\xvec_\text{n}^{(g)}\bigr\} \gets \Call{crossover}{\bigl\{\zvec^{(g)}_m\bigr\},N_\text{c}}$
     \State $ \bigl\{\xvec_\text{n}^{(g)}\bigr\} \gets \Call{mutation}{\bigl\{\xvec_\text{n}^{(g)}\bigr\}}$
     \State $\bigl\{\vvec^{(g)}_i\bigr\} \gets \bigl\{\zvec_\text{n}^{(g)}\bigr\} \bigcup  \bigl\{\xvec_\text{n}^{(g)}\bigr\} $ \Comment{Population}
     \State $\bigl\{\Fvec(\vvec^{(g)}_i)\bigr\} \gets \Fvec\Bigl(\bigl\{\vvec^{(g)}_i\bigr\}\Bigr)$ \Comment{MD/$\nnmap$ sim.}
     \State $i, d_{\vvec^*} \gets \Call{min\_norm}{\bigl\{\Fvec(\vvec^{(g)}_i)\bigr\} }$ 
     \State $\vvec^* \gets \vvec^{(g)}_i$
     \State $d_{\vvec} \gets d_{\vvec^*}$, $g \gets g+1$ \Comment{While loop check}
   \EndWhile
   \State\Return{$\vvec^*$}
 \EndFunction
 \end{algorithmic}
\end{algorithm}
\par
 For any parent selection method used, a diversity feature (secs.~\ref{subsec:iofo}
 and \ref{subsec:mofo}) must be included in the heuristics of the GA since the success of
 the approach hinges on quickly exploring the domain of $\Fvec$ so that
 $F_k\bigl(\vvec^{(g)*}\bigr)$ is sufficiently small, where $\vvec^{(g)*}$ is the best
 parameter vector \eqref{eqn:minproblem2} of the generation
 $\bigl\{\zvec^{(g)}_m,\xvec^{(g)}_n\bigr\}$. Note that a parent from one generation can
 be carried over to the next generation to promote elitism in the population.
\par
 For the case of multi-objective optimization, parent selection will be significantly more
 complex since it has to systematically select parent vectors from a Pareto front,
 $\Paretoset$, therefore satisfying the initial claim that the GA method will compute sets
 of solutions at once (sec.~\ref{subsec:mofo}).
\par
 It is clear that this optimization loop (fig.~\ref{fig:ga-oploop}) is computationally
 intensive since a full equilibrium MD simulation (requiring parallel computing) for each
 member of any population is embedded into the loop (algo.~\ref{alg:evolve}, line~9). Two
 approaches are used here to mitigate cost. First, the equilibrium MD simulation can be
 done with a reduced number of molecules/atoms because it is possible to improve the LJ
 parameters using a sufficiently small system and still obtain satisfactory results for a
 larger-size system (demonstrated later). Second, as the results of many MD simulations
 are accumulated (saved), they can be used to build an off-line neural network mapping
 $\nnmap:\vvec\in\Positives^N\rightarrow \pvec\in\Positives^M$ of the thermophysical
 properties such that $\nnmap \approx \yvec$ in some sense.
\par
 The $\nnmap$ mapping provides approximate values of thermophysical properties from the
 fitted data in a small fraction of time compared to a full MD simulation, hence this can
 be used in lieu of an MD simulation to accelerate the optimization loop with a drastic
 decrease in
 computational cost. Obviously, as regions of the domain of $\Fvec$ that are not in the
 domain of $\nnmap$ begin to be probed by the GA loop, the neural network mapping is less
 reliable in predicting thermophysical properties and the resulting population of
 parameter vectors begin to be less effective in reducing the value of the objective
 function. This can be further improved by switching back to MD simulations in the loop
 so new information is injected into the algorithm, while saving the results for a future
 update of the neural network mapping. Hence this hybrid evolutionary approach
 (algo.~\ref{alg:evolve}, line~9) has much value and can be explored/improved further.
 This work investigates the off-line use of $\nnmap$.
%
\subsection{Molecular dynamics simulation}\label{subsec:mds}
 The molecular dynamics simulation component is the central engine that computes the
 thermophysical properties when given a force field parameter set
 (fig.~\ref{fig:ga-oploop}). In a molecular dynamics simulation the potential energy of
 the system of $N_\text{a}$ atoms gives
 \begin{align}
  \nonumber
  E(\rvec_1,\ldots,\rvec_{N_\text{a}}) := & \sum_\text{all bonds} K_r\bigl(r-r_\text{eq}\bigr)^2  + 
         \sum_\text{all angles}K_{\theta}\bigl(\theta-\theta_\text{eq}\bigr)^2 
 \\ 
  \nonumber
     + & \sum_\text{all dihedrals}\Bigl[\sum_{n}V_n\bigl(1+\cos(n\phi - \phi_\text{eq})\bigr)\Bigr] 
 \\ 
  \label{eqn:e}
     + & \sum_{i<j} \Bigl[ 
   k_\text{A} \frac{q_iq_j}{r_{ij}}+ 
   4 \epsilon_{ij} \Bigl( 
     \bigl( \frac{\sigma_{ij}}{r_{ij}} \bigr)^{12} 
     - 
     \bigl( \frac{\sigma_{ij}}{r_{ij}} \bigr)^6  
                   \Bigr) 
                \Bigr],
 \end{align}
 where the terms on the right side represent bond length stretching, bond angle bending, bond torsion, 
 electrostatic interactions between pairs of atoms, and van der Waals potential, respectively. The system 
 is conservative, hence the force exerted on the $i^\text{th}$ atom by all other atoms is obtained from the 
 gradient of the potential energy,
 $\fvec_i:=-\grad\!_{\rvec_i}\!{E}$. 
\par
\begin{figure}
 \begin{center}
  \graphicspath{{figs/}}
  \includegraphics[width=3.2in]{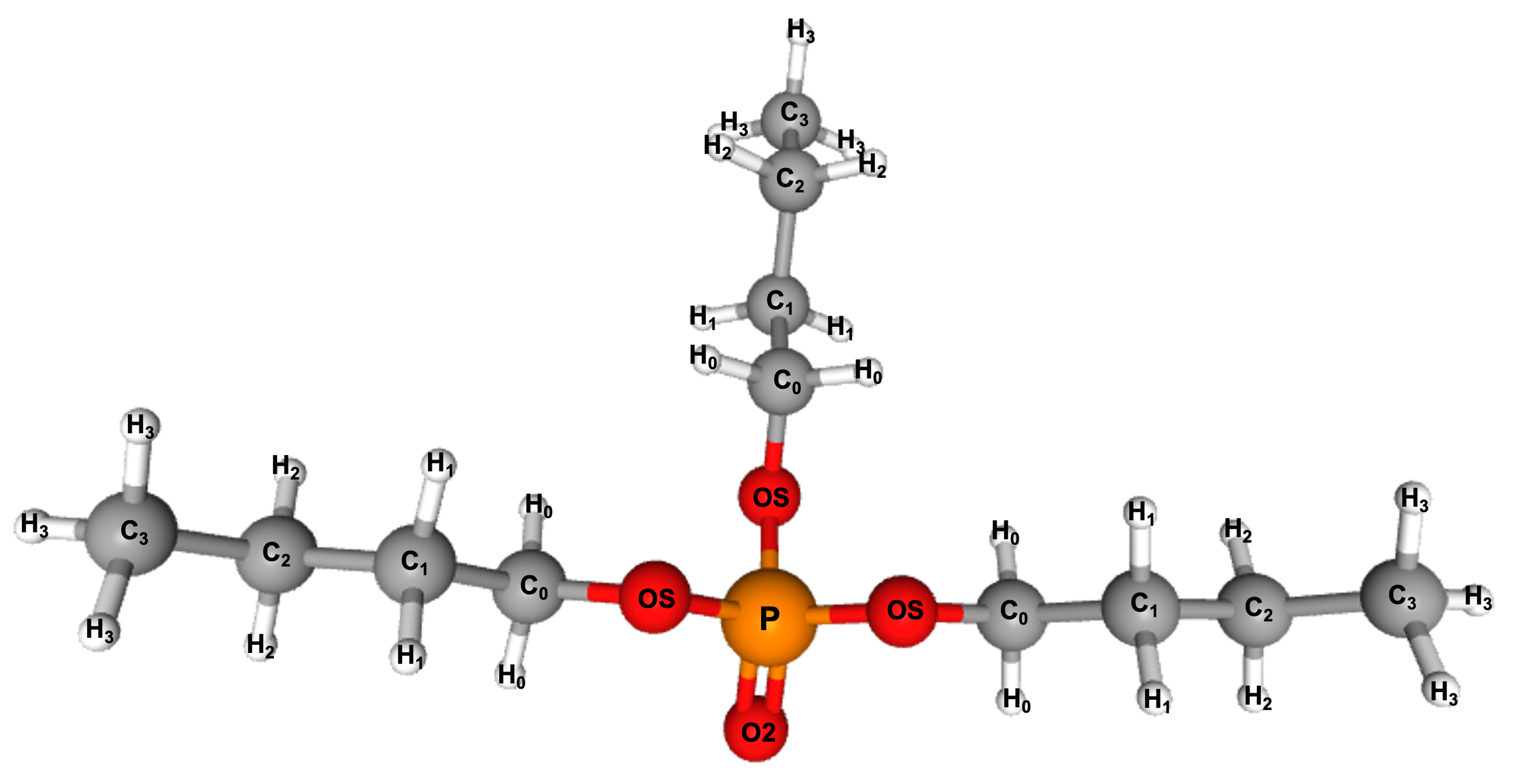}
 \end{center}
 \caption[]
  {Labeling of MD atom types on the TBP molecule. Refer to Table~\ref{tbl:lj} for LJ parameters.}
  \label{fig:tbp}
\end{figure}

 The bonded term parameters were derived from the AMBER package~\cite{amber99-website} 
 (as previously outlined in ~\cite{hatami-de_almeida25:art}) and were not changed in this work. 
 The charges of atoms, $q_i$, were taken from DFT/RHF~\cite{cui2014molecular} and were kept constant
 in this study. The LJ cross terms in the van der Waals potential, ($\sigma_{ij}$, $\epsilon_{ij}$), are 
 calculated using mixing rules for dissimilar atoms: arithmetic mean for $\sigma_{ij}$,
 and geometric
 mean for $\epsilon_{ij}$ for all force fields, otherwise $\sigma_{ii} = \sigma_i$, and 
 $\epsilon_{ii}=\epsilon_i$ (fig.~\ref{fig:tbp} and table~\ref{tbl:lj}). 
 The LJ parameters in Table~\ref{tbl:lj} are used as the parents of the first generation in the optimization loop (fig.~\ref{fig:ga-oploop}); that is, each force field column 
 (table~\ref{tbl:lj}) is a parent vector (5 parents).
 Specifically any parameter vector is formed as
 $\vvec = \{\sigma_1,\ldots,\sigma_{n_\text{a}},\epsilon_1,\ldots,\epsilon_{n_\text{a}}\} \in \Positives^N$ 
 where $n_\text{a}=11$ is the number of atom types, thus $N=2\,n_\text{a}=22$. Hence the first-generation 
 parents $\bigl\{\zvec^{(1)}_m\mid m=1,\ldots,N_\text{p}\bigr\}$ are formed from the five columns ($N_p=5$)
 of Table~\ref{tbl:lj} by concatenating the $\sigma_i$ and $\epsilon_i$ columns for each force field model.
\begin{table*}[pos=h]
  \caption{Lennard-Jones parameters for MD simulation atom types. Radial distance
    parameter for the $i^\text{th}$ atom, $\sigma_i$ in units of [\si{\angstrom}], and
    the atomic attractive pair potential well, $\epsilon_i$, in units of
    [\si{\calorie\per\mol}]. The zero pair potential point occurs at radial distance
    $\sigma_i$. However the zero force point occurs at $1.12\,\sigma_i$, hence this
    radial distance separates the repulsion from the attractive region around an atom.
    This distance ($1.12\,\sigma_i$) is also where the minimum pair potential energy
    occurs, \emph{i.e.} $-\epsilon_i$.}
 \label{tbl:lj}
 \begin{center}
 \begin{tabular}[c]{c|cc|cc|cc|cc|cc}
 \thickhline
 Atom & \multicolumn{2}{c|}{AMBER} & \multicolumn{2}{c|}{OPLS} & \multicolumn{2}{c|}{OPLS2005} & 
 \multicolumn{2}{c|}{GAFF} & \multicolumn{2}{c}{OpenFF}  \\
 \cline{2-3} \cline{4-5} \cline{6-7} \cline{8-9} \cline{10-11}
 Type & $\sigma_i$ & $\epsilon_i$ & $\sigma_i$ & $\epsilon_i$ & $\sigma_i$ & $\epsilon_i$ & $\sigma_i$ & 
 $\epsilon_i$ & $\sigma_i$ & $\epsilon_{i}$ \\
\hline
 O2 & 2.960 & 210    & 2.960 & 210  & 2.980 & 200    & 3.7261 & 50     & 3.0398 & 210.2  \\
 P  & 3.742 & 200    & 3.742 & 200  & 3.740 & 200    & 3.7000 & 200    & 3.7418 & 200    \\
 OS & 3.000 & 170    & 3.000 & 170  & 2.850 & 139.9  & 3.000  & 170    & 3.0251 & 168.5  \\
 C0 & 3.400 & 109.4  & 3.500 & 66   & 3.500 & 64.8   & 3.3997 & 109.4  & 3.3795 & 108.8  \\
 C1 & 3.400 & 109.4  & 3.500 & 66   & 3.500 & 64.8   & 3.3997 & 109.4  & 3.3795 & 108.8  \\
 C2 & 3.400 & 109.4  & 3.500 & 66   & 3.500 & 64.8   & 3.3997 & 109.4  & 3.3795 & 108.8  \\
 C3 & 3.400 & 109.4  & 3.500 & 66   & 3.500 & 64.8   & 3.3997 & 109.4  & 3.3795 & 108.8  \\
 H0 & 2.650 & 15.7   & 2.500 & 30   & 2.500 & 29.9   & 2.4714 & 15.7   & 2.6445 & 15.8   \\
 H1 & 2.650 & 15.7   & 2.500 & 30   & 2.500 & 29.9   & 2.6495 & 15.7   & 2.5832 & 16.4   \\
 H2 & 2.650 & 15.7   & 2.500 & 30   & 2.500 & 29.9   & 2.6495 & 15.7   & 2.5832 & 16.4   \\
 H3 & 2.650 & 15.7   & 2.500 & 30   & 2.500 & 29.9   & 2.6495 & 15.7   & 2.5832 & 16.4   \\
 \thickhline
 \end{tabular}
 \end{center}
\end{table*}
\par
 MD simulations were performed with the Large-scale Atomic Molecular Massively Parallel
 Simulator (LAMMPS)~\cite{plimpton1995fast}, using the velocity Verlet algorithm to
 integrate the equations of motion, with a time step of \SI{1}{\femto\second} and
 three-dimensional periodic boundary conditions. The particle-particle particle-mesh
 (PPPM) method was used for long-range electrostatics. For van der Waals and electrostatic
 forces, non-bonded interactions between atoms in TBP, separated by three bonds
 (1--4~interactions) used scaling factors of 1/2. To limit hydrogen atom bonds, the SHAKE
 algorithm with a tolerance of \SI{e-5}{\angstrom} was chosen.
\par
 The energy of the system of molecules was initially minimized using the steepest descent
 method at \SI{0}{\kelvin}. Following this, the system was gradually heated, increasing
 its temperature from \SI{0}{\kelvin} to \SI{398.15}{\kelvin} in \SI{0.5}{\nano\second}.
 Subsequently, the system was cooled to room temperature over \SI{0.5}{\nano\second} and
 allowed to reach equilibrium at \SI{298.15}{\kelvin} and \SI{1}{\bar} pressure for
 \SI{1.5}{\nano\second}.
 The Nos\'e-Hoover thermostat isothermal-isobaric (NPT) ensemble was used to achieve
 temperature, pressure, and mass density convergence while scaling the system volume.
 Finally for any property but mass density, a \SI{2}{\nano\second} production run using
 canonical (NVT) ensemble was performed under identical conditions to produce the final
 simulation results and compute the ensemble average TBP properties of interest (more on
 the MD simulation protocol used here can be found in ~\cite{hatami-de_almeida25:art}). This
 heating-quenching procedure was applied to any MD simulation performed in the
 optimization loop (fig.~\ref{fig:ga-oploop}), therefore representing an intense
 computational effort as claimed before.
\par
 We selected the experimental values of five thermophysical properties to drive the
 optimization of LJ parameters. That is, we used three thermodynamic properties: mass
 density, heat of vaporization (HOV), and electric dipole moment (EDM) calculated via
 equilibrium MD (EMD) ensemble average described earlier. In addition, we used two
 transport properties, namely, shear viscosity, calculated via the Green-Kubo formalism,
 and self-diffusion coefficient, calculated via the Einstein relation, both within
 equilibrium MD simulations.
\par
 The $k^\text{th}$ thermophysical property, $y_k(\vvec)$, is one of the ensemble average
 properties~\cite{hatami-de_almeida25:art}, namely, $\langle \rho \rangle_\text{NPT}$,
 $\bigl\langle \Delta H_\text{vap} \bigr\rangle_\text{NPT}$, $p_D$, $\eta$, and $D$, that
 is, mass density, heat of vaporization, electric dipole moment, shear viscosity, and
 self-diffusion coefficients, respectively. Note that the evaluation of \eqref{eqn:obj}
 is expensive since it involves the computation of equilibrium ensemble averages for a
 given vector of LJ parameters, $\vvec$. In other words a full equilibrium MD simulation
 for every $\vvec$ in the population of a generation.
\par
 These properties are computed during each MD simulation run for a given set of LJ
 parameters in the optimization loop (fig.~\ref{fig:ga-oploop}). Therefore the cost of MD
 simulations can be reduced by using small systems of molecules, here we find 48 molecules
 to be sufficient. Our conjecture was that a suitable small system of molecules could be
 used while optimizing the LJ parameters, and once a desired parameter set is obtained,
 it can be tested with a few runs on a larger system of molecules for verifying the
 accuracy of thermophysical properties obtained.
\par
 Each MD simulation involved a system of \num{48} TBP molecules and was executed using
 LAMMPS on \num{32} CPUs. Over the course of \num{15} generations, approximately
 \num{155} MD simulations are performed per optimization scenario
 (sec.~\ref{subsec:single}, \ref{subsec:two}, \ref{subsec:three}, \ref{subsec:five}).
 Each MD run typically
 requires 2--4 hours, resulting in a cumulative computational time of approximately
 \num{20} days (wall-clock time) per scenario on a high-performance computing (HPC)
 cluster. In total,
 nine optimization scenarios were investigated in this study, encompassing single- and
 multi-objective cases.
\par
 In summary, we start with the LJ parameters of existing force fields
 (table~\ref{tbl:lj}) as the first generation of parent parameter vectors, and attempt to
 improve upon the parameters by minimizing the error of the predicted thermophysical
 properties relative to experimental data. Therefore for future analysis, it is important
 to note the average values of the characteristic length, and energy well depth LJ
 parameters for the first-generation parents $\zvec^{(1)}_m$ (table~\ref{tbl:ljref}).
 These average values will be instrumental in providing insight on the origin of parameter
 variation.
\begin{table}[pos=h]
 \small
 \setlength{\tabcolsep}{2.0pt}
 \caption []
 {Average values of LJ parameters per atom type for all force fields in Table~\ref{tbl:lj}.
    For comparison, van der Waals radii, $r_\text{vdW}$ of unbound atoms are listed.
  }
 \label{tbl:ljref}
 \begin{center}
 \begin{tabular}{cccc}
 \toprule
  Atom & \multicolumn{2}{c}{} \\
  Type & $\bar{\sigma}_i$  & $\bar{\epsilon}_i$ [\si{\calorie\per\mol}] 
       & $r_\text{vdW}$ [\si{\angstrom}] \\
 \midrule
  O2 & 3.133 (0.298)  & 176    (63.1) & 1.50 \\
  P  & 3.733 (0.166)  & 200    (0.0)  & 1.90 \\
  OS & 2.975 (0.0633) & 163.7  (11.9) & 1.50 \\
  C  & 3.436 (0.0529) & 91.7   (21.5) & 1.77 \\
  H  & 2.571 (0.0705) & 21.5   (6.9) & 1.20 \\
 \bottomrule
 \multicolumn{3}{l}{Standard deviation in parenthesis.}
 \end{tabular}
 \end{center}
\end{table}
%
\subsubsection{Neural network property mapping}\label{subsec:methodNN}
 The usage of the optimization loop (fig.~\ref{fig:ga-oploop}) generates artifact data
 that can be put to good use to accelerate the loop iterations, and reduce computational
 cost. This can be done by building a mapping from parameter vectors
 $\vvec_i\in\Positives^{N}$ to corresponding thermophysical properties computed by MD,
 $\yvec(\vvec_i)\in\Positives^M$. A neural network method is suitable to construct such
 a mapping as described here.
\par
 The data collected from all past MD simulations is denoted by the set of input, output
 pairs $\bigl\{\bigl(\vvec_i, \yvec(\vvec_i)\bigr) \mid i=1,\ldots,S \bigr\}$, where
 $S=1143$ is the number of samples (total number of successful MD simulations).
 Pre-processing of
 both input and output data is crucial because quantities' magnitude can significantly
 impact training (calibrating) the network \cite{han2022data}. Therefore, normalizing or
 standardizing
 the data must be applied. In the dataset utilized, LJ parameters and physical properties
 underwent normalization via two techniques: logarithmic transformation and min-max
 scaling \cite{buzsaki2014log, bland1996transformations}. Logarithmic transformation is
 commonly applied to positively skewed data. Thus, all data were subjected to a
 logarithmic base \num{10} normalization to condense larger values, resulting in a more
 uniform distribution. Following this, min-max normalization was employed to rescale the
 data from their original domain to a new range, typically between \num{0} and \num{1}
 \cite{al2006normalization}.
\par
\begin{figure}
 \begin{center}
  \graphicspath{{figs/}}
  \includegraphics[width=2in]{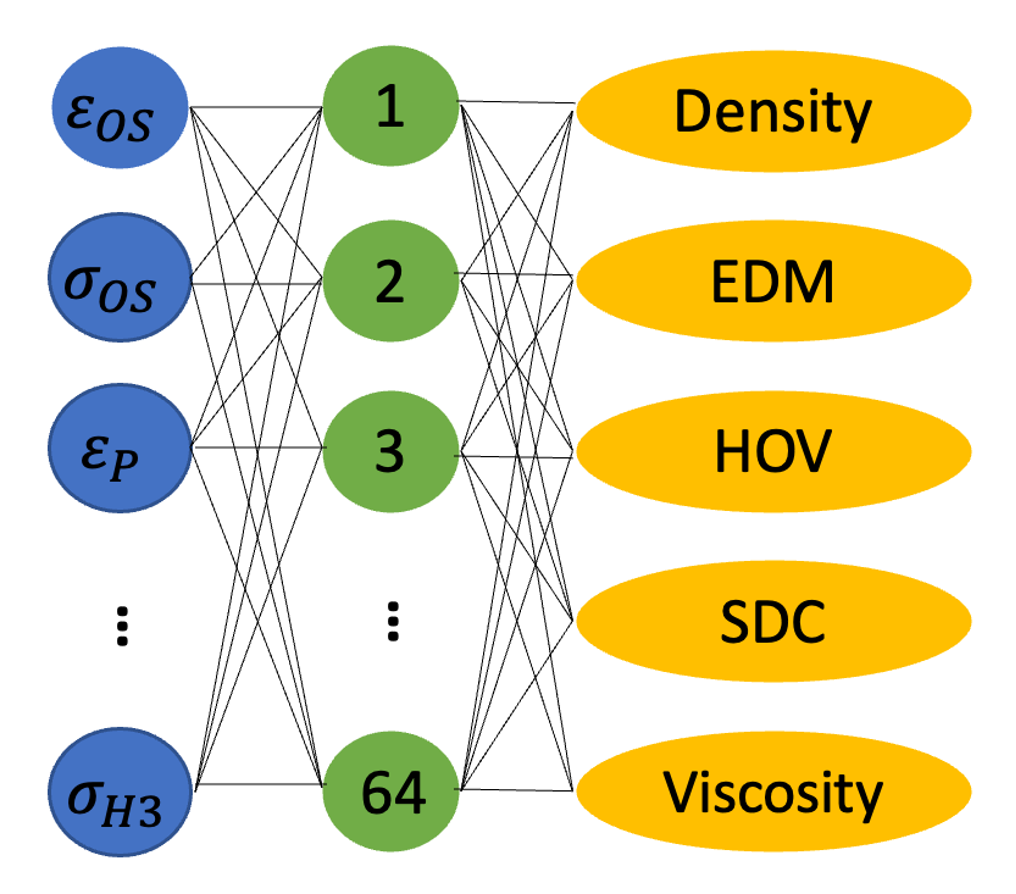}
 \end{center}
 \caption 
  []
  {Schematic representation of the neural network mapping. Blue, green, and yellow circles
   represent: input layer (22 nodes), hidden layer (64 neurons), and output layer
   (5 nodes), respectively. ReLU were employed as the activation function for all
   layers, accelerating the training process while maintaining the ability to model
   nonlinear relationships.}
 \label{fig:NN}
\end{figure}

 The data set is split into training, testing, and validation sets, ensuring a robust
 evaluation. We reserved \num{10}\% of the data to validating the neural network mapping.
 Following training, the mapping accuracy is rigorously assessed using a held-out set.
 A training limit of \num{440} epochs is set, complemented by early stopping after
 \num{100} epochs when without improvement on the validation set. Furthermore, striking a
 balance between network depth and over-fitting, the adopted model incorporates a single
 hidden layer consisting of \num{64} neurons (fig.~\ref{fig:NN}). Training spans \num{440}
 epochs with a batch size of \num{16}. To prevent over-fitting, a dropout rate of
 \num{0.1} was applied after the hidden layer. Rectified linear units (ReLU) serve as the
 activation function for all layers. Compared to sigmoid non-linearity, ReLU has been
 demonstrated to expedite training while maintaining a nonlinear mapping capability. The
 architecture of the neural network significantly impacts fitting accuracy.
\par
 To evaluate the mapping accuracy, the coefficient of determination for the $k^\text{th}$
 thermophysical property
 \begin{equation}\label{eqn:R2}
   R^2_k = 1 -
   \frac{\sum_{i=1}^{S} \Bigl( \mu_k(\vvec_i) - y_k(\vvec_i)  \Bigr)^2}
   {\sum_{i=1}^{S}
     \Bigl( y_k(\vvec_i) - \overline{y_k} \Bigr)^2 },
          \quad k=1,\ldots,M,
 \end{equation}
 and the overall mean square error
 \begin{equation}\label{eqn:MSE}
 \text{MSE} = \frac{1}{S} \sum_{i=1}^{S}
     \norm{\nnmap(\vvec_i) - \yvec(\vvec_i)}^2,
 \end{equation}
 indicators are computed throughout the training process where
 $\overline{y_k} :=
  \frac{1}{s} \sum_{i=1}^S y_k(\vvec_i)$
 stands for the arithmetic average value of the $k^\text{th}$ property predicted by the
 MD simulation for all samples.
 Over the training process, the adaptive moment estimation \cite{kingma2014adam} optimizer
 is used to minimize the MSE cost function and enhance the coefficient of determination.
\par
 With $\nnmap$ at hand, it can be used as an alternative to MD simulations in the
 optimization loop (fig.~\ref{fig:ga-oploop}) allowing for a much larger population size
 of parameter vectors, a much larger number of generations, and much shorter
 computational
 times. A sufficient number of previous MD simulations needs to be available to build
 $\nnmap$, and the optimization loop needs to be monitored for diminishing returns when
 using the neural network mapping in the loop. It may be necessary to use a hybrid
 approach where
 $\nnmap$ needs to be retrained with new, additional MD data. Hence this combination
 MD/$\nnmap$ in the optimization loop (algo.~\ref{alg:evolve}, line~9) is attractive and
 accelerates the overall simulation; retraining was not implemented in this work.
%
\subsection{Individual-objective function optimization}\label{subsec:iofo}
 Investigating the scalar-valued optimization problem obtained from each component of a
 vector-valued objective function optimization deserves much attention for many reasons.
 First, in our application, it provides needed insight into the relationship of LJ
 parameters and thermophysical properties. Second, practical scalar optimization problems
 have either one global min/max solution or countable local min/max solutions which is a
 much simpler problem to work with when compared to a Pareto-optimal uncountable set of
 solutions. Third, the data generated from studying all components individually can be
 used for accelerating the solution method of the corresponding vector-value optimization
 problem (algo.~\ref{alg:evolve}; $\nnmap$ mapping). Last, it is a progressive approach
 to implement many pieces of the computational loop we developed for multi-objective
 functions
 (fig.~\ref{fig:ga-oploop}).
\par
 In this work we start by computing an optimal solution for scalar-valued objective
 function optimization problems obtained from the components of \eqref{eqn:minproblem1},
 that is, find the best LJ parameter vector $\vvec^*_{k^\prime} \in \Positives^N$ such
 that
 \begin{equation}\label{eqn:minproblem-indiv}
     F_{k^\prime}(\vvec^*_{k^\prime}) = \min_{\vvec\,\in\,\Positives^N} 
     \biggl(\frac{y_{k^\prime}(\vvec) - y_{k^\prime}^\text{(exp)}}{y_{k^\prime}^\text{(exp)}} \biggr)^2,
 \end{equation}
 for an individual objective function $k^\prime$ without regards to others
 $k\ne k^\prime$; this is then repeated for every $k^\prime = 1,\ldots,M$. An approximate
 solution to \eqref{eqn:minproblem-indiv} can be computed via a classical GA
 (fig.~\ref{fig:ga-oploop}, algo.~\ref{alg:evolve}) applied to the $k^\prime$ component
 of $\Fvec$. The particular implementation of the selection, crossover and mutation
 algorithms follows.
\par
 Parent selection is the process that decides which LJ parameter vectors of a population will survive and 
 continue on to the next generation of the computational loop. The selection implementation used in this 
 work is the so called binary tournament. It entails choosing a tournament pool, a fitness test, and 
 generating a mating (parent) pool as a result. For simple GA algorithms applied to scalar-valued 
 objective functions, the tournament pool is the entire population; not the case for vector-valued objective
 functions (sec.~\ref{subsec:mofo}).

 To select the parents for the next generation, $\zvec^{(g+1)}_m$, from the current
 population $\vvec^{(g)}_i$, two random elements from the population are chosen and
 compared for corresponding values of $F_{k^\prime}\bigl(\vvec^{(g)}_i\bigr)$, the
 element with lowest value wins (algo.~\ref{alg:bin-tour}). Various tournaments are
 realized (algo.~\ref{alg:bin-tour}, line~4) until a set of parents, $\Zset^{(g)}$, with
 the desired size of mating pool, $N_\text{p}$, is obtained. According to the algorithm
 proposed, a parent and its offspring from a previous generation can continue as fit
 parents to the next generation, therefore allowing for elitism. The random selection does
 not always pick the fittest parent, therefore this assures some degree of diversity into
 the mating pool. The \textsc{selection} function pseudo-code (algo.~\ref{alg:bin-tour})
 for scalar-valued objective function has the required signature of input parameters
 where the second argument, $\Fvec$, has only one component, namely, $F_k$ for some $k$
 (compare to algo.~\ref{alg:evolve}, line~5).
\begin{algorithm}
 \caption{Binary tournament parent selection. Gen. $g$.}
 \label{alg:bin-tour}
 \begin{algorithmic}[1]
 \Function{selection}{$\{\vvec_i\}, F_k, N_\text{p}$}
   \Assert $\#\{\vvec_i\} > N_\text{p}$
   \State $\Zset = \emptyset$ \Comment{Parents set holder}
   \While{$\#\!\Zset \ne N_\text{p}$} 
     \State $\zvec_1, \zvec_2 \gets \Call{random2}{\{\vvec_i\}}$ \Comment{Pick two vec's}
     \If{$F_k(\zvec_1) < F_k(\zvec_2)$} \Comment{Tournament}
       \State $\Zset \gets \Zset \bigcup \{\zvec_1\}$    \Comment{Winner 1}
     \Else
       \State $\Zset \gets \Zset \bigcup \{\zvec_2\}$    \Comment{Winner 2}
     \EndIf
   \EndWhile
   \State\Return{$\Zset$}  \Comment{No duplicates mating pool}
 \EndFunction
 \end{algorithmic}
\end{algorithm}
\par
 Crossover of the parents in the mating pool, $\Zset^{(g)}$, is the next step in the
 evolution algorithm (algo.~\ref{alg:evolve}, line~6) after parent selection. Crossover to
 create $N_\text{c}$ children can be done by various mechanisms previously proposed in the
 literature. Here the simulated binary crossover (SBX) \cite{deb2007self} method,
 originally designed for binary encoded variables is used. The LJ parameter values are
 used directly in this work, hence no encoding is applied. The process involves generating
 two offspring from each pair of parents. For each LJ parameter, SBX uses a probability
 distribution, $\beta$, to determine how the parameter values of the offspring are
 inherited from the parents. For instance, this probability reads
 \begin{equation}\label{eqn:beta}
     \beta = \abs{ \frac{ x_{2,i} - x_{1,i} }{ z_{2,i} - z_{1,i} } }
     \quad \forall \quad i=1,\ldots,N,
 \end{equation}
 for children, $\xvec_1$ and $\xvec_2$, and their parents, $\zvec_1$, and 
 $\zvec_2$. The value of $\beta$ is determined using a polynomial probability distribution 
 defined as
 \begin{align*}
  \beta = 
  \begin{cases}
     2 \, u^{\frac{1}{\eta_c +1}} & \text{if} \ u \leq \frac{1}{2},
   \\
    \Bigl( \frac{1}{2(1 - u)} \Bigr)^{ \frac{1}{\eta_c +1} } & \text{otherwise} ,
  \end{cases}
 \end{align*}
 where $u$ is a randomly generated real number ranging from 0 to 1, and $\eta_c$ is  the 
 distribution index, controlling the strength of the crossover; it is a 
 non-negative real number. A higher $\eta_c$ value biases the probability towards generating offspring 
 closer to the parents, allowing a focused search. Conversely, a smaller $\eta_c$ value promotes the 
 generation of children farther from the parents, encouraging a more diverse breeding.
 The children are generated from \eqref{eqn:beta} as follows     
 \begin{align}\label{eqn:betax1}
  \xvec_1 = 0.5 \Bigl( (1+\beta)\,\zvec_1 + (1-\beta)\,\zvec_2 \Bigr),
     \\ \label{eqn:betax2}
  \xvec_2 = 0.5 \Bigl( (1-\beta)\,\zvec_1 + (1+\beta)\,\zvec_2  \Bigr).
 \end{align}
 The default values for $\eta_c$ and $u$ ($u=0.9$, $\eta_c=15$) in the \texttt{Pymoo}~\cite{blank2020pymoo} 
 package have been used in this study. $N_\text{p}=5$ parents are used in crossover which generates 10 
 different couples producing 20 children, where $N_\text{c}=10$ are randomly selected. The foregoing enables 
 the implementation of the \textsc{crossover} function (algo.~\ref{alg:bin-xover}) in the optimization loop 
 (algo.~\ref{alg:evolve} line~6). Because of elitism in the parent selection process, discussed previously, 
 crossover may use a parent and its own offspring. This inbreeding random event was not controlled in our 
 simulations.
\begin{algorithm}
 \caption{Binary crossover algorithm. Gen. $g$.}
 \label{alg:bin-xover}
 \begin{algorithmic}
 \Require $\beta$
 \Function{crossover}{$\{\zvec_m\}, N_\text{c}$}
   \Assert $\#\{\zvec_m\} < N_\text{c} $
   \State $\Xset = \emptyset$ \Comment{Children set holder}
     \State $\Bset = \Call{binomial2}{\{\zvec_m\}}$ \Comment{Create set of all pairs}
   \While{$\#\!\Xset \ne N_\text{c}$} 
     \State $\zvec_1, \zvec_2 \gets \Call{random}{\Bset}$ \Comment{Pick two vec's}
     \State $\xvec_1, \xvec_2 \gets \Call{bin\_cross}{\zvec_1,\zvec_2,\beta}$ 
                     \Comment{\eqref{eqn:betax1}, \eqref{eqn:betax2}}
     \State $\Xset \gets \Xset \bigcup \{\xvec_1,\xvec_2\}$   
   \EndWhile
   \State\Return{$\Xset$}  \Comment{No duplicates offspring}
 \EndFunction
 \end{algorithmic}
\end{algorithm}
\par
 Continuing the flow-down of the evolution algorithm (algo.~\ref{alg:evolve}, line~7),
 before children LJ parameter vectors are finally created, a mutation is applied only to
 offspring using a similar probability distribution as in SBX. That is, on the
 offspring created by the crossover operator, the GA applies a polynomial mutation
 operator.
 If the value of an LJ parameter is $p \in [a, b]$, the mutated value, $p^\prime$, reads
 \begin{equation}\label{eqn:primed}
  p^\prime = 
  \begin{cases}
      p + \sigma_\text{L}\,(p - a) & \text{if} \ u \leq \frac{1}{2}, 
      \\
      p + \sigma_\text{R}\,(b - p) & \text{otherwise},
  \end{cases}
 \end{equation}
 for a randomly generated number $u$ within [0, 1]. A user-defined index parameter denoted as $\eta$ 
 (set to 15) is used to set the constants $\sigma_\text{L}$ and $\sigma_\text{R}$ according to 
 \begin{align*}
     \sigma_\text{L} &= (2u)^{\frac{1}{1+\eta}} - 1  , 
     \\
     \sigma_\text{R} &= 1 - 2(1-u)^{\frac{1}{1+\eta}} .
 \end{align*}
In this study, we allowed the LJ parameters to vary at will for research purposes,
that is, in practice we used for the distance parameter of any atom type,
$\sigma\in [0,\SI{10}{\angstrom}]$, and for the pair potential well parameter,
$\epsilon\in[0.1, \SI{1e+4}{\calorie\per\mole}]$.
\par
 The mutation scheme (algo.~\ref{alg:mut}) is used in the \textsc{mutation} function 
 (algo.~\ref{alg:evolve}, line~7) of the optimization loop, which leads to the next step 
 to form the new population by joining offspring and parent vectors
 (algo.~\ref{alg:evolve}, line~8).
\begin{algorithm}
 \caption{Offspring mutation algorithm. Gen. $g$.}
 \label{alg:mut}
 \begin{algorithmic}
 \Require $\eta$
 \Function{mutation}{$\{\xvec_n\}, \eta$}
   \ForAll{$\xvec \in \{\xvec_n\}$}
     \State $u \gets \Call{random}{[0,1]}$ 
     \State $\xvec^\prime \gets \Call{primed}{\xvec, u, \eta}$  \Comment{\eqref{eqn:primed}}
     \State $\xvec \gets \xvec^\prime$ \Comment{Mutate in place}
   \EndFor
   \State\Return{$\{\xvec_n\}$}
 \EndFunction
 \end{algorithmic}
\end{algorithm}
\par
 Generations of a population are evolved until either the number of 
 generations reach a maximum, $N_\text{g} = 15$, or \eqref{eqn:minproblem-indiv} reaches a minimum 
 tolerance ($\varepsilon$, algo.~\ref{alg:evolve}, line~3) for any vector of a given population. Notice that 
 applying the GA algorithm (algo.~\ref{alg:evolve}) on a specific component of $\Fvec$ will solve 
 \eqref{eqn:minproblem-indiv} upon convergence which is forced by the tournament selection step
 (algo.~\ref{alg:bin-tour}, line~6). The appeal of this solution method is its algebraic nature only involving
 evaluations of the scalar-valued objective function (no differentiation). On the other hand, its weakness 
 is its heuristics and dependency on the initial population choice. If a random population of LJ parameter 
 vectors is chosen, to begin with, the method may require too many iterations, and as discussed before become
 prohibitively expensive since a full equilibrium MD simulation is needed (algo.~\ref{alg:evolve}, line~9) 
 for every parameter vector in the population for every generation.
%
\subsection{Multi-objective function optimization}\label{subsec:mofo}
 Unlike individual-objective GA optimization with a classical minimization/maximization
 problem, solving multi-objective optimization (MOO) is based on two steps,
 namely, search for a Pareto-optimal set, and decide on a solution
 (sec.~\ref{subsec:mps}). Our stated MOO minimization problem
 \eqref{eqn:paretoset}--\eqref{eqn:minproblem2} also uses the general loop
 (fig.~\ref{fig:ga-oploop},
 algo.~\ref{alg:evolve}) which applies to both scalar-valued function optimization
 (sec.~\ref{subsec:iofo}) and vector-valued function optimization. The significant change
 for multi-objective function optimization is in the \textsc{selection} function now
 designed to force parents to be picked from an approximate Pareto-optimal set
 (sec.~\ref{subsec:mps}). All other heuristic algorithms described earlier for
 individual-objective function optimization (sec.~\ref{subsec:iofo}) apply to the
 multi-objective case and will be reused without change.
%
\subsubsection{Non-dominated partial fronts ranking}\label{subsec:ndpfr}
%
%
 While problem~\eqref{eqn:minproblem-indiv} seeks a solution in the whole $\Positives^N$, a solution for \eqref{eqn:minproblem2} is sought in the Pareto-optimal subset 
 $\Paretoset\subset\Positives^N$~\eqref{eqn:paretoset}.
 This makes the search for solutions more involved than before (sec.~\ref{subsec:iofo}) for two reasons. First the search needs to be restricted to the subset $\Paretoset$ of all
 possible parameter vectors. Second, the subset has an uncountable number of solutions,
 hence a way to compute a finite number of representative, spatially distributed solution
 candidates over the subset needs to be in place. Geometrically, this can be visualized in
 2D as obtaining solutions whose images, $\Fvec(\vvec)$, form a plane curve, called the
 Pareto front (fig.~\ref{fig:pa-sol}); in higher dimension $\Positives^N$ the fronts
 generalize to $(N-1)$-dimensional hyper-surfaces.
\par
 This section describes a useful algorithm (algo.~\ref{alg:multi-front}), \textsc{pareto\_fronts}, used to 
 prepare data needed for parent selection later. It ranks the population of parameter vectors according to 
 hierarchical (or partial) Pareto-optimal fronts, where low rank means that the set of vectors is closer to 
 the Pareto-optimal set solution.
\begin{algorithm}
 \caption{Multi-front Pareto-optimal sorting. Gen. $g$.}
 \label{alg:multi-front}
 \begin{algorithmic}[1]
 \Function{pareto\_fronts}{$\{\vvec_i\}$, $\Fvec$}
 \State{\# First front, dominated sets, domination counters}
   \State $\Frontset[1] = \emptyset$ \Comment{Init. first front set}
   \ForAll{$\vvec \in \{\vvec_i\}$}
   \State $S_{\vvec} = \emptyset$  \Comment{Save vec's dominated by $\vvec$}
   \State $n_{\vvec} = 0$  \Comment{Count \# of vectors that dominate $\vvec$}
     \ForAll{$\uvec \in \{\vvec_i\}$}
       \If{$\Fvec(\vvec)\prec\Fvec(\uvec)$} \Comment{see \eqref{eqn:paretoset}}
         \State $S_{\vvec} \gets S_{\vvec} \bigcup \{\vvec\}$
       \Else
         \State $n_{\vvec} \gets n_{\vvec} + 1$ \Comment{$\vvec$ dominated by $\uvec$}
       \EndIf
     \EndFor
     \If{$n_{\vvec} = 0$} \Comment{If $\vvec$ is non-dominated}
       \State $\Frontset[1] \gets \Frontset[1] \bigcup \{\vvec\}$
     \EndIf
   \EndFor
%
   \Assert $\#\Frontset[1] > 0$
 \State{\# Proceede to remaining fronts}
   \State $r = 1$ \Comment{Init. rank counter}
   \While{$\Frontset[r]\ne\emptyset$}
     \State $Q = \emptyset$ \Comment{Init. tmp storage of next front}
     \ForAll{$\vvec \in \Frontset[r]$}
       \ForAll{$\uvec \in S_{\vvec}$} \Comment{$\uvec$ dominated by $\vvec$}
         \State $n_{\uvec} \gets n_{\uvec} - 1$ \Comment{Subtract front}
         \If{$n_{\uvec} = 0$} \Comment{$\uvec$ in the next front}
           \State $Q \gets Q \bigcup \{\uvec\}$
         \EndIf
       \EndFor
     \EndFor
     \State $r \gets r + 1$
     \State $\Frontset.\text{append}(Q)$
   \EndWhile
     \State\Return{$\Frontset$} \Comment{Ranked list of fronts (subsets)}
 \EndFunction
 \end{algorithmic}
\end{algorithm}
\par
 The population of LJ parameter vectors at some generation, $\bigl\{\vvec^{(g)}_i\bigr\}$,
 are ranked (grouped, sorted) according to hierarchical Pareto-optimal fronts. By
 selectively choosing parents from this sorted population the generations of population
 produced next may eventually converge to $\Paretoset$. The fronts are hierarchical sets
 where each front is the image of non-dominated vectors \eqref{eqn:paretoset} when
 compared to all fronts of higher rank (algo.~\ref{alg:multi-front});
 the rank of a front is its index, \emph{i.e.} rank of $\Frontset_1$ is 1. That is,
 for a given population of parents and children at generation $g$, the population set,
 $P^{(g)}$, can be partitioned into disjoint fronts (algo.~\ref{alg:multi-front}, similar
 to \cite{deb-etal2002:art}), $P^{(g)} = \cup_r \Frontset^{(g)}_r$, where
 $\cap_r \Frontset^{(g)}_r = \emptyset$, and $\Frontset^{(g)}_1 \approx \Paretoset$ for
 large enough $g$.
\par
\begin{figure}
 \begin{center}
  \graphicspath{{figs/}}
  \includegraphics[width=3.2in]{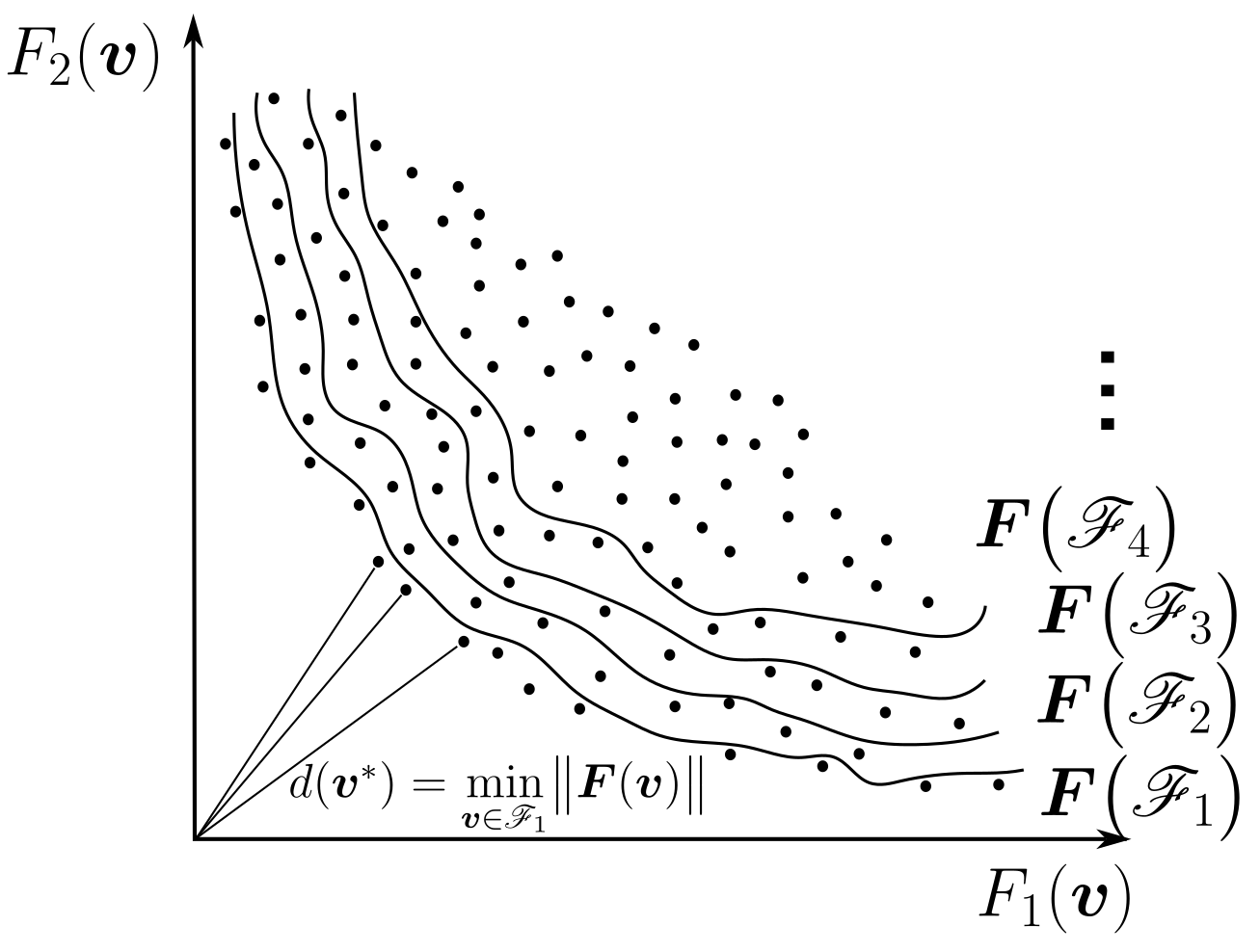}
 \end{center}
 \caption[]
  {Sketch of partial Pareto-optimal front plane curves for a two-objective functions
   optimization at generation $g$. For large enough $g$,
   $\Frontset^{(g)}_1\approx\Paretoset$, and the solution LJ parameter vector
   $\vvec^{(g)*}\in\Frontset^{(g)}_1$ is selected as the shortest Euclidean distance
   from the origin in objective space; multiple vectors may satisfy this condition. In
   higher-dimensional Euclidean objective spaces, curves generalize to hyper-surfaces.}
 \label{fig:pareto-fronts}
\end{figure}

 The partitioning is created by first finding the Pareto-optimal vectors associated to
 the entire population, $\Frontset_1$, (algo.~\ref{alg:multi-front}, lines~4--14), then
 removing these vectors from the population (algo.~\ref{alg:multi-front} lines~20--23)
 and repeating the process to create the next front, until no element is left in the
 population (algo.~\ref{alg:multi-front}, lines~16--25). The image of these hierarchical
 fronts can be visualized in objective-function space as plane curves for a 2D sketch
 (fig.~\ref{fig:pareto-fronts}). The implementation of a function,
 \textsc{pareto\_fronts} (algo.~\ref{alg:multi-front}, line~1), returns a rank-ascending,
 ordered list of subsets of vectors, $\Frontset=[\Frontset_1,\Frontset_2,\ldots]$, from
 the original population, $\{\vvec_i\}$, and the vector-valued objective function,
 $\Fvec$, passed as arguments.
%
\subsubsection{Crowding distance quantifier}\label{subsec:cdq}
 Without a controlling mechanism, the points along any partial Pareto-optimal front,
 $\Frontset_r$, obtained from \textsc{pareto\_fronts} (sec.~\ref{subsec:ndpfr}) as the
 generations are evolved, will not be spread uniformly (fig.~\ref{fig:pareto-fronts}). In
 general cases, they will tend to bunch up and crowd on portions of the front. This can be
 mitigated by adding a diversity heuristics to encourage the selection of parents
 distributed along the front when creating new generations.
\par
 For that purpose, given a previously computed partial front (say at generation $g$), and
 the vector-valued objective function evaluation rule, a function,
 \textsc{crowding\_distance}$(\Frontset_r,\Fvec)$, can be implemented to return a
 quantifier of how crowded the objective space around the image of a front vector is. To
 this end, each LJ parameter vector in the partial front set, $\Frontset_r$, can be
 visited and associated with an objective space crowding distance quantifier. This
 crowding distance is a measure of the density of neighboring vectors around a particular
 vector in the front (algo.~\ref{alg:crowding-distance}, similar to
 \cite{deb-etal2002:art}). The distance is a characteristic length in normalized
 objective-function space computed by inspecting two next-neighbor values of the
 objective function components (algo.~\ref{alg:crowding-distance}, line~4) for every
 vector $\vvec$ in its corresponding front $\Frontset$
 (algo.~\ref{alg:crowding-distance}, lines~7--8). The end vectors in the front
 $\Frontset$ are assigned large values of crowding distances
 (algo.~\ref{alg:crowding-distance}, line~6) so they have precedence over all other
 vectors in the same front when/if considered for the next generation thereby encouraging
 maximal distribution; note the loop in line~7 (algo.~\ref{alg:crowding-distance}) skips
 the first and last vectors of the list. The calculation of the crowding distance can be
 done by inspecting each $k^\text{th}$ component of the objective function and sorting the
 vectors in the front in descending values of objective function
 (algo.~\ref{alg:crowding-distance}, lines~4--5); a list of distances must be sorted
 accordingly (algo.~\ref{alg:crowding-distance} line~5) so the crowding distance
 accumulation (algo.~\ref{alg:crowding-distance}, line~8) is correctly computed.
\par
 The pseudo-code for \textsc{crowding\_distance} returns a list of vectors,
 $\Frontset_r=[\vvec_{r,1},\vvec_{r,2},\ldots]$, and a corresponding list of crowding
 distances $\lambda_{\vvec,r}=[\lambda_{\vvec_{r,1}},\lambda_{\vvec_{r,2}},\ldots]$
 (algo.~\ref{alg:crowding-distance}, line~9). Larger crowding distances indicate a less
 crowded environment, hence vectors in the front with larger values of the distance are
 preferred to be selected (sec.~\ref{subsec:pswcd}) to go into the next generation.
\begin{algorithm}
    \caption{Crowding distance assignment. For partial Pareto-optimal front $\Frontset_r$.}
 \label{alg:crowding-distance}
 \begin{algorithmic}[1]
 \Function{crowding\_distance}{$\Frontset, \Fvec$}
   \State $\Frontset[] \gets \Frontset$ \Comment{Set to list}
   \State $\lambda_{\vvec} \gets []*$\Call{len}{$\Frontset$} \Comment{Init. distance of all $\vvec \ \text{in} \ \Frontset[]$}
   \For{$F_k \ \text{\bfseries in} \ \Fvec$}
     \State $\Frontset, \lambda_{\vvec} \gets$ \Call{sort}{$\Frontset$,$\lambda_{\vvec}$,$F_k$} 
                                \Comment{Sort in $\downarrow F_k(\vvec)$}
     \State $\lambda_{\vvec_0} = \infty$; $\lambda_{\vvec_{-1}} = \infty$; \Comment{Select end vectors}
     \For{$\vvec_i \ \text{\bfseries in} \ \Frontset[1:-1]$}
     \State ${\small \lambda_{\vvec_i} \mathrel{+}= \bigl(F_k(\vvec_{i+1}) - F_k(\vvec_{i-1})\bigr)/
                                      \bigl(F^\text{max}_k-F^\text{min}_k\bigr)}$
     \EndFor
   \EndFor
     \State\Return{$\Frontset,\lambda_{\vvec}$} \Comment{Two matching lists}
 \EndFunction
 \end{algorithmic}
\end{algorithm}
%
\subsubsection{Parent selection: rank and crowding distance}\label{subsec:pswcd}
 We now describe geometric heuristics (called non-dominated sorting genetic algorithm,
 NSGA-II, selection) applied to vector-valued objective function optimization and
 generate an algorithm for parent selection, \textsc{selection} function, for our general
 optimization loop (algo.~\ref{alg:evolve}, line~5).
\par
 The mechanism for selecting parents is still the binary tournament described earlier
 (algo.~\ref{alg:bin-tour}) but with modifications on how to create the tournament pool
 and how to make the fitness test. The tournament pool will be created from a reduced,
 non-dominant ranked population (sec~\ref{subsec:ndpfr}) by prioritizing vectors that
 belong to low rank fronts (low values of $r$ in $\Frontset_r$), and breaking the tie of
 vectors in the same rank by their crowding distance metric
 $\lambda_{\vvec,r}$ (larger distance is preferred for distribution over the Pareto
 front). A \textsc{selection} function (algo.~\ref{alg:nsga-ii}) for multi-objective
 optimization will differ from its counterpart for scalar-objective function
 (algo.~\ref{alg:bin-tour}) in the size and constituents of the tournament pool, and in
 the way a winner of the tournament is selected.
\par
 Thus the strategy for parent selection is a two-step process, first gather vectors from
 non-dominated subsets to create a tournament pool, second apply a binary tournament with
 winning based on rank and crowding distance of the opponents, hence producing a mating
 pool. As a matter of comparison, the tournament pool of the scalar-valued objective
 function optimization \textsc{selection} is the whole population, $\{\vvec_i\}$,
 (algo.~\ref{alg:bin-tour}, line~1), whereas for the vector-valued counterpart, the
 tournament pool is more complex (algo.~\ref{alg:nsga-ii}, lines~5--20). Likewise the
 binary tournament in both algorithms differ on the comparison operators, for
 scalar-valued objective function, a single numeric comparison (algo.~\ref{alg:bin-tour},
 line~6) is used, whereas for the vector-valued counterpart, rank and crowding distance
 comparisons are used (algo.~\ref{alg:nsga-ii}, line~25).
\begin{algorithm}
 \caption{Non-dominated sorting parent selection for NSGA-II. Generation $g$.}
 \label{alg:nsga-ii}
 \begin{algorithmic}[1]
 \Function{selection}{$\{\vvec_i\}, \Fvec, N_\text{p}$}
   \State $N_\text{t} \gets \text{int}\bigl(\#\{\vvec_i\}/2\bigr)$ \Comment{Tournament pool size}
   \Assert $\#\{\vvec_i\} > N_\text{p} \ \text{and} \ N_\text{t} \ge N_\text{p}$
   \State $\Frontset \gets \Call{pareto\_fronts}{\{\vvec_i\},\Fvec}$ \Comment{Fronts list}
   \State $\Sset = \emptyset$ \Comment{Init. tournament pool}
   \State \# Get prospective parents by front rank
   \For{$\Frontset_r\ \text{\bfseries in} \ \Frontset$}
     \If{$\#\Sset + \Call{len}{\Frontset_r} \le N_\text{t}$} 
       \State $\overrightarrow{\Frontset}_r, \lambda_{\vvec,r} \gets \Call{crowding\_distance}{\Frontset_r,\Fvec}$
       \State $r \gets \Frontset.\text{index}(\Frontset_r)$ \Comment{Rank}
       \State $\{(\vvec_j,r,\lambda_{\vvec_j})\} \gets \Call{zip}{\overrightarrow{\Frontset}_r, \lambda_{\vvec,r}}$ \Comment{Tuples}
       \State $\Sset \gets \Sset \bigcup \{(\vvec_j,r,\lambda_{\vvec_j})\} $ 
     \EndIf
   \EndFor
   \State \# Get prospective parents by crowding distance
     \If{$\#\Sset < N_\text{t}$} 
     \State $r \gets r+1$ \Comment{Get the next front rank}
     \State $\overrightarrow{\Frontset}_r, \lambda_{\vvec,r} \gets \Call{crowding\_distance}{\Frontset_r,\Fvec}$
     \State $\overrightarrow{\Frontset}_r,\lambda_{\vvec,r} \gets \Call{sort}{\overrightarrow{\Frontset}_r, \lambda_{\vvec,r}}$ \Comment{Descending $\lambda_{\vvec,r}$}
     \State $K \gets N_\text{t}-\#\Sset$ \Comment{\# of vacant spots}
     \State $\{(\vvec_j,r,\lambda_{\vvec_j})\} \gets \Call{zip}{\overrightarrow{\Frontset}_r[:K], \lambda_{\vvec,r}[:K]}$
     \State $\Sset \gets \Sset \bigcup \{(\vvec_j,r,\lambda_{\vvec_j})\} $
   \EndIf
   \State \# Proceed to binary tournament in $\Sset$
   \State $\Zset = \emptyset$ \Comment{Parents set holder}
   \While{$\#\!\Zset \ne N_\text{p}$} 
     \State $(\zvec_1,r_1,\lambda_{\zvec_1}), (\zvec_2,r_2,\lambda_{\zvec_2}) \gets \Call{random2}{\Sset}$
     \If{$(\zvec_1,r_1,\lambda_{\zvec_1}) \prec_\Sset (\zvec_2,r_2,\lambda_{\zvec_2})$} \Comment{\eqref{eqn:precpool}}
       \State $\Zset \gets \Zset \bigcup \{\zvec_1\}$    \Comment{Winner 1}
     \Else
       \State $\Zset \gets \Zset \bigcup \{\zvec_2\}$    \Comment{Winner 2}
     \EndIf
   \EndWhile
   \State\Return{$\Zset$}  \Comment{No duplicates mating pool}
 \EndFunction
 \end{algorithmic}
\end{algorithm}
\par
 Specifically, on the tournament pool creation, it is smaller than the whole population.
 Typically, say half of the population size will be allocated for the tournament pool,
 $N_\text{t}:=(N_\text{p}+N_\text{c})/2$, or another value as long as
 $N_\text{t} > N_\text{p}$ (algo.~\ref{alg:nsga-ii}, line~2). The tournament set,
 $\Sset$, elements consist of tuples formed by three ordered components, namely, an LJ
 vector, its front rank, and its crowding distance (algo.~\ref{alg:nsga-ii}, line~11).
 Filling up the tournament pool is accomplished in two parts (algo.~\ref{alg:nsga-ii},
 lines~6 and 13, respectively). First, the tournament pool, $\Sset$, is filled with
 tuples entered into the set in ascending order of the rank of the fronts
 (algo.~\ref{alg:nsga-ii}, lines~7--12) until the number of tuples is less or equal to
 $N_\text{t}$. Second, since it is unlikely that an integer number of fronts (and their
 vectors) exactly matches the desired number of elements in the tournament pool, the
 additional available entries (algo.~\ref{alg:nsga-ii}, line~18) in $\Sset$, are filled
 with tuples fetched from the next non-included front $\Frontset_{r+1}$ sorted in
 descending order of crowding distance (algo.~\ref{alg:crowding-distance} lines~14--20).
 By virtue of the first part, the $r+1$ front will have enough vectors to deliver
 $\#\Sset=N_\text{t}$, and this completes the creation of the tournament pool.
\par
 Next a binary tournament (algo.~\ref{alg:nsga-ii}, line~23) is carried out between any
 two randomly picked tuples of $\Sset$ (line~24). That is, the values of the rank and
 crowding distance are compared as follows:
 \begin{align}\notag
 &(\zvec_1,r_1,\lambda_{\zvec_1}) \prec_\Sset (\zvec_2,r_2,\lambda_{\zvec_2})
   =
\\
 & \begin{cases} \label{eqn:precpool}
  \zvec_1 & \text{if} \quad r_1 < r_2 \quad \text{or} \quad
                      \bigl(r_1 = r_2 \ \text{and} \ \lambda_{\zvec_1} \ge \lambda_{\zvec_2}\bigr).
  \\
   \zvec_2 & \text{otherwise} ,
  \end{cases}
 \end{align}
 with each winner added to the set of parents $\Zset$ which is returned at the end of the function
 (algo.~\ref{alg:nsga-ii}, line~29). Various implementations of this algorithm 
 (algo.~\ref{alg:nsga-ii}) exists, we have used \cite{blank2020pymoo}, NSGA-II, with parent population of
 $N_\text{p}=5$ and offspring of $N_\text{c}=10$. These values are limited by computational resources 
 available at the time of this work. Larger values can be used in the future when larger resources are 
 available. Finally, the subsequent steps in the optimization loop (algo.~\ref{alg:evolve}, lines~6--12) 
 follow exactly the same operations described in the GA method (sec.~\ref{subsec:iofo}).
%
\subsubsection{Parent selection: rank and reference distance}\label{subsec:pswrp}
 With increasing dimension of the objective function space, the parent selection with
 crowding distance (sec.~\ref{subsec:pswcd}) becomes computationally expensive and not
 very effective in generating offspring that are both well distributed over the
 Pareto-optimal set and convergent~\cite{deb-jain2014:art, jain-deb2014:art}. For
 parameter spaces of dimension $M=4$ and higher, there is room for a more focused
 approach to select a tournament pool based on Pareto-optimal vectors wherein points in
 objective function space are pre-defined to guide the selection of parents. The choice
 of points is made on a normalized version of the objective function space as we describe
 next.
\par
 We first explain the normalization step. Consider the population
 $\{\vvec_i\mid i=1,\ldots,N_\text{p} + N_\text{c}\}$ at some generation $g$. Evaluating
 the components of the objective function with minimum values
 $\theta_k:=\underset{\vvec\,\in\{\vvec_i\}}{\min}F_k(\vvec), \ k=1,\ldots,M$, and making a
 change of origin $F^\prime_k := F_k - \theta_k$, one can define the normalized objective
 function:
 $\widetilde{F}_k := F^\prime_k/\underset{\vvec\,\in\{\vvec_i\}}{\max}\bigl(F^\prime_k(\vvec)\bigr)$.
 This normalization is done for each generation of the population, therefore it is
 self-adaptive.
\par
\begin{figure}
 \begin{center}
  \graphicspath{{figs/}}
  \includegraphics[width=3.2in]{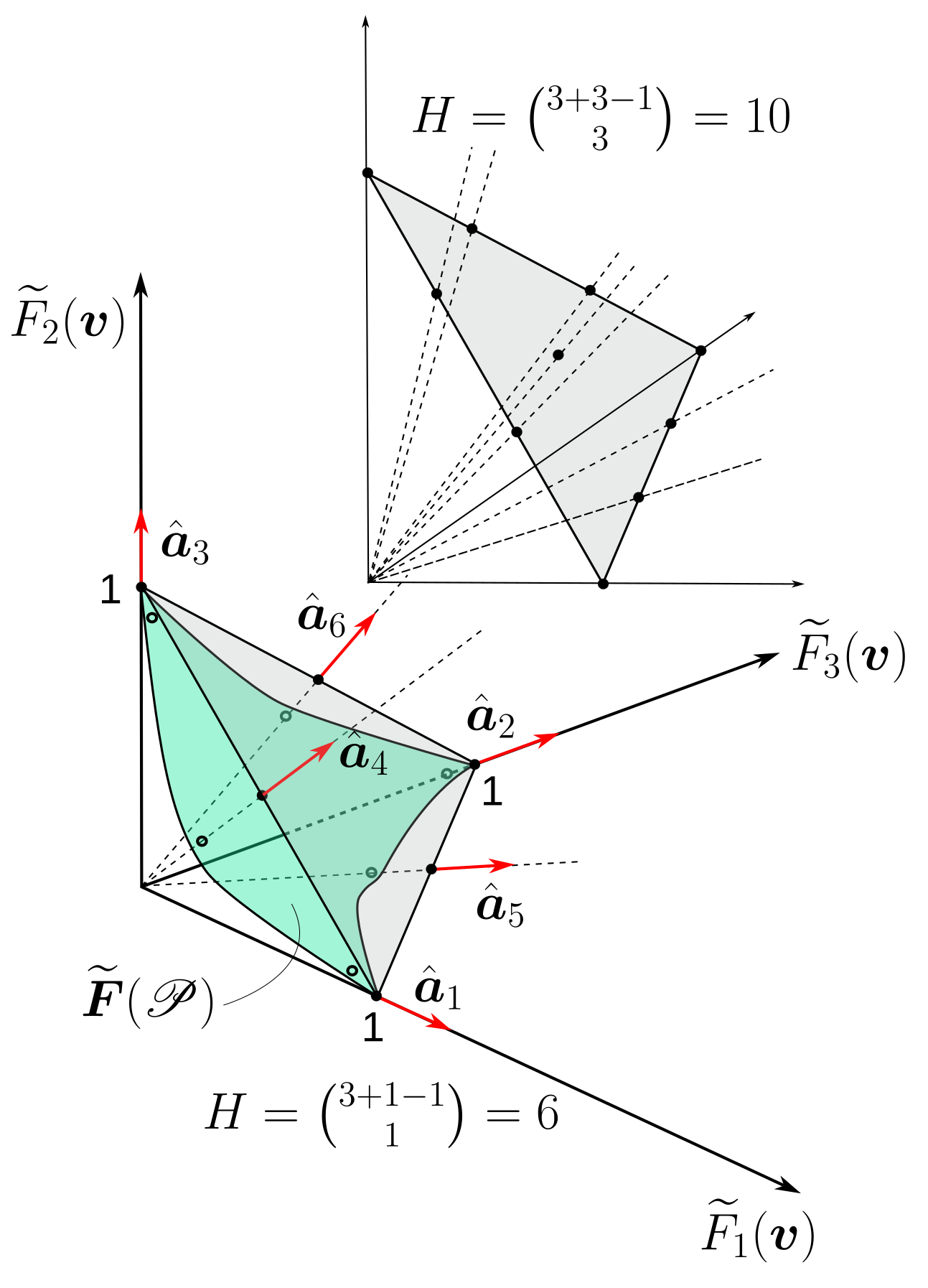}
 \end{center}
 \caption[]
  {Sketch of reference points (filled circles) in normalized objective space (3-D example) for parent 
   selection. Cases with 2 divisions and 3 divisions (insert). The normalized Pareto-optimal hyper-surface 
    front, $\widetilde{\Pset}$, will stay roughly between the origin and the 2-D reference plane with 
    possible intersection away from the corner points. Open circles on $\widetilde{\Pset}$ indicate 
    expected computational points attracted by the reference points and guiding directions.}
  \label{fig:ref-pts}
\end{figure}

 One can now provide a number of divisions, $\delta$, along the normalized objective
 function axis interval $[0,1]$ (fig.~\ref{fig:ref-pts}). Therefore $\delta+1$ points are
 generated on the axes, and the binomial formula \eqref{eqn:binomial} can be used to
 generate the total number of reference points to be uniformly distributed on a $M-1$
 hyper-plane of reference points
 \begin{equation}\label{eqn:binomial}
     H = \binom{M + \delta - 1}{\delta} = \frac{({M+\delta-1})!}{(M-1)!\,\delta!}.
 \end{equation}
 In view of the normalization, the $M-1$ Pareto-optimal (curved) hyper-surface front will
 be roughly located between the origin and the hyper-plane in the normalized objective
 function space (fig.~\ref{fig:ref-pts}).
\par
 With given reference points, the goal is to compute non-dominated parents to be
 associated with guiding lines from the origin passing through the reference points
 (fig.~\ref{fig:ref-pts}). This can be done by finding the population vectors with images
 closest to the reference guiding lines and giving them preference. Hence the orthogonal
 distance to the guiding lines is used as a quantifier for ranking the vectors in the
 population. That is, the closer to the guiding lines the more relevant the population
 vector is. In a sense this is a similar approach as in NSGA-II (algo.~\ref{alg:nsga-ii})
 where the crowding distance is replaced by a new reference distance. Unlike NSGA-II, the
 distribution of the population in normalized objective space is pre-selected and
 uniformly distributed, while in NSGA-II the distribution is computed on the fly and not
 necessarily uniform. Note that with the reference hyper-plane approach the population
 size is best to be equal or nearly equal to the number of reference points, hence it is
 recommended that
\begin{equation}\label{eqn:Hrecommended}
   N_\text{p}+N_\text{c} \approx H.
\end{equation}
\par
 To compute the reference distance, if $\avec_h$ is a reference point in normalized
 objective function space, a guiding line passing through it and the origin can be
 written as any colinear vector $\alpha\avec_h$, $\alpha > 0$, where the unit vector
 along the line pointing away from the origin is denoted
 $\hat{\avec}_h:=\avec_h/\norm{\avec_h}$. The orthogonal distance in normalized objective
 function space from an arbitrary population vector image, $\widetilde{\Fvec}(\vvec)$,
 and a guiding line, is 
 $\norm{\Ptensor_h\widetilde{\Fvec}(\vvec)}$ where $\Ptensor_h:=\Itensor-\hat{\avec}_h\otimes\hat{\avec}_h$
 is the orthogonal projection tensor on the line. This distance in normalized objective
 space can be computed for all vectors $\vvec$ of a population for all reference lines
 (algo.~\ref{alg:ref-distance}). The \textsc{reference\_distance}$(\Frontset, \{\hat{\avec}_h\}, \widetilde{\Fvec})$ function
 (algo.~\ref{alg:ref-distance}, \textsc{ref\_dist} for short) finds the closest reference
 guiding line for each vector of a given partial Pareto-optimal front. Note that the
 arguments of this function are the partial Pareto-optimal fronts, the set of directions
 at the reference points, $\{\hat{\avec}_h\mid h=1,\ldots,H\}$, and the normalized
 vector-valued objective function, respectively. The first argument is obtained as before
 (sec.~\ref{subsec:ndpfr}) using \textsc{pareto\_fronts}, and the last two arguments need
 to be obtained as described earlier and encapsulated into a
 \textsc{normalization}$(\{\vvec_i\}, \Fvec, H)\rightarrow (\{\hat{\avec}_h\}, \widetilde{\Fvec})$
 function which we do not describe here for brevity but we do indicate where it is
 called/needed in the resulting algorithm (algo.~\ref{alg:nsga-iii}, line~7).
\par
 The return of the \textsc{reference\_distance} function (algo.~\ref{alg:ref-distance},
 line~13) is similar to the NSGA-II crowding distance function but with an additional list
 containing the reference direction vectors, $\hat{\avec}_{\vvec}$, associated to the
 list of LJ
 parameter vectors, $\Frontset$, and the corresponding list of distances,
 $\lambda_{\vvec}$. These lists are ordered and have size equal to the number of LJ vectors
 in a given partial Pareto-optimal front.
\par
 With the \textsc{reference\_distance} function at hand, it can replace the crowding
 distance function in the parent selection algorithm described earlier
 (algo.~\ref{alg:nsga-ii}) to create a modified reference-based algorithm named
 NSGA-III~\cite{deb-jain2014:art} (algo.~\ref{alg:nsga-iii}--\ref{alg:nsga-iiic}). The
 modification fits into the same binary tournament structured described earlier
 (sec.~\ref{subsec:pswcd}) with some changes to take into account the usage of
 pre-specified reference points in the normalized objective function space.
\begin{algorithm}
 \caption{Reference distance assignment. For partial Pareto-optimal front $\Frontset[i]$.}
 \label{alg:ref-distance}
 \begin{algorithmic}[1]
   \Function{reference\_distance}{$\Frontset, \{\hat{\avec}_h\}, \widetilde{\Fvec}$}
   \State $\Frontset[] \gets \Frontset$ \Comment{Set to list}
   \State $\lambda_{\vvec} \gets []*$\Call{len}{$\Frontset$} \Comment{Init. distance of all $\vvec \ \text{in} \ \Frontset[]$}
   \State $\hat{a}_{\vvec} \gets []*$\Call{len}{$\Frontset$} \Comment{Init. ref. $\hat{\avec}_h$ of all $\vvec \ \text{in} \ \Frontset[]$}
   \For{$\vvec_j \ \text{\bfseries in} \ \Frontset$} 
     \State $\lambda_{\vvec_j} \gets \infty$
     \For{$\hat{\avec}_h \ \text{\bfseries in} \ \{\hat{\avec}_h\}$} \Comment{$h=1,\ldots,H$}
       \State $\Ptensor_h \gets \Itensor - \hat{\avec}_h\otimes\hat{\avec}_h$
       \State $d \gets \norm{\Ptensor_h\widetilde{\Fvec}(\vvec_j)}$ 
       \If {$d < \lambda_{\vvec_j}$}
         \State $\lambda_{\vvec_j} \gets d$
         \State $\hat{a}_{\vvec_j} \gets \hat{\avec}_h$
       \EndIf
     \EndFor
   \EndFor
   \State\Return{$\Frontset,\lambda_{\vvec}, \hat{a}_{\vvec}$} \Comment{Three matching lists}
 \EndFunction
 \end{algorithmic}
\end{algorithm}
\par
 Notably the tournament pool set, $\Sset$, must be modified and it now consists of an
 extended tuple as compared to before by adding the specific reference direction
 $\hat{\avec}_{h,j}$ associated to the LJ vector $\vvec_j$ of the tuple
 (algo.~\ref{alg:nsga-iii}, line~14). The first part of filling-up the tournament pool
 set, $\Sset$, (algo.~\ref{alg:nsga-iii}, line~9) follows similarly as before in NSGA-II,
 using partial Pareto-optimal fronts up to $\Frontset_r$, but with the extended tuple
 (algo.~\ref{alg:nsga-iii}, lines~10--15).
\par
 The second part of filling up $\Sset$ (algo.~\ref{alg:nsga-iiic}, line~17) is more
 complex than in NSGA-II because the idea is to pick the remaining LJ vectors to complete
 a pool size of $N_\text{t}$ so that they are the closest to the reference directions
 least populated in $\Sset$, that is, least number of associated vectors obtained in the
 first filling-up part. To accomplish this, first we compute the number of vectors in the
 tournament pool associated to each of the reference directions after completing the first
 part of the filling-up process (algo.~\ref{alg:nsga-iiic}, lines 19--23). Next we
 proceed with filling up the vacant spots in $\Sset$ (algo.~\ref{alg:nsga-iiic}, line~27)
 by finding any direction least populated, line~28, then finding the vector(s) in the next
 partial Pareto-optimal front, $\Frontset_{r+1}$, associated to the direction least
 populated (algo.~\ref{alg:nsga-iiic}, line~29), and finally selecting the one closest to
 the direction least populated, line~31. The data associated to this vector, namely, its
 components, rank, reference distance, and reference direction are packaged into a tuple
 (algo.~\ref{alg:nsga-iiic}, lines~32--34) and inserted into $\Sset$, line~34. The data
 is removed from the corresponding lists (algo.~\ref{alg:nsga-iiic}, line~35), and the
 counter of vectors associated to reference directions incremented
 (algo.~\ref{alg:nsga-iiic}, line~36). If there are no vectors associated to a particular
 reference direction, the direction is removed from consideration
 (algo.~\ref{alg:nsga-iiic}, line~38). The aforementioned group of operations is referred
 to as niching prospective parents (line~17).
\begin{algorithm}
 \caption{Non-dominated sorting parent selection for NSGA-III. Generation $g$.}
 \label{alg:nsga-iii}
 \begin{algorithmic}[1]
 \Require $\delta$
 \Function{selection}{$\{\vvec_i\}, \Fvec, N_\text{p}$}
   \State $N_\text{t} \gets \text{int}\bigl(\#\{\vvec_i\}/2\bigr)$ \Comment{Half of population}
   \State $M = \Call{len}{\vvec_0}$
   \State $H = \Call{binomial}{M+\delta-1,\delta}$
   \Assert $\#\{\vvec_i\} > N_\text{p} \ \text{and} \ N_\text{t} \ge N_\text{p}$
   \State $\Frontset \gets \Call{pareto\_fronts}{\{\vvec_i\},\Fvec}$ \Comment{Fronts list}
     \State $\{\hat{\avec}_h\}, \widetilde{\Fvec} \gets \Call{normalization}{\{\vvec_i\},\Fvec, H}$
   \State $\Sset = \emptyset$ \Comment{Init. tournament pool}
   \State \# Get prospective parents by front rank
   \For{$\Frontset_r\ \text{\bfseries in} \ \Frontset$}
     \If{$\#\Sset + \Call{len}{\Frontset_r} \le N_\text{t}$} 
       \State $\overrightarrow{\Frontset}_r, \lambda_{\vvec,r}, \hat{a}_{\vvec,r} \gets \Call{ref\_dist}{\Frontset_r, \{\hat{\avec}_h\}, \widetilde{\Fvec}}$
       \State $r \gets \Frontset.\text{index}(\Frontset_r)$ \Comment{Rank}
       \State $\{(\vvec_j,r,\lambda_{\vvec_j},\hat{\avec}_{h,j})\} \gets \Call{zip}{\overrightarrow{\Frontset}_r, \lambda_{\vvec,r}, \hat{a}_{\vvec,r}}$
       \State $\Sset \gets \Sset \bigcup \{(\vvec_j,r,\lambda_{\vvec_j},\hat{\avec}_{h,j})\} $
     \EndIf
   \EndFor
   \State \# To be continued (algo.\ref{alg:nsga-iiic})
  \algstore{bkbreak}
 \end{algorithmic}
\end{algorithm}
\par 
 At this point after the second part of the filling-up process of the tournament pool set
 is completed, the cardinality of the set should be $N_\text{t}$ or at the minimum greater
 or equal to $N_\text{p}$ (algo.~\ref{alg:nsga-iiic}, line~39). The tournament for
 selection of the parents (mating pool) follows as before for NSGA-II using the rank and
 the reference distance to the associated reference direction of tournament pool elements
 to obtain $N_\text{p}$ parents mating pool for the next generation
 (algo.~\ref{alg:nsga-iiic}, lines~42--47).
\begin{algorithm}[h]
 \caption{Algorithm~\ref{alg:nsga-iii} continued.}
 \label{alg:nsga-iiic}
 \begin{algorithmic}[1]
 \algrestore{bkbreak}
   \State \# Get prospective parents by niche
   \If{$\#\Sset < N_\text{t}$} 
     \State $\rho \gets []*H$ \Comment{Count \# $\vvec$ assoc. to ref. pts.}
     \For{$\hat{\avec}_h \ \text{\bfseries in}\ \{\hat{\avec}_h\}$}
       \For{$(\vvec_j,r,\lambda_{\vvec_j},\hat{\avec}_{h,j}) \ \text{\bfseries in}\ \Sset$}
         \If{$\hat{\avec}_{h,j} = \hat{\avec}_h$}
           \State $\rho[s] \gets \rho[s] + 1$
         \EndIf
       \EndFor
     \EndFor
     \State $r \gets r+1$ \Comment{Get the next front rank}
     \State $\overrightarrow{\Frontset}_r, \lambda_{\vvec,r}, \hat{a}_{\vvec,r} \gets \Call{ref\_dist}{\Frontset_r,\{\hat{\avec}_h\}, \widetilde{\Fvec}}$
     \State $K \gets N_\text{t} -\#\Sset$ \Comment{\# of vacant spots}
     \For{$k \ \text{\bfseries in} \ \Call{range}{K}$}
       \State $s \gets \Call{argmin}{\rho}$ \Comment{Least populated}
       \State $\{j\} \gets \Call{where}{\hat{a}_{\vvec,r}=\hat{\avec}_s}$
       \If{$\{j\}\ne\emptyset$}
          \State $j^* \gets \Call{argmin}{\lambda_{\vvec,r}[\{j\}]}$ \Comment{Closest}
          \State $\vvec_{j^*} \gets \overrightarrow{\Frontset}_r[j^*]$
          \State $\lambda_{\vvec_{j^*}} \gets \lambda_{\vvec,r}[j^*]$
          \State $\Sset \gets \Sset \bigcup \{(\vvec_{j^*},r,\lambda_{\vvec_{j^*}},\hat{\avec}_s)\} $
          \State $\overrightarrow{\Frontset}_r.\text{pop}(\vvec_{j^*}); \lambda_{\vvec,r}.\text{pop}(\lambda_{\vvec_{j^*}}); \hat{a}_{\vvec,r}.\text{pop}(\hat{\avec}_s)$ 
          \State $\rho[s] \gets \rho[s] + 1$
        \Else
         \State $\{\hat{\avec}_h\}.\text{pop}(\hat{\avec}_s)$ \Comment{Empty ref. direction}
       \EndIf
     \EndFor
   \EndIf
   \Assert $\#\Sset > N_\text{p}$
   \State \# Proceed to binary tournament in $\Sset$
   \State $\Zset = \emptyset$ \Comment{Parents set holder}
   \While{$\#\!\Zset \ne N_\text{p}$} 
     \State $(\zvec_1,r_1,\lambda_{\zvec_1}), (\zvec_2,r_2,\lambda_{\zvec_2}) \gets \Call{random2}{\Sset}$
     \If{$(\zvec_1,r_1,\lambda_{\zvec_1}) \prec_\Sset (\zvec_2,r_2,\lambda_{\zvec_2})$} \Comment{\eqref{eqn:precpool}}
       \State $\Zset \gets \Zset \bigcup \{\zvec_1\}$    \Comment{Winner 1}
     \Else
       \State $\Zset \gets \Zset \bigcup \{\zvec_2\}$    \Comment{Winner 2}
     \EndIf
   \EndWhile
   \State\Return{$\Zset$}  \Comment{No duplicates mating pool}
 \EndFunction
 \end{algorithmic}
\end{algorithm}
\par
 Finally, the subsequent steps in the optimization loop (algo.~\ref{alg:evolve},
 lines~6--12: crossover, mutation, MD/$\nnmap$ simulation) follow exactly the same
 operations described in the NSGA-II method (sec.~\ref{subsec:pswcd}), and when
 $\delta = 2$ with $M=5$ (all thermophysical properties included), then $H=15$ from
 \eqref{eqn:binomial}. Hence choosing $N_\text{p}=5$ and $N_\text{c}=10$, making the
 population size $N_\text{p} + N_\text{c} = H$, exactly one LJ parameter vector in the
 population is associated to each of the $H$ reference points in objective function space.
%

%
%
%
\section{Results and discussion}\label{sec:rd}
 Equipped with the algorithms previously described (sec.~\ref{sec:oc}),
 we analyzed thermodynamic and transport properties through several approaches starting
 with individual property objective functions \eqref{eqn:minproblem-indiv}
 (sec.~\ref{subsec:iofo}) which allowed for insights on the thermophysical properties
 sensitivity to specific LJ parameters (sec.~\ref{subsec:single}). Next we looked at
 simultaneous two (sec.~\ref{subsec:two}) and three (sec.~\ref{subsec:three}) properties
 using the NSGA-II for contrasting the optimization of thermodynamic properties
 and transport properties respectively. Finally (sec.~\ref{subsec:five}), the optimization
 study of all properties together ($M=5$) used the NSGA-III and, for comparison, NSGA-II,
 algorithms.
 In all sections, full MD simulations were performed in the optimization loop. Once
 sufficient data was obtained, a neural network fit was built off-line and used to replace
 the MD simulation in the loop (sec~\ref{subsec:NN}). A hybrid approach using both MD and
 a simultaneously on-demand trained neural network fit was not implemented in this work.
\par
 We use the Pymoo~\cite{blank2020pymoo} library in Python to set up the algorithms
 described earlier (sec.~\ref{sec:oc}). Pymoo is a framework that provides scalar- and
 vector-valued objective function optimization techniques. The initial 5 parent LJ
 parameter vectors (first generation) are taken from the five columns in
 Table~\ref{tbl:lj}. The MD simulations were run on a high-performance cluster with
 LAMMPS~\cite{plimpton1995fast} as parallel tasks on \num{32} CPUs. Generally, each MD
 simulation required a few hours to complete, and each optimization case (15 generations
 with population of 15) took approximately one month of wall-clock time.
%
\subsection{Individual objective function (thermophysical properties)}\label{subsec:single}
\begin{figure*}
 \begin{center}
  \graphicspath{{figs/}}
  \includegraphics[width=6.2in]{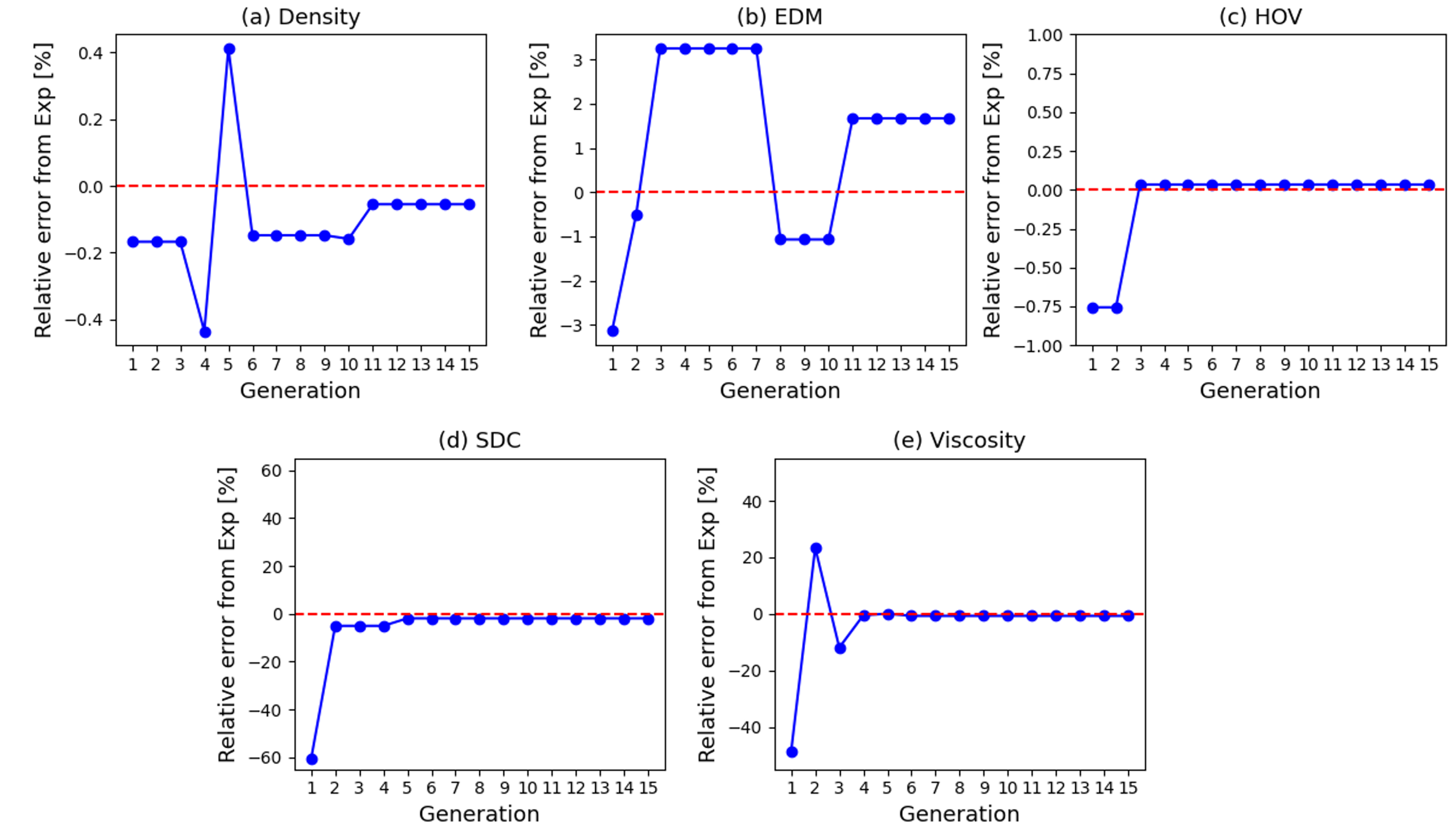}
 \end{center}
 \caption
  []
  {Percentage relative deviation from experimental values for each thermophysical property
   for the best parameter vector per generation of individual objective function GA
   optimization (fig.~\ref{fig:ga-oploop}). There are 15 equilibrium MD simulations done
   per generation point per property. Experimental values listed in the caption of
   Table~\ref{tbl:one-obj}.}
 \label{fig:sobj}
\end{figure*}

 Applying the optimization loop described earlier (sec.~\ref{subsec:oploop}), and
 using a single objective function~\eqref{eqn:minproblem-indiv} for each property to
 perform a GA optimization (sec.~\ref{subsec:iofo}), thermophysical properties were
 computed as the objective function value, $F_k(\vvec)$, was reduced for optimized values
 of the LJ parameters, $\vvec^*_k$ (table~\ref{tbl:one-obj-optlj}). Within 15 generations,
 the percentage relative deviation from experimental values of the predicted
 thermophysical properties, $\pm\sqrt{F_k(\vvec^*_k)}$, reduced to an excellent result
 (fig.~\ref{fig:sobj}); note
 that all properties in the first generation are under-predicted. In later generations all
 property values improved, in particular,
 transport properties had a substantial gain in accuracy as compared to the initial values
 in the first generation (table~\ref{tbl:lj}) \cite{hatami-de_almeida25:art}.
\begin{table*}
 \caption[]
 {Optimized LJ parameters, $\vvec^*_k, \ k=1,\ldots,5$ (in units of
  [\si{\calorie\per\mol}] and [\si{\angstrom}], respectively), using a single objective
  function $F_k(\vvec^*_k), \ k=1,\ldots,5$.}
 \label{tbl:one-obj-optlj}
 \begin{center}
 \begin{tabular}{ccccccccccc}
 \toprule
     Atom & \multicolumn{2}{c}{Density, $F_1(\vvec^*_1)$} & \multicolumn{2}{c}{EDM, $F_2(\vvec^*_2)$} & \multicolumn{2}{c}{HOV, $F_3(\vvec^*_3)$} & \multicolumn{2}{c}{SDC, $F_4(\vvec^*_4)$}  & \multicolumn{2}{c}{Viscosity, $F_5(\vvec^*_5)$} 
 \\ \cline{2-3} \cline{4-5} \cline{6-7} \cline{8-9} \cline{10-11}
 Type &  $\sigma_i$ & $\epsilon_i$ & $\sigma_i$ & $\epsilon_i$ & $\sigma_i$ & $\epsilon_i$ & $\sigma_i$ & $\epsilon_{i}$ & $\sigma_i$ & $\epsilon_{i}$
 \\ \midrule
     O2 & 2.960 & \red{112 ($\downarrow$)}  & 3.037 & \red{210 ($\uparrow$)}  & 3.039 & \red{210 ($\uparrow$)}  & 2.965 & \red{143 ($\downarrow)$}  & 3.181 & \red{214 ($\uparrow)$}   \\
  P  & 3.742 & 200  & 3.742 & 200  & 3.741 & 200  & 3.740 & 200  & 3.742 & 200   \\
  OS & 3.000 & 172  & 3.025 & 170  & 3.025 & 168  & 2.840 & 139  & 2.999 & 169   \\
 \hline
  C0 & 3.499 & 65   & 3.401 & 109  & 3.379 & 108  & 3.500 & 65   & 3.493 & 63    \\
  C1 & 3.507 & 65   & 3.401 & 109  & 3.379 & 108  & 3.379 & 64   & 3.496 & 495   \\
  C2 & 3.393 & 110  & 3.400 & 109  & 3.379 & 108  & 3.376 & 66   & 3.379 & 109   \\
  C3 & 3.492 & 110  & 3.400 & 110  & 3.379 & 108  & 3.380 & 65   & 3.510 & 109   \\
 \hline
  C$_\text{ave}$ 
     & 3.473 & 88   & 3.401 & 109  & 3.379 & 108  & 3.409 & \red{65 ($\downarrow$)}  & 3.470  & \red{194 ($\uparrow$)}  \\
 \hline
  H0 & 2.499 & 30   & 2.650 & 16   & 2.644 & 15   & 2.501 & 15   & 2.492 & 15    \\
  H1 & 2.503 & 15   & 2.650 & 16   & 2.583 & 16   & 2.613 & 17   & 2.505 & 29    \\
  H2 & 2.649 & 30   & 2.650 & 16   & 2.584 & 16   & 2.501 & 16   & 2.589 & 16    \\
  H3 & 2.660 & 29   & 2.651 & 16   & 2.583 & 16   & 2.500 & 30   & 2.579 & 17    \\
 \hline
  H$_\text{ave}$
     & 2.578 & \red{26 ($\uparrow$)}  & 2.650 & \red{16 ($\downarrow$)} & 2.599 & \red{16 ($\downarrow$)} & 2.529 & 20 & 2.541 & 19  \\
 \bottomrule
 \multicolumn{11}{l}{Arrows indicate a significant ($\approx$20\%) increase or decrease compared to Table~\ref{tbl:ljref}.} \\
 \multicolumn{11}{l}{C$_\text{ave}$ row is the average of the carbon atoms parameters.} \\
 \multicolumn{11}{l}{H$_\text{ave}$ row is the average of the hydrogen atoms parameters.}
 \end{tabular}
 \end{center}
\end{table*}
\par
 It is instructive to note that typical force fields (table~\ref{tbl:lj}) have values of
 the LJ parameters $\sigma_i$ and $\epsilon_i$ that are the same for different atom types.
 For example, the \texttt{C0} carbon has the same LJ parameters as for all other carbon
 atoms. Similarly for hydrogen atoms. By allowing all parameters to change per atom via an
 optimization algorithm aimed at making 
 better predictions of thermophysical properties, we have a chance to specialize the LJ potential
 for the molecule at hand.
\par
 For each individual objective function GA optimization \eqref{eqn:minproblem-indiv}, all thermophysical 
 properties are computed
 for comparison to their correspondent experimental values. This provides information on how one
 individual objective function optimization affects the properties not being optimized.
 The optimization of LJ parameters (table~\ref{tbl:one-obj-optlj}) based on an individual 
 property objective function had a detrimental effect in every other property not being 
 optimized (table~\ref{tbl:one-obj}).
 This adverse effect is acute when optimizing transport properties, in particular when
 trying to optimize LJ parameters for shear viscosity prediction (viscosity row in table~\ref{tbl:one-obj})
 wherein all other properties display their largest errors as compared to the simulations in all other rows.
\begin{table*}[pos=h]
 \setlength{\tabcolsep}{3.0pt}
 \caption [] {Single objective function GA optimization results for TBP properties.
 Red color for the property being optimized with experimental values chosen for
 comparison~\cite{hatami-de_almeida25:art}:
 \SI{0.9729}{\gram\per\centi\meter\cubed}, \SI{3.242}{\debye}, \SI{24.42}{\kilo\calorie\per\mol},
 \SI{2.29e-6}{\centi\meter\squared\per\second} and \SI{3.363}{\milli\pascal\second}. While
 each optimization achieves excellent precision for its targeted property, improvement in
 one property leads to a deterioration of all others relative to experimental values.}
 \label{tbl:one-obj}
 \begin{center}
 \begin{tabular}[c]{ccccccc}
 \toprule
  Property & \# of TBP & {Density [\si{\gram\per\centi\meter\cubed}]} 
                          & {EDM [\si{\debye}]} 
                          & {HOV [\si{\kilo\calorie\per\mol}]}
                          & {SDC [\SI{e-6}{\centi\meter\squared\per\second}]} 
                          & {Viscosity [\si{\milli\pascal\second}]} 
\\ \midrule
Density & 48 & \red{  \num{0.9722} (\num{-0.05}\%) }  & 
                            \num{2.955} (\num{-8.8}\%)  &  
                             \num{24.90} (\num{1.9}\%)    &  
                             \num{0.406} (\num{-82.2}\%)   &  
                             \num{1.62} (\num{-51.8}\%)    \\
                          & 512 & \red{  \num{0.9723} (\num{-0.03}\%) }  & 
                            \num{2.935} (\num{-9.4}\%)  &  
                             \num{24.95} (\num{2.1}\%)    &  
                             \num{0.532} (\num{-76.7}\%)   &  
                             \num{1.9} (\num{-43.3}\%)    
\\
EDM & 48 & \num{0.971 } (\num{-0.2}\%)  & 
                       \red{ \num{3.225 } (\num{-0.5}\%) } & 
                        \num{23.67} (\num{-3.0}\%)  & 
                        \num{0.494} (\num{ -78.4}\%) & 
                        \num{1.5} (\num{-55.3}\%)
\\
HOV& 48 & \num{0.994 } (\num{2.1}\%) & 
                         \num{3.20} (\num{-1.2}\%)  & 
                        \red{ \num{24.42} (\num{0.00}\%) } &
                         \num{0.384} (\num{-83.2}\%) & 
                         \num{1.839} (\num{ -45.3}\%)
\\
SDC& 48 & \num{0.938} (\num{-3.5}\%) &
                         \num{3.155} (\num{-2.6}\%)  & 
                         \num{17.17} (\num{-29.6}\%)   &
                        \red{ \num{2.246} (\num{-1.9}\%) }  & 
                         \num{0.799} (\num{-76.2}\%)
\\
Viscosity& 48 & \num{1.072} (\num{10.0}\%)   &
                         \num{3.59} (\num{10.8}\%)  & 
                         \num{36.9} (\num{51.2}\%)   &
                         \num{0.106} (\num{-95.3}\%)  & 
                        \red{ \num{3.3626} (\num{-1.1}\%)} 
\\\bottomrule
 \end{tabular}
 \end{center}
\end{table*}
\par
 The best values (within 15 generations) of the LJ parameters, $\vvec^*_k$, obtained for each $k^\text{th}$ 
 single objective function optimization show (table~\ref{tbl:one-obj-optlj}) that the GA changes the values 
 of $\sigma_i$ and $\epsilon_i$ for carbon and hydrogen atoms that otherwise have the same values 
 (table~\ref{tbl:lj}). This provides the flexibility and opportunity for exploiting optimization, and allows
 for the understanding of the origin of the deficiency of LJ parameter values.
 In the present analysis we are interested in the most salient changes in LJ parameters and their impact on
 thermophysical properties. We note (table~\ref{tbl:one-obj-optlj}) that changes in distance parameters were 
 all significantly less than \SI{20}{\percent}, and only the interaction energy well parameter of a few atom 
 types were changed at the level of $\approx$\SI{20}{\percent}. The latter are those we will explore in our 
 analysis.
\par
 The optimized LJ parameters (table~\ref{tbl:one-obj-optlj}) for mass density alone showed
 that the only notable changes for improvement of this property were a reduction
 (-\SI{64}{\calorie\per\mole}) of the phosphoryl oxygen \texttt{O2} interaction potential
 well depth, and an increase (+\SI{4.5}{\calorie\per\mole}) of the average interaction
 potential well depth of the hydrogen atoms (table~\ref{tbl:one-obj-optlj}). The distance
 parameters were not significantly changed, therefore given the large number of hydrogen
 atoms on the butyl chains, the increased van der Waals attraction contributed to tighter
 packing resulting in an increase of mass density, and an excellent agreement with the
 experimental value at later generations (fig.~\ref{fig:sobj}a). In addition, a
 simulation with these optimal parameters for mass density is performed with a much
 larger number of TBP molecules (512 molecules, table~\ref{tbl:one-obj}, second row for
 density) demonstrating that the usage of a smaller system for optimization of parameters
 does scale to larger systems. Another observation, looking at the mass density row of
 results (table~\ref{tbl:one-obj}), all other properties had their values worsened by
 optimizing mass density.
\par
 The optimized LJ parameters (table~\ref{tbl:one-obj-optlj}) for EDM alone showed a
 reduction (-\SI{5.5}{\calorie\per\mole}) of the
 average interaction potential well depth for hydrogen atoms
 (table~\ref{tbl:one-obj-optlj}) at the second generation (fig.~\ref{fig:sobj}b). In
 addition, an increase for the correponding \texttt{O2} energy
 of \SI{34}{\calorie\per\mole} also resulted from the optimization. Since these changes
 are the opposite of what is needed to improve the mass density prediction, they worsen
 the value of mass density as shown in the EDM row of the mass density column. The GA
 optimization of EDM was unique (fig.~\ref{fig:sobj}b) in that the second generation had
 the best prediction as compared to any other generation.
\par
 The optimized LJ parameters (table~\ref{tbl:one-obj-optlj}) for HOV delivered a near
 match to the experimental value, \SI{0}{\percent} deviation (table~\ref{tbl:one-obj}).
 Interestingly, the optimized LJ parameter for HOV are essentially the same as those
 obtained for EDM, therefore an improvement in predicting HOV should lead to an
 improvement in predicting EDM which can be seen in the HOV row at the EDM column, where
 the relative error for EDM compared to the experimental result is -\SI{1.2}{\percent},
 which is an improvement over the initial error -\SI{3}{\percent} at the start of the
 optimization (fig.~\ref{fig:sobj}b). Conversely, since optimization of EDM and HOV are
 aligned, improving HOV leads to a deterioration of mass density, as corroborated by the
 data (table~\ref{tbl:one-obj}).
\par
 The optimized LJ parameters for SDC alone delivered an impressive small relative error of 
 -\SI{1.9}{\percent} (table~\ref{tbl:one-obj}) which is much improved from the starting value of 
 -\SI{60}{\percent} (fig.~\ref{fig:sobj}d) at the first generation. The reasons for improvement were a
 decrease of the interaction energy potential well depth of the phosphoryl oxygen by 
 \SI{33}{\calorie\per\mole}, and a decrease of the average interaction energy potential well depth
 for carbon atoms by \SI{27}{\calorie\per\mole} (table~\ref{tbl:one-obj-optlj}). This decrease of 
 interaction energy, in particular on carbon atoms in all butyl groups is consistent with increasing the 
 mobility of the molecules and consequently giving rise to a larger self-diffusion coefficient 
 (fig.~\ref{fig:sobj}d). It comes with no surprise that all thermodynamic properties predicted by a more 
 mobile molecule have higher deviations from experimental values than those obtained with LJ parameter in 
 the first generation (SDC row in table~\ref{tbl:one-obj}).
\par
 Finally, the shear viscosity optimal LJ parameters obtained (table~\ref{tbl:one-obj-optlj}, righ-most 
 column) produce an outstanding viscosity value as compared to the experimental average, \SI{-1.10}{\percent} 
 relative error (table~\ref{tbl:one-obj}). As with the prediction of SDC, the viscosity result is slightly 
 underpredicted, consistent with our past research~\cite{hatami-de_almeida25:art} wherein transport
 properties are underpredicted whether equilibrium or non-equilibrium MD simulations are performed.
 In contrast to SDC results, the optimal values, $\vvec^*_5$, differ significantly from the average initial 
 population (table~\ref{tbl:ljref}) in the exact opposite way as with $\vvec^*_4$ for SDC, that is, an 
 increase of the \texttt{O2} interaction potential energy well of \SI{38}{\calorie\per\mole} is observed, 
 and similarly an increase of the \texttt{C}$_\text{ave}$ interaction potential energy well of 
 \SI{102}{\calorie\per\mole}. Therefore what improves the prediction of shear viscosity is the opposite to 
 what improves the prediction of SDC on the same most sensitive LJ parameters, $\epsilon_\text{O2}$, and
 $\epsilon_{\text{C}_\text{ave}}$. That is, a more mobile molecule experience higher SDC, but less viscous 
 shear. The optimization of the LJ parameters for viscosity leads to the most significant prediction 
 degration of thermophysical properties not being optimized (table~\ref{tbl:one-obj}, last row).
\par
 In summary, with the exception of EDM and HOV, optimization of LJ parameters for
 predicting an individual thermophysical property leads to the degradation of all other
 properties not being optimized. This also holds true for the transport properties
 themselves, SDC and shear viscosity, where LJ parameters that lead to a more mobile
 molecule, that is, better SDC prediction, also lead to a less viscous fluid worsening the
 prediction of shear viscosity. The LJ distance parameters of the initial population do
 not change appreciably when optimizing the prediction of the thermophysical properties,
 and only the atomic attractive pair potential well of the atom types \texttt{O2} and
 hydrogens significantly change for thermodynamic properties. As for transport properties,
 the interaction energy of \texttt{O2} and carbons are the most affected. The energy
 changes are in the order of tens of \unit{\calorie\per\mole}.
%
\subsubsection{Sensitivity of thermophysical properties}\label{subsec:sensitivity}
\par
\begin{figure*}
 \begin{center}
  \graphicspath{{figs/}}
  \includegraphics[width=5in]{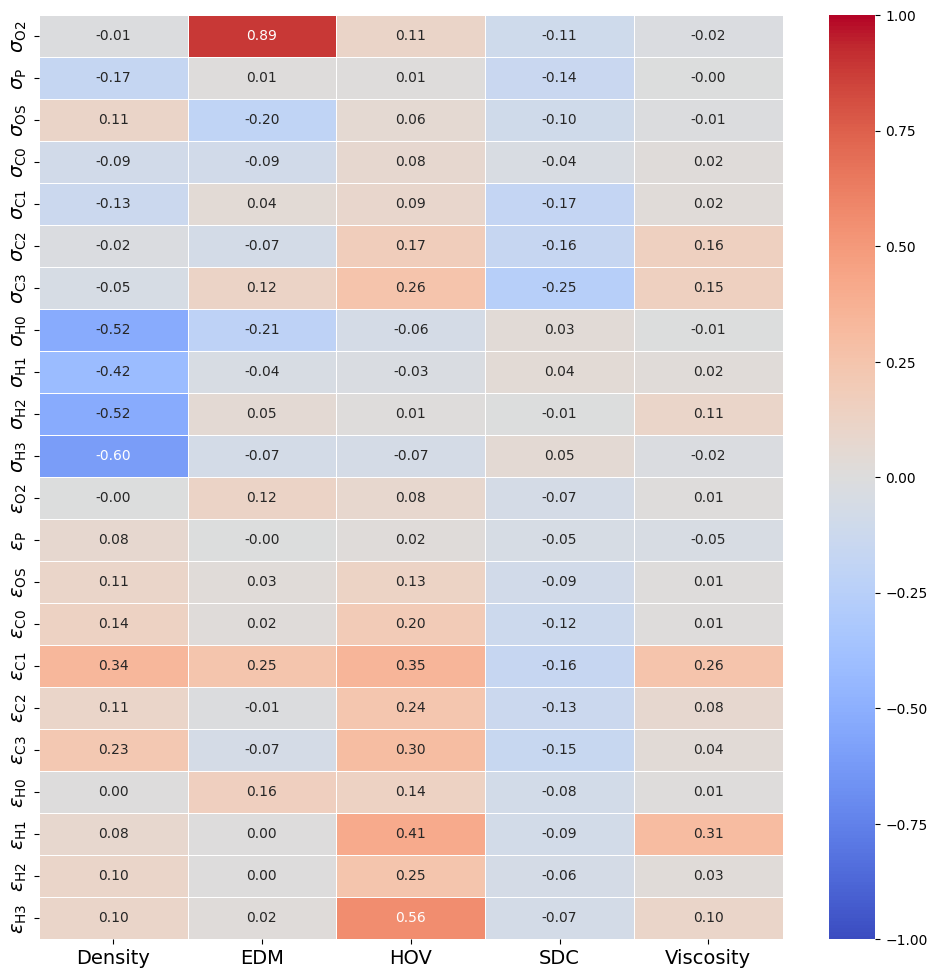}
 \end{center}
 \caption 
  []
  {Correlation coefficients \eqref{eqn:pearson} between LJ parameters and thermophysical properties.}
  \label{fig:heatmap}
\end{figure*}

 The previous analysis (sec.~\ref{subsec:single}) was made relative to the initial parent 
 population which is a significant one since the parameters stem from well known force 
 fields (table~\ref{tbl:lj}). However a complementary
 intrinsic analysis can be performed using the sensitivity of the thermophysical properties
 with respect to the LJ parameters. This can be presented in terms of the Pearson 
 correlation coefficients:
 \begin{align}\label{eqn:pearson}
 r_{jk} = \frac{
     \sum_{n=1}^{S}\bigl((\vvec_n)_j - \overline{(\vvec_n)_j}\bigr)  
                   \bigl(y_k(\vvec_n) - \overline{y_k}\bigr)
 }
 {
  \sqrt{
  \sum_{n=1}^{S}\bigl((\vvec_n)_j - \overline{(\vvec_n)_j}\bigr)^2 
  \sum_{n=1}^{S}\bigl(y_k(\vvec_n) - \overline{y_k}\bigr)^2}
  }, 
 \end{align}
 where $-1 \le r_{jk} \le 1$ quantifies the correlation of the $j^\text{th}$ LJ parameter,
 $(\vvec)_j$, to the $k^\text{th}$ property when a $S$-sample of
 $\Positives^N\times \Positives^M$ pairs, $\bigl(\vvec,\yvec(\vvec)\bigr)$, are available.
 The $S$-sum in \eqref{eqn:pearson} is over all successful MD simulations performed in the
 course of this
 study ($S=1143$). The numerator of \eqref{eqn:pearson} calculates the covariance between
 the two variables, while the denominator normalizes the covariance by dividing by the
 product of the standard deviations of the two variables.  The correlation coefficients
 \eqref{eqn:pearson} can be arranged in a table (fig.~\ref{fig:heatmap}) in accord to a
 specific LJ parameter and TBP property pair.
 The color of each table cell indicates the strength and direction of the correlation,
 positive correlations signify that an increase or decrease of LJ parameter values cause
 a similar variation of the associated TBP properties. Conversely negative correlations
 produce the opposite variation.
\par
 For mass density (fig.~\ref{fig:heatmap}), a reduction of the distance parameter,
 $\sigma$, for all atom types (closer packing) results in an increase of mass density
 with the exception of $\sigma_\text{OS}$ which has a relatively small magnitude
 correlation.
 Conversely an increase in the attractive pair potential well parameter, $\epsilon$, for
 all atom types (closer packing) leads to an increase in mass density with the exception
 of $\epsilon_\text{O2}$ for which its sensitivity is very small in magnitude.
\par
 It is notable that mass density (fig.~\ref{fig:heatmap}) exhibits a weak dependency on
 the LJ pair potential well parameters in general. However this dependency for the
 headgroup \texttt{O2} atom type and tailgroup for hydrogen types
 support the variation behavior described earlier (table~\ref{tbl:one-obj-optlj}) wherein
 the $\epsilon$ parameter for these atom types varied the most during parameter
 optimization since
 they are the least sensitive. The direction of the variation is also supported by the
 sensitivity results, since the correlation coefficient for $\epsilon_{\text{O2}}$ is
 small and negative ($\approx -\num{0.0}$), its value decreased significantly in the
 optimization of mass density (table~\ref{tbl:one-obj-optlj}, 112($\downarrow$)). The
 same conclusion can be made for the hydrogen tailgroup $\texttt{H0}$, $\texttt{H1}$,
 $\texttt{H2}$, and $\texttt{H3}$, with small correlation coefficients (\num{0.0},
 \num{0.08}, \num{0.10} and \num{0.10}) leading to a significant increase in the average
 $\epsilon_{\text{H}_{\text{ave}}}$ for mass density optimization
 (table~\ref{tbl:one-obj-optlj}, 26($\uparrow$)). Conversely the correlation coefficients
 for the hydrogen tailgroup associated to the distance parameter $\sigma$, \num{-0.52},
 \num{-0.42}, \num{-0.52} and, \num{-0.60} (fig.~\ref{fig:heatmap}, mass density), were
 the most significant leading to a small change in the LJ distance parameter variation
 during optimization.
\par
 The sensitivity results helped clarify the improvement of mass density predictions using
 the individual-objective-function GA optimization which primarily activated the
 variation of
 the interaction energy well parameter of \texttt{O2} and \texttt{H$_\text{ave}$}
 (table~\ref{tbl:one-obj-optlj}) because they are weakly sensitive allowing for larger
 parameter variations.
 \par
 For EDM one expects variations of atomic partial charges and intra-molecular force field
 parameters to be the most relevant (not taken into account in this study). This shows in
 the LJ parameter sensitivity results obtained in this study where in general the
 correlation
 coefficients are relatively small with the exception of the highly sensitive LJ distance
 parameter of the \texttt{O2} atom type
 (correlation coefficient of \num{0.89}, fig.~\ref{fig:heatmap}, EDM column) because it
 is the most electronegative exposed atom in the TBP molecule, and because the distance
 of this atom to the molecule center of mass has a significant contribution to the
 molecular electric dipole moment. Accordingly, $\sigma_{\text{O2}}$ has a positive
 correlation with EDM which is consistent with its definition. The sensitivity results
 indicate that in general EDM is not directly sensitive to variations in other parameters
 (fig.~\ref{fig:heatmap}, EDM column) and the great majority of interaction energy well
 parameters are positively correlated with EDM. Earlier we showed
 (sec.~\ref{subsec:single}) that the decrease of the interaction energy well parameters of
 the butyl hydrogens through $\epsilon_{\text{H}_\text{ave}}$
 (table~\ref{tbl:one-obj-optlj}, EDM column) activated in the GA optimization was
 significant so was the increase in $\epsilon_\text{O2}$. Since both these quantities are
 positively correlated with EDM, the increase in the latter was the dominant factor in
 improving the result for the optimized EDM value. 
\par
 Although the significant variations in parameter optimization for the prediction of EDM
 and HOV are aligned (observed earlier in sec.~\ref{subsec:single}), the sensitivity of
 HOV with respect to LJ parameters differs significantly from EDM (fig.~\ref{fig:heatmap},
 compare columns EDM and HOV). This is consistent with the distinct nature of these two
 properties. HOV is the only property where all interaction energy well parameters are
 positively correlated, and the highest positively correlation is for the terminal
 $\epsilon_{\text{H3}}$. This implies that the interaction strength of this specific
 hydrogen site plays an important role in determining the energy required for phase
 transition from liquid to vapor. Here again (as compared to EDM), the optimization of
 HOV implied a decrease in $\epsilon_{\text{H}_\text{ave}}$
 (table~\ref{tbl:one-obj-optlj}, HOV column) which decreases HOV, however this is
 compensated by the increase in $\epsilon_\text{O2}$ to deliver an almost exact value for
 HOV as compared to the experimental result (table~\ref{tbl:one-obj}).
\par
 The SDC is the only property that has a negative correlation for all LJ interaction
 energy well parameters (fig.~\ref{fig:heatmap}, SDC column, $\epsilon$), moreover this
 correlation is relatively weak, $\lvert r_{\epsilon,\text{SDC}}\rvert < 0.2$. Therefore,
 decreasing $\epsilon$ increases SDC and this behavior supports the observation in the
 optimization results (table~\ref{tbl:one-obj-optlj}, SDC column). Despite the weakness
 of the correlation, the energy variations, \SI{-33}{\calorie\per\mole} for
 $\epsilon_\text{O2}$ and \SI{-27}{\calorie\per\mole} for
 $\epsilon_{\text{C}_\text{ave}}$, were sufficient to optimize the prediction of SDC to
 an excellent value as compared to the experimental reference (table~\ref{tbl:one-obj}).
\par
 Contrary to SDC, shear viscosity shows a positive correlation for the overwhelming
 majority of the interaction potential well parameters (fig.~\ref{fig:heatmap}, viscosity
 column, $\epsilon$).  This is consistent with the significant increase in energy observed
 in the optimization of shear viscosity for the parameters: $\epsilon_\text{O2}$ and
 $\epsilon_{\text{C}_\text{ave}}$. Similarly to SDC, the LJ parameter sensitivities for
 viscosity are relatively weak, nevertheless the increases in energy of
 \SI{38}{\calorie\per\mole} for $\epsilon_\text{O2}$ and \SI{102}{\calorie\per\mole} for
 $\epsilon_{\text{C}_\text{ave}}$, were sufficient to optimize the predicted shear
 viscosity to an equally excellent value compared to the experimental reference
 (table~\ref{tbl:one-obj}). The highest sensitivity coefficients obtained were for
 $\epsilon_\text{C1}$ and $\epsilon_\text{H1}$, and those were the same atom types for
 which the parameter optimization varied the most relative to others of the same group
 (\SI{495}{\calorie\per\mole} for carbon and \SI{29}{\calorie\per\mole} for hydrogen). We
 note that the optimzized LJ parameter $\epsilon_\text{C1}$ is somewhat unrealistic but
 this is a consequence of allowing parameters to vary at will (sec.~\ref{subsec:iofo})
 for the sake of evaluating the algorithm proposed.
\par
 In summary the sensitivity results corroborated the trends observed in the parameter
 optimization of individual thermophysical properties (sec.~\ref{subsec:single}) and shed
 light on the correlation magnitudes. It is found that the most significantly activated
 parameters during optimization (table~\ref{tbl:one-obj-optlj}) were not necessarily the
 most sensitive. This point to the fact that the sensitivity information, when gathered
 after the fact, could be recycled back into the optimization genetic algorithm heuristics
 to improve the search mechanism for better or more diverse LJ parameter vector options
 (a future research topic).
%
\subsection{Two objective functions (transport properties)}\label{subsec:two}
\begin{figure}
 \begin{center}
  \graphicspath{{figs/}}
  \includegraphics[width=3in]{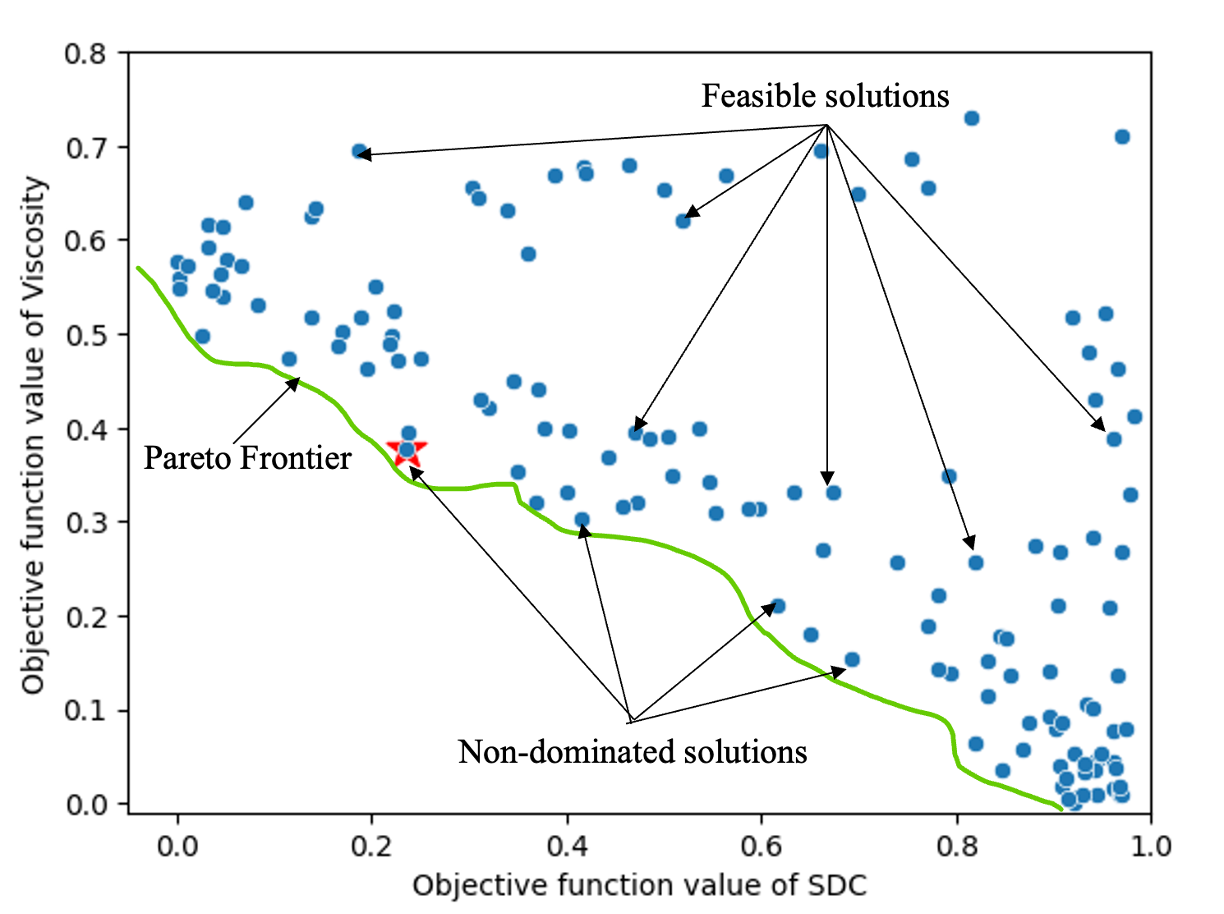}
 \end{center}
 \caption 
 []
 {Two-dimensional objective function space for the cumulative populations during
  optimization of EDC and shear viscosity. The non-convex Pareto optimal front shows
  Pareto points obtained through the NSGA-II method. The red star marks the selected
  optimal solution with objective function values of 0.2357 and 0.378 for SDC and shear
  viscosity respectively, corresponding to errors of \num{-48.3}\% and \num{-61.5}\%,
  respectively, relative to experimental data.}
 \label{fig:two}
\end{figure}

 In this section, we discuss the results of a combined MOO applied to two transport
 properties: SDC and shear viscosity. Thermodynamic properties are left alone without any
 control allowing for a focused investigation on transport properties. The computed
 objective function space (fig.~\ref{fig:two})
 presents a non-convex Pareto front that collects the best parameter vectors and
 trade-offs between
 SDC and viscosity. Unlike convex optimization landscapes, search in non-convex spaces
 presents intricate challenges and an extra layer of complexity to the decision-making
 process. The inherent difficulty arises from the fact that navigating the
 non-convex region of the Pareto optimal front poses challenges in locating solutions
 that truly represent the best compromise between the two objectives. That is, a rough
 front may hide an optimal solution that could be missed during the search.
 A larger number of generations and populations of solutions would alleviate this
 difficulty but would also increase the computational cost.
\par
 The optimization process provides results encompassing the non-dominated set of
 solutions derived from the NSGA-II algorithm (algo.~\ref{alg:nsga-ii}). A single solution
 from this pool is obtained using the Euclidean distance between each solution and the
 ideal point (origin). The solution with the minimum Euclidean
 distance is chosen as the best selection, a favorable trade-off between SDC and
 viscosity, and denoted by a red star on the objective function space (fig.~\ref{fig:two}).
 The optimized LJ parameters obtained (table~\ref{tbl:two-obj-optlj}) are similar to those
 of the OPLS2005 force field (table~\ref{tbl:lj}) except for a reduction of
 \SI{23}{\kilo\calorie\per\mole} on the \texttt{O2} atom pair potential well parameter,
 $\epsilon_\text{O2}$. Compared to the average property values of the initial parent
 population, the greatest variations were in the interaction potential well energy for
 the carbon and hydrogen butyl atoms (arrows in table~\ref{tbl:two-obj-optlj}). The
 sensitivity of the corresponding LJ parameters
 for these atom types for SDC and shear viscosity is exactly the opposite 
 (fig.~\ref{fig:heatmap})
\begin{table}
 \caption[]
 {Optimized LJ parameters, $\vvec^*$, using two objective functions.
  $F_k(\vvec^*), \ k=4,5$
  for joint SDC and shear viscosity coefficients, respectively.}
 \label{tbl:two-obj-optlj}
 \begin{center}
 \begin{tabular}{ccc}
 \toprule
 Atom & \multicolumn{2}{c}{$\vvec^*$ for $F_4$, $F_5$} \\
 Type &  $\sigma_i$ [\si{\angstrom}] & $\epsilon_i$ [\si{\calorie\per\mol}] \\
 \midrule
  O2  & 2.980 & 177  \\
  P   & 3.740 & 200  \\
  OS  & 2.850 & 140  \\
 \hline
  C0  & 3.500 & 65   \\
  C1  & 3.500 & 65   \\
  C2  & 3.500 & 65   \\
  C3  & 3.500 & 65   \\
 \hline
 C$_\text{ave}$ 
      & 3.500 & 65 ($\downarrow$) \\
 \hline
  H0  & 2.500 & 30   \\
  H1  & 2.500 & 30   \\
  H2  & 2.500 & 30   \\
  H3  & 2.500 & 30   \\
 \hline
 H$_\text{ave}$ 
      & 2.500 & 30 ($\uparrow$)   \\
 \bottomrule
 \multicolumn{3}{l}{Arrows indicate $\ge 20\%$ change relative to table~\ref{tbl:ljref}.}
 \end{tabular}
 \end{center}
\end{table}
%
%
\par
 With the optimized parameters in hand they are employed to derive the thermodynamic and
 transport properties (table~\ref{tbl:two-obj}) for systems comprising of \num{48} TBP
 molecules.
\begin{table}[pos=h]
 \setlength{\tabcolsep}{3.0pt}
 \caption [] 
 {Two objective functions NSGA-II optimization results for TBP properties.
  Red color for the properties being optimized with experimental values as in Table~\ref{tbl:one-obj}. 
  The optimization is not effective in optimizing the intended properties simultaneously.}
 \label{tbl:two-obj}
 \begin{center}
 \setlength{\abovetopsep}{0pt}
 \setlength{\belowrulesep}{0pt}
 \begin{tabular}[c]{l|c}
 \thickhline
 Obj. Func. & $F_4$, $F_5$ \\
 \# of TBP & 48  \\
 \hline
 {Density [\si{\gram\per\centi\meter\cubed}]} & \num{0.9705} (\num{-0.22 }\%) \\
 {EDM [\si{\debye}]} & \num{3.193 } (\num{-1.51 }\%) \\
 {HOV [\si{\kilo\calorie\per\mol}]} & \num{21.646} (\num{-11.35 }\%) \\
 \hline
 {SDC [\SI{e-6}{\centi\meter\squared\per\second}]} & \red{\num{1.184} (\num{-48.3}\%)} \\
 {Viscosity [\si{\milli\pascal\second}]} & \red{\num{1.295} (\num{-61.5}\%)} \\
 \thickhline
 Overall deviation & \num{24.6}\%
 \end{tabular}
 \end{center}
\end{table}
 The resulting values for SDC and viscosity are
 \SI{1.184e-6}{\centi\meter\squared\per\second} and \SI{1.295}{\milli\pascal\second}
 respectively. Both properties exhibit a high error of
 \num{-48.3}\% and \num{-61.5}\% relative to experimental data. Compared to the initial
 error (first generation in figs.~\ref{fig:sobj}d, e) and the trend described in the
 single objective function results (table~\ref{tbl:one-obj-optlj}), it is consistent that
 the error in the value of SDC is smaller than the starting value for the single
 objective function, and the error in the value of viscosity is higher than the starting
 value. This follows from the sole reduction of $\epsilon_\text{O2}$ (as mentioned earlier) in the two-objective-function optimization.
 The results do not look discouraging in view of the fact that only \num{15} generations
 of a population of 15 parameter vectors (5 parents, and 10 children) were used
 for optimization. Also note that the accuracy of the thermodynamic properties remains
 satisfactory (table~\ref{tbl:two-obj}).
 \subsection{Three objective functions (thermodynamic properties)}\label{subsec:three}
\begin{figure}
 \begin{center}
  \graphicspath{{figs/}}
  \includegraphics[width=3in]{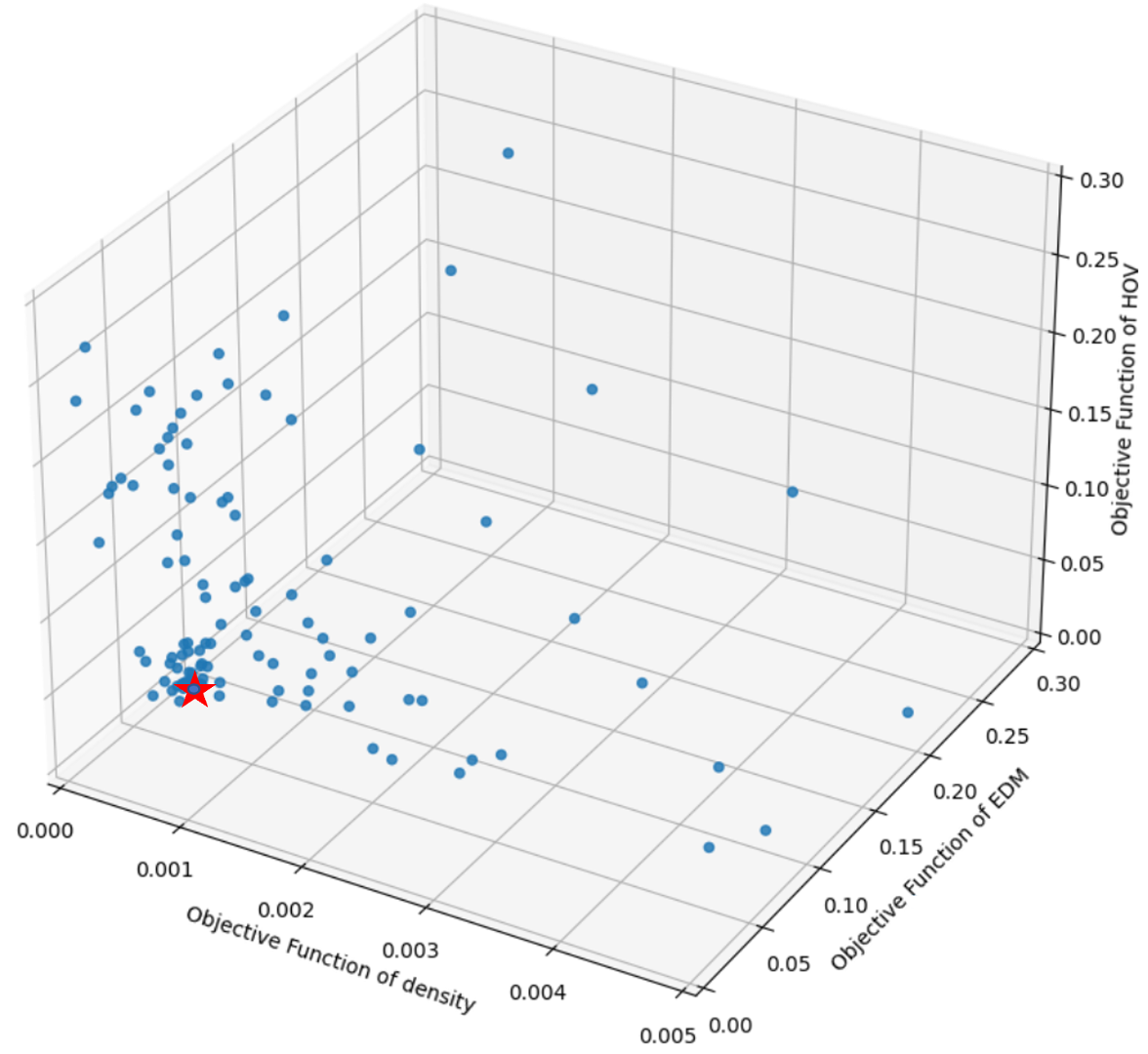}
 \end{center}
 \caption
  {Three-dimensional objective functional space for the cumulative populations during
   optimization of mass density, EDM, and HOV.
   The red color star marks the selected optimal solution with objective function values of
   \num{0.0001}, \num{0.0839}, and \num{0.0001}, respectively.
   The thermodynamic properties exhibit very low relative errors of \num{0.749}\%,
   \num{0.034}\%, and \num{0.035}\% compared to experimental data, while non-optimized
   SDC and shear viscosity show high relative errors (\num{-78.03}\% and
   \num{-40.56}\%).}
  \label{fig:three}
\end{figure}

 Similarly to the two-objective function investigation previously described, here we
 discuss MOO results from a three-objective optimization scenario, encompassing the
 thermodynamic properties, mass density, HOV, and EDM for TBP while
 leaving transport properties alone without any control.
 Following the approach of previous scenarios, the optimization process yields a
 comprehensive representation of results, incorporating the non-dominated set of solutions
 derived from the NSGA-II algorithm. To determine the optimal solution, we choose
 the solution in objective function space with smallest Euclidian distance from the origin
 (red star in fig.~\ref{fig:three}).
\par
 Utilizing the LJ parameters corresponding to this optimal solution
 (table~\ref{tbl:three-obj-optlj}), we derive the thermodynamic and transport properties
 (table~\ref{tbl:three-obj}). The resulting values for mass density, HOV, and EDM are
 \SI{0.9799}{\gram\per\centi\meter\cubed},
 \SI{3.243}{\debye}, and \SI{24.43}{\kilo\calorie\per\mol}, respectively. While these
 properties exhibit very low relative errors of \num{0.749}\%, \num{0.034}\%, and
 \num{0.035}\% compared to experimental data, the accuracy of SDC and viscosity
 is unsatisfactory given the high relative errors (\num{-78.03}\% and \num{-40.56}\%,
 respectively). Here again, 15 generations of populations of 15 parameter vectors
 (5 parents and 10 children) were used giving an optimal vector
 (table~\ref{tbl:three-obj-optlj}) that has been slightly corrected in every parameter
 component as compared to the initial average parent vector (table~\ref{tbl:ljref}).
\par
 This exercise emphasizes the importance of considering transport and thermodynamic 
 properties together during the optimization, prompting the exploration of a scenario 
 involving five objective functions in the subsequent analysis.
\begin{table}
 \caption[]
 {Optimized LJ parameters, $\vvec^*$, using three objective functions.
  $F_k, \ k=1,\ldots,3$, for mass density, EDM, and HOV, respectively.}
 \label{tbl:three-obj-optlj}
 \begin{center}
 \begin{tabular}{ccc}
 \toprule
 Atom & \multicolumn{2}{c}{$\vvec^*$ for $F_1$, $F_2$, $F_3$} \\
 Type &  $\sigma_i$ [\si{\angstrom}] & $\epsilon_i$ [\si{\calorie\per\mol}] \\
 \midrule
  O2  & 2.960 & 210   \\
  P   & 3.742 & 205   \\
  OS  & 2.832 & 170   \\
 \hline
  C0  & 3.358 & 67    \\
  C1  & 3.399 & 109   \\
  C2  & 3.368 & 109   \\
  C3  & 3.397 & 109   \\
 \hline
 C$_\text{ave}$ 
      & 3.381 & 98.5 \\
 \hline
  H0  & 2.650 & 16    \\
  H1  & 2.484 & 41    \\
  H2  & 2.646 & 29    \\
  H3  & 2.647 & 16    \\
 \hline
 H$_\text{ave}$ 
      & 2.607 & 25.5 \\
 \bottomrule
 \end{tabular}
 \end{center}
\end{table}
%
%
\begin{table}[pos=h]
 \setlength{\tabcolsep}{2.0pt}
 \caption []
 {Three objective functions NSGA-II optimization results for TBP properties.
  Red color for the properties being optimized with experimental values as in
  Table~\ref{tbl:one-obj}. While the optimization is effective to accurately predict all
  thermodynamic properties involved in the objective functions simultaneously, it does not
  improve the prediction of transport properties absent in the optimization method.}
 \label{tbl:three-obj}
 \begin{center}
 \begin{tabular}[c]{l|c}
 \thickhline
  Obj. Func. & $F_1$, $F_2$, $F_3$ \\
  \# of TBP & 48 \\
  \hline
  {Density [\si{\gram\per\centi\meter\cubed}]} & \red{ \num{0.9799 } (\num{0.74 }\%) } \\
  {EDM [\si{\debye}]} & \red{ \num{3.243  } (\num{0.034 }\%) } \\
  {HOV [\si{\kilo\calorie\per\mol}]} & \red{ \num{24.43 } (\num{0.035 }\%) } \\
  \hline
  {SDC [\SI{e-6}{\centi\meter\squared\per\second}]} & \num{0.503 } (\num{ -78.03 }\%) \\
  {Viscosity [\si{\milli\pascal\second}]} & \num{1.99 }  (\num{-40.56 }\%) \\
 \thickhline
  Overall deviation & \num{23.9}\%
 \end{tabular}
 \end{center}
\end{table}
%
 \subsection{Five objective functions (thermophysical properties)}\label{subsec:five}
 This section is carried out with a five-objective optimization approach utilizing both the
 NSGA-II and NSGA-III algorithms (as detailed in sec.~\ref{subsec:ndpfr}) using exclusively
 MD simulations in the loop. Simultaneously
 considering mass density, EDM, HOV, SDC, and shear viscosity calls for a compromise on
 the accuracy to which all these properties can be predicted.
\par
\begin{figure}
 \begin{center}
  \graphicspath{{figs/}}
  \includegraphics[width=3.3in]{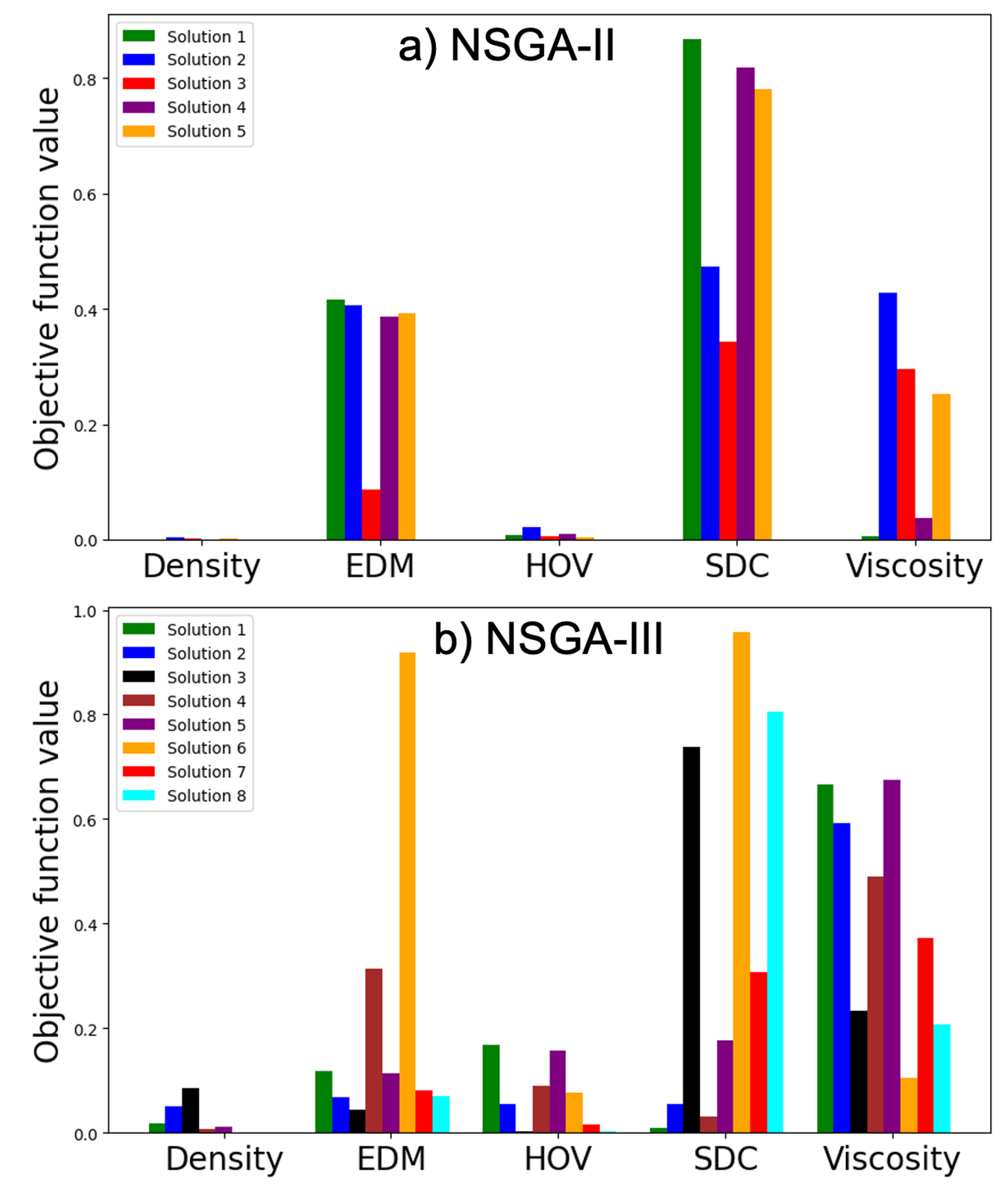}
 \end{center}
  \caption[]
  {Pareto front solutions obtained in the final generation ($g=15$) of different
   algorithms for a five-objective-function optimization: a) NSGA-II, b) NSGA-III using
    MD simulations (no neural network fit acceleration used).
   The red color solution provides the shortest Euclidean distance from the origin in the
   objective function space (fig.~\ref{fig:pareto-fronts}).}
  \label{fig:all}
\end{figure}

 A visual representation of Pareto front solutions obtained in the last
 generation by the NSGA-II (fig.\ref{fig:all}a) and NSGA-III (fig.\ref{fig:all}b)
 algorithms shows trade-offs inherent in various LJ parameter sets across five objective
 functions. None of the displayed solutions achieves exceptional accuracy by
 minimizing the objective function for all five properties simultaneously across both
 algorithms. Instead, a trade-off emerges, where enhancing one property may necessitate
 concessions in others. In this scenario, we highlight solutions by red bars
 (fig.\ref{fig:all}) as significant trade-off solutions identified through the NSGA-II
 and NSGA-III algorithms. The optimal solution Euclidean distances from the origin are
 the smallest found which indicates a balanced approach that aligns well with the
 optimization objectives.
\par
 The LJ parameters obtained from the selected solutions of both NSGA-II and NSGA-III
 (table~\ref{tbl:five-obj-optlj}) are utilized to calculate the thermodynamic and transport
 properties of the system comprising 48 TBP molecules (table~\ref{tbl:five-obj}). A
 notable compromise is represented by the solution from NSGA-II, where the relative
 deviation from experimental data for three thermodynamic properties is less than
 \num{\pm 10}\%, although SDC and viscosity values still exhibit a high relative error of
 \num{-58}\%, and \num{-54}\% compared to experiments. Likewise, the LJ
 parameters associated with the selected solution from NSGA-III predict SDC and
 shear viscosity with deviations of \num{-55.4}\%, and \num{-61}\% respectively, compared 
 to experimental values; similarly to NSGA-II, thermodynamic properties are predicted
 successfully with NSGA-III. It is evident that both NSGA-II and NSGA-III produced similar
 results  (table~\ref{tbl:five-obj}) and could not enhance the transport property
 predictions to the same degree as the thermodynamic properties were predicted.
 In addition, the previously identified contention of varying the interaction potential
 well parameters for $\texttt{C}_\text{ave}$ and $\texttt{H}_\text{ave}$ in opposite
 directions is present for this case (compare to table~\ref{tbl:two-obj-optlj}).
\begin{table}
\caption[]
 {Optimized LJ parameters, $\vvec^*$, using five objective functions.
 $F_k,\ k=1,\ldots, 5$, for mass density, EDM, HOV, SDC, and
 shear viscosity, respectively. Here two MD NSGA results are provided.}
 \label{tbl:five-obj-optlj}
 \begin{center}
 \begin{tabular}{ccccc}
 \toprule
      & \multicolumn{4}{c}{$\vvec^*$ for $F_1,\ldots,F_5$} \\
 Atom & \multicolumn{2}{c}{NSGA-II} & \multicolumn{2}{c}{NSGA-III} \\
 Type & $\sigma_i$ [\si{\angstrom}] & $\epsilon_i$ [\si{\calorie\per\mol}] & $\sigma_i$ & $\epsilon_{i}$ \\
 \midrule
  O2 & 2.963 & 211    & 2.938 & 206   \\
  P  & 3.742 & 200    & 4.000 & 200   \\
  OS & 3.000 & 170    & 2.846 & 140   \\
 \hline
  C0 & 3.500 & 65     & 3.500 & 65    \\
  C1 & 3.221 & 283    & 3.500 & 65    \\
  C2 & 3.576 & 65     & 3.500 & 65    \\
  C3 & 3.388 & 65     & 3.500 & 68    \\
 \hline
  C$_\text{ave}$ & 3.421 & 119.5 ($\uparrow$)  & 3.500 & 65.8 ($\downarrow$) \\
 \hline
  H0 & 2.498 & 30     & 2.500 & 30    \\
  H1 & 2.504 & 16     & 2.167 & 30    \\
  H2 & 2.498 & 16     & 2.545 & 31    \\
  H3 & 2.500 & 4      & 2.49  & 16    \\
 \hline
  H$_\text{ave}$ & 2.500 & 16.5 ($\downarrow$) & 2.426 & 26.8 ($\uparrow$)  \\
 \bottomrule
 \multicolumn{5}{l}{Arrows indicate $\ge 20\%$ change relative to table~\ref{tbl:ljref}.}
 \end{tabular}
 \end{center}
\end{table}
\begin{table}[pos=h]
 \setlength{\tabcolsep}{3.0pt}
 \setlength{\extrarowheight}{1.5pt}
 \caption []
 {Five objective functions NSGA-II and NSGA-III optimization results for TBP properties.
  Red color for the properties being optimized with experimental values as in
  Table~\ref{tbl:one-obj}. While optimization is effective to accurately predict
  thermodynamics properties, it is lacking in improving the prediction of
  transport properties.}
 \label{tbl:five-obj}
 \begin{center}
 \scriptsize
 \begin{tabular}[c]{l|ccc}
 \thickhline
 \rule[-1.2ex]{0pt}{1.2ex}
  $F_1,\ldots,F_5$ & NSGA-II & NSGA-III & MD with $\vvec^*_\text{NSGA-III}$ \\
 \hline
  \# of TBP & 48 & 48 & 512 \\
 \hline
  {Density [\si{\gram\per\centi\meter\cubed}]} & \red{ \num{1.022} (\num{5}\%) } &
                                                 \red{ \num{0.9875} (\num{1.5}\%) } &
                                                 \red{ \num{0.9881} (\num{1.6}\%) } \\
  {EDM [\si{\debye}]} & \red{ \num{3.14} (\num{-2.9}\%) } &
                        \red{ \num{3.23} (\num{-0.36}\%) } &
                        \red{ \num{3.20} (\num{-1.3}\%) } \\
  {HOV [\si{\kilo\calorie\per\mol}]} &
                         \red{ \num{22.55} (\num{-7.6}\%) } &
                         \red{ \num{21.29 } (\num{-12.8 }\%) } &
                         \red{ \num{21.16 } (\num{-13.3 }\%) }  \\
  \hline
  {SDC [\SI{e-6}{\centi\meter\squared\per\second}]} &
                         \red{  \num{0.953} (\num{-58}\%)  } &
                         \red{  \num{1.02 } (\num{-55.4 }\%)  } &
                         \red{  \num{1.19 } (\num{-47.9 }\%)  } \\
  {Viscosity [\si{\milli\pascal\second}]} &
                         \red{  \num{1.53} (\num{-54}\%) } &
                         \red{  \num{1.31 } (\num{-61.1 }\%) } &
                         \red{  \num{1.34 } (\num{-59.9 }\%) } \\
 \thickhline
     Overall deviation & \num{25.5}\% & \num{26.2}\% &  \num{24.8}\%
 \end{tabular}
 \end{center}
\end{table}
\par
 The performance of both algorithms used in this section in finding LJ parameters that
 accurately replicate the values of all five thermophysical properties over \num{15}
 generations of optimization falls short of expectations. However, there are positives:
 first, thermodynanic properties are well predicted, second prediction of transport
 properties do not worsen as new generations are created, and last, the number of
 generations and size of the population of parameter vectors used was small. The latter
 suggests that larger simulations could allow for a better probing of the objective
 function space and potentially better results for transport properties. 
\par
 Hence at this
 point this exercise is encouraging since much larger computational resources will become
 available and accessible in the future (possibly as early as the time this work is
 published). Another positive is that the scaling of the number of TBP
 molecules used in the optimization from 48 to 512 (table~\ref{tbl:five-obj} rightmost
 column) shows that the results with 48 molecules are practically unchanged when a much
 larger MD simulation is performed with 512 molecules. This shows that a parallel
 optimization implementation of the algorithm proposed here (fig.~\ref{fig:ga-oploop})
 could unlock the current size restriction on population and generation
 sizes even with the use of the MD simulation in the optimization loop.
\par
 Finally, much has been developed recently on high-dimensional data fitting using neural
 networks. The next section (sec.~\ref{subsec:NN}) describes details on using the data
 accumulated in the optimization loop to train a neural network to predict thermophysical
 properties and potentially replace an MD simulation for some population vectors,
 thereby accelerating the simulation.
\subsubsection{Neural Network Property Fit}\label{subsec:NN}
 To this point the neural network (NN) property mapping (sec.~\ref{subsec:methodNN})
 has not been used to accelerate and improve the genetic algorithm we proposed
 (algo.~\ref{alg:evolve}, fig.~\ref{fig:ga-oploop}). Here we describe the implementation
 and results of using a thermophysical property data fitting neural network in the
 optimization loop with the NSGA-III algorithm, \emph{i.e.} no additional MD simulations
 were performed past this point and the optimization algorithm relied solely on the
 $\nnmap$ mapping.
\par
\begin{figure}
 \begin{center}
  \graphicspath{{figs/}}
  \includegraphics[width=3.3in]{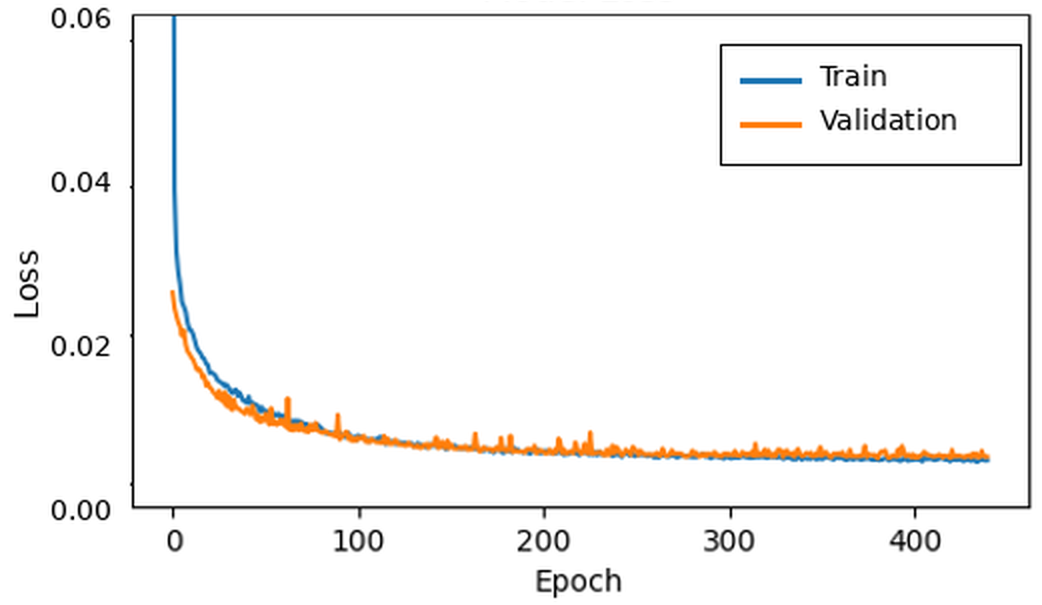}
 \end{center}
 \caption 
  []
  {MSE loss error~\eqref{eqn:MSE} as a function of epoch for the training (blue) and 
   validation (orange) datasets.}
  \label{fig:loss}
\end{figure}

 To construct the NN property map (sec~\ref{subsec:methodNN}) its coefficients and weights
 need to be computed (a process called training) based on a training set that included
 1143 MD simulations. To evaluate the quality of the NN mapping during training, the MSE
 loss function \eqref{eqn:MSE} (fig.~\ref{fig:loss}) of the training and validation
 datasets were scrutinized. The training and validation results showed similar patterns,
 that is, loss decreased as the model improved, reaching a minimum loss error at around
 \num{440} epochs.
\par
\begin{figure*}
 \begin{center}
  \graphicspath{{figs/}}
  \includegraphics[width=5in]{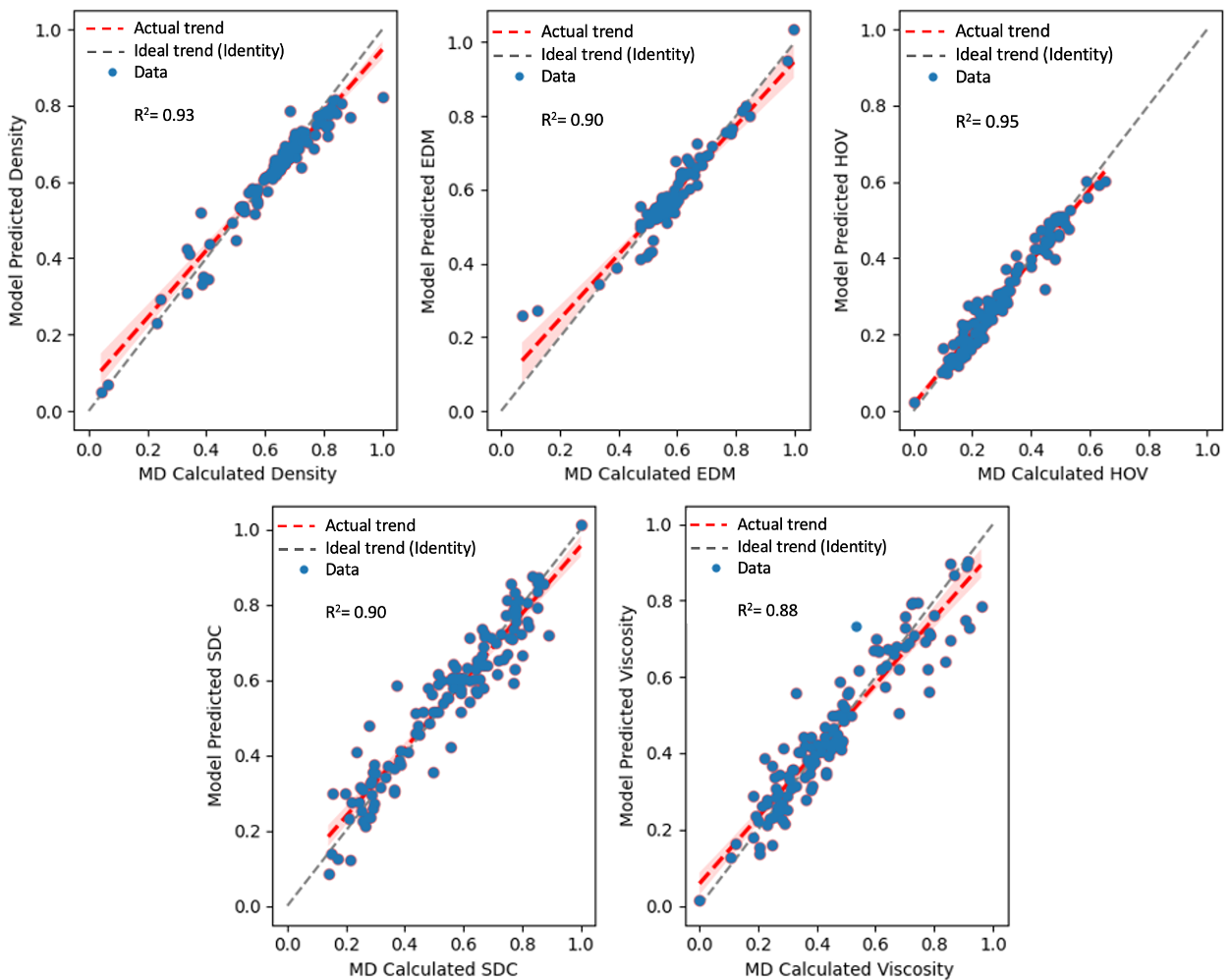}
 \end{center}
 \caption 
 []
 {Data regression of the NN mapping predicted vs. MD-calculated values of TBP
  properties for the validation dataset with \num{127} values for each property.
  The shaded red region marks the \num{95}\% confidence interval around the regression
  line.
  }
  \label{fig:regression}
\end{figure*}

 Data regression was made (fig.~\ref{fig:regression}) comparing the calibrated NN mapping
 prediction of TBP properties on \num{127} randomly chosen MD simulations that
 were held out as \num{10}\% of the total dataset for validation. The agreement of NN
 predicted properties versus the MD simulations is in general very good for all
 thermodynamic properties, and
 reasonably good for transport properties where shear viscosity is the least accurate.
 In fact, it can be seen that both transport properties are noticeably more difficult
 to fit with a neural network model than thermodynamic properties are; this could be
 potentially addressed in another study using a deeper neural network.
 The crossover point of under-prediction and over-prediction regions generally occurs
 at the middle of the dataset with the exception of HOV which happens to have the best
 $R^2$ value, hence best clustering around the identity line.
\par
 With $\nnmap$ constructed, it can be used in the parameter optimization loop in lieu of
 MD simulations. Choosing the number of divisions along each objective axis $\delta=2$
 \eqref{eqn:binomial}, and a parent population size $N_\text{p}=15$, offspring size
 $N_\text{c}=200$, and $N_\text{g}=1000$ generations (maximum), as basic parameters,
 we present the outcome of integrating the trained NN mapping with the NSGA-III algorithm
 for optimizing force field parameters and predicting thermophysical properties. We do
 not perform any additional MD simulation while running the genetic algorithm
 optimization with $\nnmap$ in the loop.
\par
 As presented earlier, the optimization simultaneously considers five objective functions
 associated to the thermophysical properties: mass density, EDM, HOV, SDC, and shear
 viscosity. The entire optimization process, with the aforementioned parameters
 (\emph{i.e.} much larger population and generations than previously used in the other
 sections) was completed in approximately \num{5} hours of wall-clock time computing. We
 established a tolerance of \num{0.001} for changes in the objective function values. If
 the alteration in these values remained below this threshold for \num{50} consecutive
 generations, the algorithm stopped. In our case, the optimization process terminated
 after \num{503} generations, as the objective function values exhibited a change of less
 than \num{0.001} over the preceding \num{50} generations. We note here that MD
 simulations and the NN mapping runtimes are not to be compared. Without MD simulation
 data, a NN mapping cannot be constructed, therefore the cost of constructing $\nnmap$ is
 as high as the MD simulations. However, the cost of building the mapping can be amortized
 by replacing MD simulations with $\nnmap$ evaluations in the optimization loop
 particularly when larger populations and generations of parameter vectors are used.
\par
\begin{figure}
 \begin{center}
  \graphicspath{{figs/}}
  \includegraphics[width=3.3in]{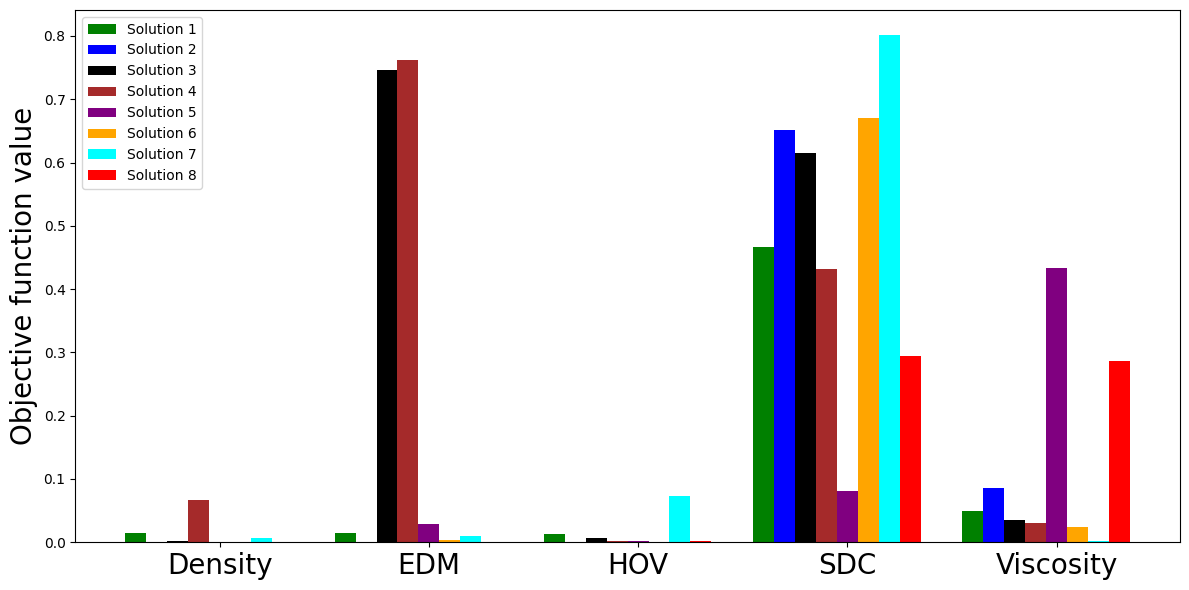}
 \end{center}
  \caption[]
   {Pareto front solutions obtained in the final generation ($g=503$) of the NN NSGA-III
    algorithm. The solution (table~\ref{tbl:nn-five-obj-optlj}) marked with red bars, has
    the shortest Euclidean distance from the origin of the objective function space
    (fig.~\ref{fig:pareto-fronts}).}
  \label{fig:five-nn}
\end{figure}

 To gain visual insight into the solution options, the Pareto front solutions computed in
 the final generation by the NSGA-III algorithms (fig.\ref{fig:five-nn}) using the NN
 mapping, illuminate the quality and trade-offs inherent in various LJ parameter sets
 across the five objectives. The optimized parameter vector
 (table~\ref{tbl:nn-five-obj-optlj}) produces the objective function values of
 \num{8.78e-06}, \num{3.44e-05}, \num{2.95e-02}, \num{0.58}, and \num{0.16}
 (fig.~\ref{fig:five-nn}, red bars). The LJ parameters of the optimized solution have been
 varied much more than in previous cases when using the NN NSGA-III optimization algorithm
 (table~\ref{tbl:nn-five-obj-optlj}) which was an expected outcome after 503 generations
 with a population of 215 parameter vectors. This variation can be readily seen when
 comparing to the results obtained when using the MD NSGA-III
 (table~\ref{tbl:five-obj-optlj}) algorithm. Despite the differences, when using the NN
 mapping, the optimal solution predicts thermophysical property values:
 \SI{0.9718}{\gram\per\centi\meter\cubed} (\num{-0.1}\%),
 \SI{3.26}{\debye} (\num{0.5}\%), 
 \SI{23.2}{\kilo\calorie\per\mol} (\num{-4.9}\%),
 \SI{1.0e-6}{\centi\meter\squared\per\second} (\num{-54}\%) and
 \SI{1.56}{\milli\pascal\second} (\num{-53}\%) for density, EDM, HOV, SDC,
 and viscosity, respectively (table~\ref{tbl:nn-five-obj}, first column) so that they
 are similar to those presented by the MD NSGA-III (table~\ref{tbl:five-obj},
 second column) counterpart; in fact both solutions are on the Pareto front.
\begin{table}
 \caption[]
  {Optimized LJ parameters, $\vvec^*$, using five objective functions.
   $F_k,\ k=1,\ldots,5$,
   for mass densith, EDM, HOV, SDC, and shear viscosity, respectively.
   Here MD simulations were replaced by the NN data fit.}
 \label{tbl:nn-five-obj-optlj}
 \begin{center}
 \begin{tabular}{ccc}
 \toprule
      & \multicolumn{2}{c}{$F_1(\vvec^*),\ldots,F_5(\vvec^*)$} \\
 Atom & \multicolumn{2}{c}{ NN NSGA-III} \\
 Type &  $\sigma_i$ [\si{\angstrom}] & $\epsilon_i$ [\si{\calorie\per\mol}] \\
 \midrule
  O2  & 2.807                & 78 ($\downarrow$)  \\
  P   & 3.360                & 198   \\
  OS  & 2.135 ($\downarrow$) & 267 ($\uparrow$)   \\
  \hline
  C0  & 2.850 & 35    \\
  C1  & 3.921 & 97    \\
  C2  & 4.280 & 95    \\
  C3  & 2.502 & 152   \\
  \hline
  C$_\text{ave}$ & 3.388 & 94.8 \\
  \hline
  H0  & 2.234 & 4     \\
  H1  & 2.202 & 3     \\
  H2  & 2.151 & 45    \\
  H3  & 2.705 & 56    \\
  \hline
  H$_\text{ave}$ & 2.323 & 27 ($\uparrow$)   \\
 \bottomrule
 \multicolumn{3}{l}{Arrows indicate $\ge 20\%$ change relative to table~\ref{tbl:ljref}.}
 \end{tabular}
 \end{center}
\end{table}
%
%
\begin{table}[pos=h]
 \small
 \setlength{\tabcolsep}{3.0pt}
 \caption []
 {Five objective functions NSGA-III optimization results for TBP properties. Red color for
  the properties being optimized with experimental values as in Table~\ref{tbl:one-obj}.
  While optimization is effective to accurately predict thermodynamics properties, it is
  lacking in improving the prediction of transport properties.}
 \label{tbl:nn-five-obj}
 \begin{center}
 \begin{tabular}[c]{lcc}
 \thickhline
 $F_1,\ldots,F_5$ & NN NSGA-III & MD NSGA-III \\
 \hline
 \# of TBP & --- & 48 \\
 \hline
 {Density [\si{\gram\per\centi\meter\cubed}]} & \red{ \num{0.9718 } (\num{-0.1}\%) }  &
                                                \red{ \num{ 0.99 } (\num{1.6 }\%) }  \\
 {EDM [\si{\debye}]} & \red{ \num{3.26} (\num{0.5 }\%) }  &
                       \red{  \num{3.68 } (\num{13.5}\%)  } \\
 {HOV [\si{\kilo\calorie\per\mol}]} & \red{ \num{23.2 } (\num{-4.9 }\%) }  &
                                      \red{ \num{21.87 } (\num{-10.4 }\%) } \\
 \hline
 {SDC [\SI{e-6}{\centi\meter\squared\per\second}]} & \red{ \num{1.0 } (\num{-54.2 }\%) } &
                                                     \red{ \num{1.48 } (\num{-35.3 }\%) } \\
 {Viscosity [\si{\milli\pascal\second}]} & \red{ \num{1.56 } (\num{-53.4 }\%) } &
                                           \red{  \num{1.58 } (\num{-53.0 }\%) } \\
 \thickhline
     Overall deviation & \num{22.6}\% & \num{22.8}\%
 \end{tabular}
 \end{center}
\end{table}
\par
 To further verify the consistency of the NN NSGA-III optimal solution
 (table~\ref{tbl:nn-five-obj-optlj}), we performed an MD simulation using the 
 optimal LJ parameters and \num{48} molecules, and compared the obtained thermophysical
 properties (table~\ref{tbl:nn-five-obj}).
 MD simulation using these optimal LJ parameters produced the mass density of 
 \SI{0.99}{\gram\per\centi\meter\cubed} (deviation of
 \num{1.6}\%), EDM of  \SI{3.68}{\debye} (deviation of \num{13.5}\%),
 HOV of \SI{21.87}{\kilo\calorie\per\mol} (deviation of \num{-10.4}\%),
 SDC of \SI{1.48e-6}{\centi\meter\squared\per\second} (deviation of \num{-35.3}\%) and
 viscosity of \SI{1.58}{\milli\pascal\second} (deviation of \num{-53}\%).
\par
 The deviations between MD results and NN predictions (Table~\ref{tbl:nn-five-obj}) show
 relative differences of \num{1.8}\%, \num{11}\%, \num{5}\%, \num{32}\% and \num{1.2}\%
 for density, EDM, HOV, SDC and viscosity, respectively. From the fact that the NN mapping
 is less accurate for transport properties than thermodynamic ones, these differences are
 not surprising except for the shear viscosity which is more accurate than expected; a
 lucky break. The deviations on thermodynamic properties are reasonable in absolute terms.
 This exercise shows that there is much promise in using a fit of the MD simulation data
 to allow a much larger density of parameter vectors to be probed in the objective
 function space optimization. This improves the chance of finding Pareto solutions closer
 to the origin if any exists. Therefore this initial effort points to two directions of
 improvements. First the recommendation \eqref{eqn:Hrecommended} calls for an increased
 value of $\delta\approx 6$ as compared with what was initially used here. This new value
 will spread the population of LJ parameter vectors more uniformly over the Pareto
 hyper-surface (fig.~\ref{fig:ref-pts}). Second, the neural network mapping, $\nnmap$,
 needs to be interwoven with MD simulations to bring new information in the optimization
 loop. This is a more intricate integration to be pursued in future research.
%

%
%
%
\section{Conclusions and outlook}
\label{sec:co}
 This study investigated the MOO of LJ parameters for the TBP molecule using a genetic
 algorithm approach. It integrated direct MD simulations into the optimization loop to
 compute five thermophysical properties: liquid mass density, EDM, HOV, SDC, and
 shear viscosity, and compared the results to experimentally measured values.
\par
 Computed LJ parameters of single-objective optimization for each TBP property can predict
 results with less than \SI{2}{\percent} relative error compared to experimental values
 (table~\ref{tbl:one-obj-optlj}). However optimized LJ parameters for one property are not
 optimized for another (sec.~\ref{subsec:single}). Hence, the best LJ parameters,
 \emph{i.e.}, Pareto-optimal parameter vector \eqref{eqn:minproblem2}, were computed
 developing an optimization loop algorithm (fig.~\ref{fig:ga-oploop}) for the numerical
 approximation of the Pareto-optimal set of vectors (fig.~\ref{fig:pa-sol}). Any
 Pareto-optimal vector represents a compromise for how close groups of thermophysical
 properties can be made to their experimental values.
\par
 We systematically optimized LJ parameter vectors to predict different groups of
 thermophysical properties. An LJ paramater vector can be found
 (table~\ref{tbl:three-obj-optlj}) such that thermodynamic properties are exceedingly
 accurate ($<$ \SI{1}{\percent} relative error; table~\ref{tbl:three-obj}). However
 transport properties fell short of being predicted accurately since the LJ parameters
 were not optimized for that purpose.
 Conversely, no LJ parameter vector could be found that predicted transport properties
 accurately (table~\ref{tbl:two-obj}) since the parameters that improved the prediction of
 SDC the most, were also the same that made shear viscosity worse
 (table~\ref{tbl:two-obj-optlj}, fig.~\ref{fig:heatmap}). Finally the optimized LJ
 parameters for the combined group of all thermophysical properties using multiple
 algorithms with either MD simulations in the loop or an alternative neural network
 mapping could not find parameters that delivered small deviations in all property
 predictions
 as compared to the experimental values (tables~\ref{tbl:five-obj}, \ref{tbl:nn-five-obj}).
\par
 Nevertheless the optimal solution we computed (table~\ref{tbl:nn-five-obj-optlj}) predicts
 thermophysical properties with an overall deviation from experiments of \num{22.8}\%
 which is a significant improvement over the best parameters found in our previous study
 \cite{hatami-de_almeida25:art}, P AMBER-MNDO with \num{74}\% overall relative deviation.
 While this is very encouraging, we do not recommend this set of LJ parameters 
 (table~\ref{tbl:nn-five-obj-optlj}) to be adopted in general because the optimization of
 force fields should be performed with all parameters involved. Notably, we do not change
 the bonded potential or partial electric charge parameters in this study. In addition, as
 previously explained the bounds on the LJ parameters used here were essentially unlimited
 for the sake of experimenting with the optimization loop algorithm. In the future,
 rigorous bounds on the values of parameter should be observed.
\par
 The above-mentioned results need to be considered in light of the resources available for
 the work. While this was an intense initial effort to develop an algorithm for LJ
 parameter optimization based on minimizing the error of predicted thermophysical
 properties relative to experimentally measured quantities, the population of parameter
 vectors and the number of generations in the non-dominated search genetic algorithm
 developed cannot be considered sufficient to probe the LJ parameter space. Therefore
 we conjecture that access to larger computing power to increase the population and
 generations in the genetic algorithms could improve the results; in particular when using
 parallel MD simulations for the population of LJ vectors. Note that we did perform
 parallel MD simulations with one given LJ parameter vector. However, there remains to be
 explored
 simultaneous MD simulations for many LJ parameter vectors at once. This larger simulation
 would also create additional data for building a larger input for the neural network
 mapping of thermophysical properties which could further improve the final outcome of
 optimized LJ parameters.
\par
 Finally, it is possible that the overall deviation of thermophysical property predictions
 relative to experimental values based on LJ parameter optimization can only be minimized
 to a certain point because other force field parameters need to be included. Nevertheless,
 it is instructive to find this minimum while improving the algorithms discussed here
 before scaling up to consider a full set of parameters.
%



%
\section*{Acknowledgments}
 This work was supported by the University of Massachusetts Lowell, Francis College
 of Engineering. The written analysis, documentation, and submission/revision were
 supported by Cortix Tech,
 Lowell, MA. Computational support was provided by the Unity cluster at the University
 of Massachusetts Amherst.








 

\bibliographystyle{cas-model2-names}
\bibliography{bibliography} 

%


%
%
%
%
%
%
%
%
\end{document}